\documentclass[conference]{IEEEtran}
\ifCLASSINFOpdf
\else
\fi
\usepackage{amsmath,amsfonts}
\usepackage{textcomp}
\usepackage{stfloats}
\usepackage{url}
\usepackage{verbatim}
\usepackage[normalem]{ulem} 

\usepackage{hyperref}  
\hypersetup{
    colorlinks=true,    
    linkcolor=black,    
    citecolor=black,    
    urlcolor=black,     
    filecolor=black,    
}
\usepackage{stmaryrd}  
\usepackage{graphicx}  
\usepackage{caption}
\usepackage{subcaption}
\usepackage{titlesec}  
\usepackage{soul}  
\usepackage{diagbox}
\usepackage{booktabs}  
\usepackage{makecell}  
\usepackage{multirow}  
\usepackage{xcolor}    
\usepackage{threeparttable} 
\usepackage[linesnumbered,ruled,vlined]{algorithm2e}  
\usepackage{enumitem}
\usepackage{colortbl}  
\usepackage{hhline}  
\setlist{nosep}  
\SetAlgoNlRelativeSize{-1}
\newcommand{\add}[1]{{\leavevmode\color{black}#1}}
\newcommand{\discuss}[1]{{\leavevmode\color{black}#1}}

\begin{document}
\title{VMask: Tunable Label Privacy Protection for Vertical Federated Learning via Layer Masking}
\author{
    Juntao Tan\textsuperscript{1}, Lan Zhang\textsuperscript{1}, Zhonghao Hu\textsuperscript{2}, Kai Yang\textsuperscript{3}, Peng Ran\textsuperscript{3}, and Bo Li\textsuperscript{4} \\
    \textsuperscript{1}University of Science and Technology of China, Hefei, China \\
    \textsuperscript{2}Key Laboratory of Internet and Industrial Integration and Innovation, CAICT, MIIT, Beijing, China \\
    \textsuperscript{3}Research Institute of Safety Technology, China Mobile Research Institute, Beijing, China \\
    \textsuperscript{4}Hong Kong University of Science and Technology, HongKong \\
    \texttt{tjt@mail.ustc.edu.cn, zhanglan@ustc.edu.cn, huzhonghao@caict.ac.cn} \\
    \texttt{\{yangkaiyj, ranpeng\}@chinamobile.com, bli@ust.hk} \\
}

\IEEEoverridecommandlockouts
\makeatletter\def\@IEEEpubidpullup{6.5\baselineskip}\makeatother
\IEEEpubid{\parbox{\columnwidth}{
		Network and Distributed System Security (NDSS) Symposium 2025\\
		24-28 February 2025, San Diego, CA, USA\\
		ISBN 979-8-9894372-8-3\\
		https://dx.doi.org/10.14722/ndss.2025.23xxxx\\
		www.ndss-symposium.org
}
\hspace{\columnsep}\makebox[\columnwidth]{}}

\maketitle

\begin{abstract}
Though vertical federated learning (VFL) is generally considered to be privacy-preserving, recent studies have shown that VFL system is vulnerable to label inference attacks originating from various attack surfaces.
Among these attacks, the model completion (MC) attack is  currently the most powerful one.
Existing defense methods against it either sacrifice model accuracy or incur impractical computational overhead.
In this paper, we propose VMask, a novel label privacy protection framework designed to defend against MC attack from the perspective of layer masking.
Our key insight is to disrupt the strong correlation between input data and intermediate outputs by applying the secret sharing (SS) technique to mask layer parameters in the attacker's model. 
We devise a strategy for selecting critical layers to mask, reducing the overhead that would arise from naively applying SS to the entire model.
Moreover, VMask is the first framework to offer a tunable privacy budget to defenders, allowing for flexible control over the levels of label privacy according to actual requirements. 
We built a VFL system, implemented VMask on it, and extensively evaluated it using five model architectures and 13 datasets with different modalities, comparing it to 12 other defense methods.
The results demonstrate that VMask achieves the best privacy-utility trade-off, successfully thwarting the MC attack (reducing the label inference accuracy to a random guessing level) while preserving model performance (e.g., in Transformer-based model, the averaged drop of VFL model accuracy is only 0.09\%).
VMask's runtime is up to 60,846 times faster than cryptography-based methods, and it only marginally exceeds that of standard VFL by 1.8 times in a large Transformer-based model, which is generally acceptable.

\end{abstract}


\IEEEpeerreviewmaketitle

%

\section{Introduction}

Vertical federated learning (VFL) is a decentralized learning paradigm that enables multiple participants to collaboratively train a global machine learning (ML) model without compromising their private datasets \cite{yang2019federated,liu2022vertical,li2023vertical,yang2023survey,wei2022vertical,khan2022vertical}. 
In VFL, different participants share the same sample ID space, yet each participant owns a subset of features from the total feature space. 
The only participant holding the labels is typically referred to as the \textit{active party}, while the participant that holds features is referred to as the \textit{passive party}. 
VFL has been successfully applied to various services such as disease prediction, financial risk control, and online advertising ~\cite{yang2019federated}.

VFL is generally regarded as privacy-preserving since  private datasets are kept locally, and only intermediate information, i.e., feature embeddings and their gradients, is communicated among participants.
However, recent studies have shown that VFL systems are susceptible to label inference attacks, which can originate from various attack surfaces, including feature embeddings \cite{li2022label,sun2022label}, model parameters \cite{fu2022label}, and model gradients \cite{zou2022label}, each exploiting distinct vulnerabilities.
Among these, the model completion (MC) attack \cite{fu2022label} is currently the most powerful and severe \cite{kang2022framework}.
It achieves the highest attack accuracy and poses a significant threat to VFL, as label information is a critical and highly sensitive asset (e.g., a user's credit score in financial risk control or whether a user has a disease in disease prediction), and its disclosure can severely impede the application of VFL in real-world scenarios.
While attacks from other surfaces can be effectively mitigated by existing defenses \cite{li2022label,sun2022label,kang2022framework}, practical and effective solutions to counter MC attack is still lacking \cite{kang2022framework}.
This paper addresses this gap by focusing on the urgent need for effective protection method against MC attack.




Recently, several studies have introduced various approaches to defending against the MC attack.
Fu et al. \cite{fu2022label} propose perturbing the backpropagated gradients to reduce the information in it. 
Zou et al. \cite{zou2022label} and Ghazi et al. \cite{ghazi2021labeldp} employ autoencoder and differential privacy techniques to enhance label protection.
Zou et al. \cite{zou2023mutual}, Sun et al. \cite{sun2022label}, and Zheng et al. \cite{zheng2022making} propose adding an extra regularization term to the  VFL training objective to reduce the correlation between intermediate information and private labels. 
In addition to these ML-based defense methods, other works \cite{zhang2020additively,wang2023beyond,zhou2022toward,cai2022secure,fu2022blindfl} aim to provide label protection from a cryptographic perspective.
These studies typically design protocols that use homomorphic encryption (HE) to encrypt exchanged information or secure multi-party computation (MPC) to securely train VFL models.

However, existing defense methods against MC attack fail to be practical due to three key challenges.
\textit{C1: Suboptimal trade-off between model utility and label privacy.} 
The design of ML-based defense methods typically involves a hyperparameter that balances model utility and data privacy. 
These methods often either achieve good model performance or strong data privacy, but rarely both simultaneously. 
\textit{C2: Impractical running time.} 
Cryptography-based defense methods rely on computation-intensive techniques, such as HE, in training protocols. 
While these methods provide rigorous data privacy protection, their prohibitively high computational costs pose significant barriers to practical deployment.
\textit{C3: Limited control over the label privacy level.}
Both cryptography-based and ML-based defense methods lack mechanisms that allow the defender (active party) to determine or adjust the level of label privacy based on  actual model accuracy and resource requirements.

Therefore, this study aims to address the pressing need for a practical defense against MC attack by achieving high model utility, tunable levels of label privacy, and low computational cost.
To achieve this ambitious goal, we first investigate the root cause of label leakage in MC attack. 
Our findings reveal that, following typical VFL training, the attacker's model becomes proficient at extracting representative feature embeddings that strongly predict  private labels. 
Such a well-trained model can be easily fine-tuned to achieve high attack accuracy, even with limited samples.
Building on these findings, we propose a layer masking approach to randomize layer parameters within the attacker's model.
This disrupts the strong correlation between input data and feature embeddings, effectively preventing attackers from obtaining representative feature embeddings and thereby reducing attack accuracy.
Based on this idea, we develop our design to address the  aforementioned challenges as follows:
\begin{itemize}
    \item 
    We employ secret sharing (SS) to mask (i.e., randomize) the linear layer parameters within the attacker's model. By leveraging SS, we preserve model performance (as SS maintains the integrity of linear computations) while effectively reducing attack accuracy (as the extracted feature embeddings are randomized), thereby addressing C1.

    \item Our observations indicate that not all layers equally contribute to label inference accuracy. To mitigate the overhead of naively applying SS across the entire model, we design a layer selection strategy that targets a subset of critical layers.
    By selectively applying SS to mask these critical layers, we significantly reduce attack accuracy while maintaining computational efficiency, thereby addressing C2.

    \item For C3, to enable the defender to flexibly control the level of label privacy, we design our approach to support a tunable privacy budget, which sets an upper limit on the permissible extent of label leakage.
    This control is implemented through a shadow model, structured identically to the attacker's model. 
    The shadow model is trained synchronously with the VFL model using a small auxiliary dataset.
    Throughout training, the defender can locally simulate the MC attack on the shadow model at each epoch to estimate the extent of label leakage. 
    Based on this estimation, the defender can selectively mask the most critical layers until the observed leakage falls below the privacy budget.
    This strategy ensures that label leakage does not exceed the defined privacy budget while incurring minimal additional overhead.
\end{itemize}

Putting it all together, in this study, we design a novel framework, \textbf{VMask}, to protect label privacy in \textbf{\underline{V}}FL from the perspective of layer \textbf{\underline{Mask}}ing. 
VMask can be easily integrated into a typical VFL system by incorporating two modules—layer masking and layer selection—and two additional procedures—secure model update and shadow model update.
VMask operates as follows: 
Each training epoch begins with the layer masking module, which masks critical layers identified by the layer selection module.
The secure model update procedure then trains the masked layers using SS while training the remaining layers in plaintext. 
After the VFL model is updated, the shadow model update procedure trains the shadow model with an auxiliary dataset.
Next, the MC attack is simulated on the updated shadow model to assess label leakage. 
Layers in shadow model with larger accumulated gradient norms over the training history are gradually selected for masking until the estimated leakage falls below the defined privacy budget. 
These selected layers are then passed to the layer masking module for next epoch. 
This iterative process continues until the VFL model converges.





We develop a VFL system on a cluster of Linux servers and implement VMask on it.
We comprehensively evaluate VMask using five model architectures—MLP3, LeNet5, VGG13, ResNet18, and Transformer-based DistilBERT (rarely explored in previous works)—on 13 datasets with different modalities (i.e., tabular, image, and text), and compare VMask with 12 other defense methods in terms of effectiveness and efficiency.
The results show that VMask consistently achieves the best privacy-utility trade-off, with an average model accuracy drop of at most 0.34\% in all cases (e.g., in DistilBERT, the drop is only 0.09\%), while reducing attack accuracy to a random level, similar to that achieved by a randomized attacker's model, an inevitable leakage in any VFL system.
Additionally, VMask consistently controls label leakage within the defined privacy budget across different privacy levels.
In terms of efficiency, VMask significantly outperforms cryptography-based approaches. 
For example, in the VGG13 case, VMask runs up to 60,846 times faster than HE-based methods. 
While VMask has a slightly higher running time than standard VFL, the increase is marginal and generally acceptable.
For instance, even for a large model like DistilBERT (67M parameters), VMask's running time is only 1.8 times higher than that of standard VFL.


\textcolor{red}{
}

\vspace{-12pt}
Our contributions are summarized as follows:
\begin{itemize}
    
    

    \item We propose a novel framework, VMask, that utilizes secret sharing (SS) to mask layer parameters, effectively defending against the MC attack while preserving model performance.
    Additionally, we devise a strategy to select critical layers for masking, reducing the overhead associated with SS.

    \item VMask introduces the first framework to support a tunable privacy budget for VFL, enabling defenders to flexibly adjust the level of label privacy with minimal overhead. 
    This design allows defenders to achieve an acceptable level of label privacy when resources are limited and cannot afford the expensive cryptography-based defense methods.

    
    \item 



        We extensively evaluated the effectiveness of VMask across five models and 13 datasets, comparing  it to 12 other defense methods.
        VMask consistently demonstrates the best privacy-utility trade-off, reducing MC attack accuracy to a random level while maintaining near-identical model utility.
        Additionally, VMask's running time is orders of magnitude lower than that of cryptography-based methods and only marginally higher than that of standard VFL.


\end{itemize}

\section{Background} \label{sec:background}

\subsection{VFL Setting} \label{sec:vfl_setting}

\begin{figure}
    \centering
    \includegraphics[width=0.7\linewidth]{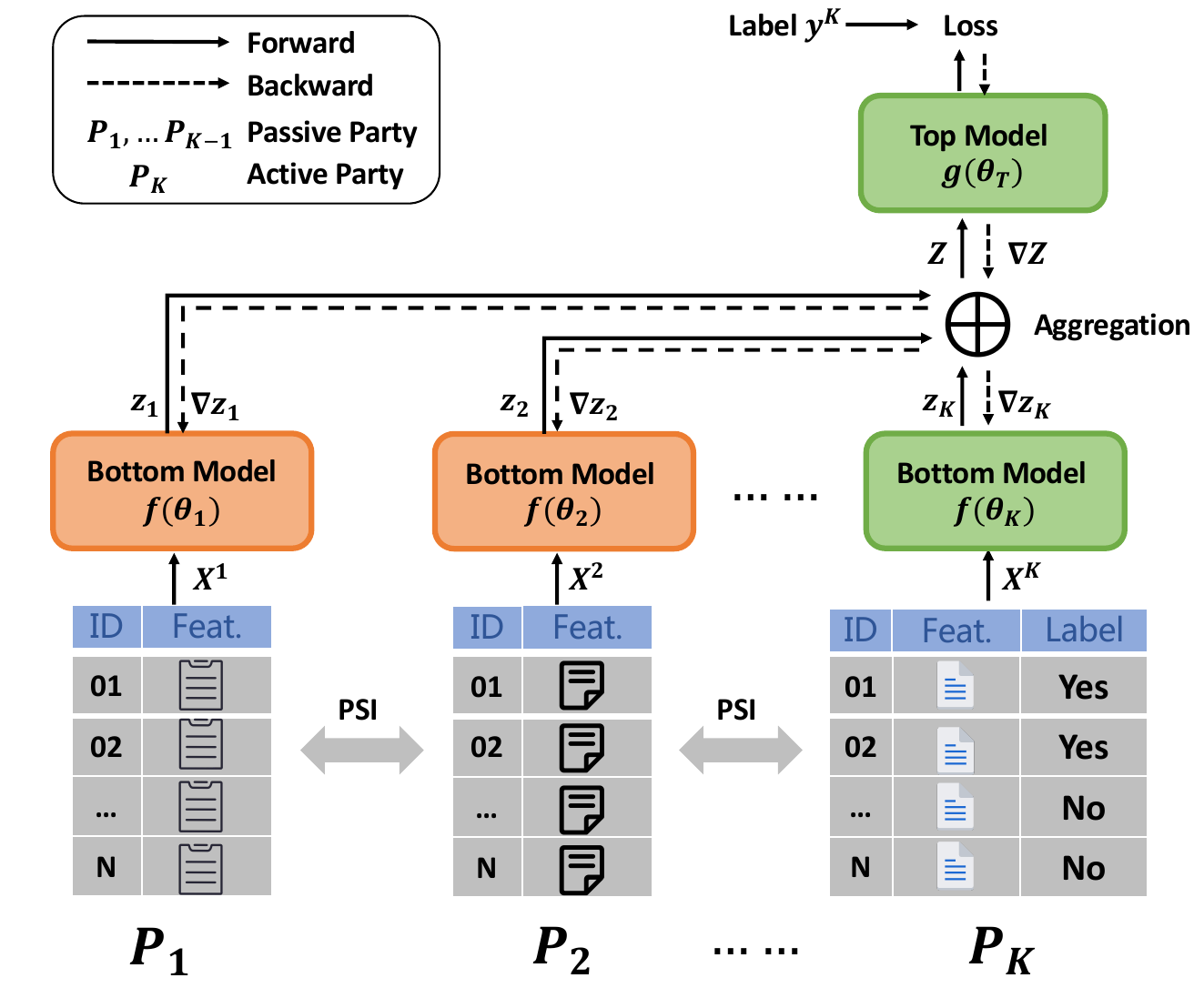}
    \caption{A typical vertical federated learning system.}
    \label{fig:vfl_setting}
    \vspace{-4pt}
\end{figure}

A typical VFL system, illustrated in Figure \ref{fig:vfl_setting}, consists of $K$ distributed participants \( P_1,P_2,\cdots,P_K \). 
Here, $P_K$ is the only active party with true labels, while the rest are passive parties. 
The dataset of $P_K$ is denoted as $D_K = \{(X_i^K, y_i^K)\}_{i=1}^N$, where $X_i^K$ and $y_i^K$ are the feature vector and label of the $i$-th sample, and $N$ is the total number of samples.
The dataset of a passive party $P_k$ is denoted as $D_k = \{X_i^k\}_{i=1}^N, k\in[1, K-1]$.
Before collaborative VFL training, each participant aligns their sample IDs using techniques such as private set intersection (PSI) \cite{raghuraman2022blazing,rindal2021vole,kolesnikov2016efficient}.

In each forward pass, each participant applies its own bottom model $f(\theta_k)$ to the local features $X^k$ to extract the local feature embedding $z_k = f(X^k; \theta_k)$, which is then sent to the active party. 
The active party applies an aggregation operator $\oplus$ (such as summation, concatenation, etc.) on all embeddings to get $Z=z_1\oplus z_2\oplus \cdots \oplus z_K$.
The aggregated embedding $Z$ is then fed into a top model $g(\theta_T)$ to produce the global output $g(Z;\theta_T)$, which is combined with label $y^K$ to compute the loss. 
The overall optimization objective is to minimize 
$L = \frac{1}{N}\sum_{i=1}^N \mathcal{L}\left(g(\oplus_{k=1}^K f(X_i^k;\theta_k);\theta_T), y_i^K\right)$, 
\noindent where $\mathcal{L}$ denotes the loss function, such as cross-entropy loss.

In each backward pass, the active party first computes $\frac{\partial L}{\partial \theta_T}$ and $\nabla Z = \frac{\partial L}{\partial Z}$, and updates the top model using $\frac{\partial L}{\partial \theta_T}$. 
Then, utilizing the chain rule, the gradient of each participant's bottom model can be calculated using $\frac{\partial L}{\partial \theta_k} 
        = \frac{\partial L}{\partial Z} \cdot \frac{\partial Z}{\partial z_k} \cdot \frac{\partial z_k}{\partial \theta_k} 
        = \frac{\partial L}{\partial Z} \cdot \frac{\partial z_k}{\partial \theta_k} 
        = \nabla Z \frac{\partial z_k}{\partial \theta_k},$
where we assume the aggregation operator $\oplus$ represents summation, leading to $\frac{\partial Z}{\partial z_k} = 1$ and $\nabla Z = \nabla z_1 = \cdots \nabla z_K$.
From the above equation, it is evident that the active party only needs to send $\nabla Z$ back to all passive parties, who can then use it to update their local bottom model independently. 
The forward and backward passes are iterated until the global VFL model converges.



\subsection{Model Completion Attack in VFL} \label{sec:vfl_label_leakage}

In the model completion (MC) attack, an honest-but-curious passive party $P_k, k\in[1,K-1]$, is controlled by the attacker. 
This attacker strictly complies with the training protocol but aims to deduce private labels from their locally trained bottom model $f(\theta_k)$.
The attacker is assumed to possess \( M \) labeled samples in their dataset \( D_k \) and initially constructs an attack dataset, \( D_A^k = \{(X_i^k, y_i^K)\}_{i=1}^M \cup \{X_i^k\}_{i=M+1}^N \). 
Subsequently, they create an inference head model \( f(\theta_H) \) with the same output dimension as the top model \( g(\theta_T) \) and attach it to the bottom model \( f(\theta_k) \). 
This combination forms the attack model \( f_A(\theta_A) = f(\theta_k) \circ f(\theta_H) \), where $\circ$ denotes model concatenation.
Using the attack dataset \( D_A^k \) and the attack model \( f_A(\theta_A) \), the attacker employs semi-supervised learning techniques \cite{berthelot2019mixmatch} to fine-tune the attack model locally. 
Once sufficiently trained, the attacker can infer the label of any sample \( X_i^k \), whether from the training or the test set, by feeding it into the attack model and using \( f_A(X_i^k;\theta_A) \) as the inferred label. 
These steps are detailed further in Algorithm ~\ref{algo:mc_attack} in Appendix \ref{appendix:mc_attack}.

\subsection{Secret Sharing} \label{sec:secret_sharing}
Multi-party computation (MPC) enables $K$ participants to collaboratively compute a function $f(x_1, x_2,\cdots, x_K)$ while ensuring that each participant's data $x_k$ remains private.
MPC protocols can be implemented using techniques such as garbled circuits (GC) and secret sharing (SS). 
This paper focuses on SS, as it is significantly more efficient than GC for numerical computations, while GC is better suited for relational operations.

The classic additive SS scheme includes two basic primitives, $\mathtt{Share(\cdot)}$ and $\mathtt{Reconstruct(\cdot)}$, as shown in Equation ~\ref{eq:secret_sharing}. 
A value $x\in \mathbb{Z}_q$ is additively secret shared among $K$ participants by generating a random secret share $\llbracket x\rrbracket_k$ for each participant $P_k, k\in[1, K]$, such that $x =(\sum_{i=1}^K \llbracket x\rrbracket_k) \mod q$, where $q$ is a large modulus (for simplicity, we exclude the modulo operation in the remainder of this paper). 
The value $x$ is reconstructed by summing all the shares among the $K$ participants. 
Given additive secret shares $\llbracket x \rrbracket$ and $ \llbracket y \rrbracket$, two basic integrity-preserving arithmetic primitives can be constructed: $\mathtt{Add(\cdot)}$ and $\mathtt{Mul}(\cdot)$. 
For $\mathtt{Add(\llbracket x \rrbracket, \llbracket y \rrbracket)}$, each participant $P_k$ computes $\llbracket x \rrbracket_k + \llbracket y \rrbracket_k$ locally, and the result is also secret shared among the participants. 
For $\mathtt{Mul(\llbracket x \rrbracket, \llbracket y \rrbracket)}$, the Beaver's triple technique \cite{beaver1992efficient} is typically used. 
Specifically, a shared triple $\llbracket a \rrbracket, \llbracket b \rrbracket, \llbracket c \rrbracket$ is generated by either a trusted third party or through oblivious transfer \cite{ishai2003extending}, where $a, b, c$ are randomly chosen from $\mathbb{Z}_q$, and $c = a \times b$. 
Subsequently, through local computation and communication of intermediate results, each participant $P_k$ receives a secret share of $x\times y$.

\vspace{-8pt}
\begin{equation}
    \label{eq:secret_sharing}
    \begin{aligned}
        \llbracket x\rrbracket_1, \llbracket x\rrbracket_2, \cdots, \llbracket x\rrbracket_K \gets\mathtt{Share}(x), \\
        x \gets \mathtt{Reconstruct}(\llbracket x\rrbracket_1, \llbracket x\rrbracket_2, \cdots, \llbracket x\rrbracket_K )
    \end{aligned}
\end{equation}

These two basic arithmetic primitives enable the construction of more complex and nonlinear operations, such as exponential functions, division, and logarithms, through linear approximation techniques. 
The outputs of each operation remain secret-shared until the \(\mathtt{Reconstruct}(\cdot)\) primitive is invoked. 
Furthermore, these operations can be easily extended to handle cases where \(x\) and \(y\) are multidimensional vectors.

\section{Insight and Main Ideas of VMask} \label{sec:motivation}

\begin{figure*}[t!]
	\centering
	\begin{subfigure}[b]{0.23\textwidth}
		\includegraphics[width=\textwidth]{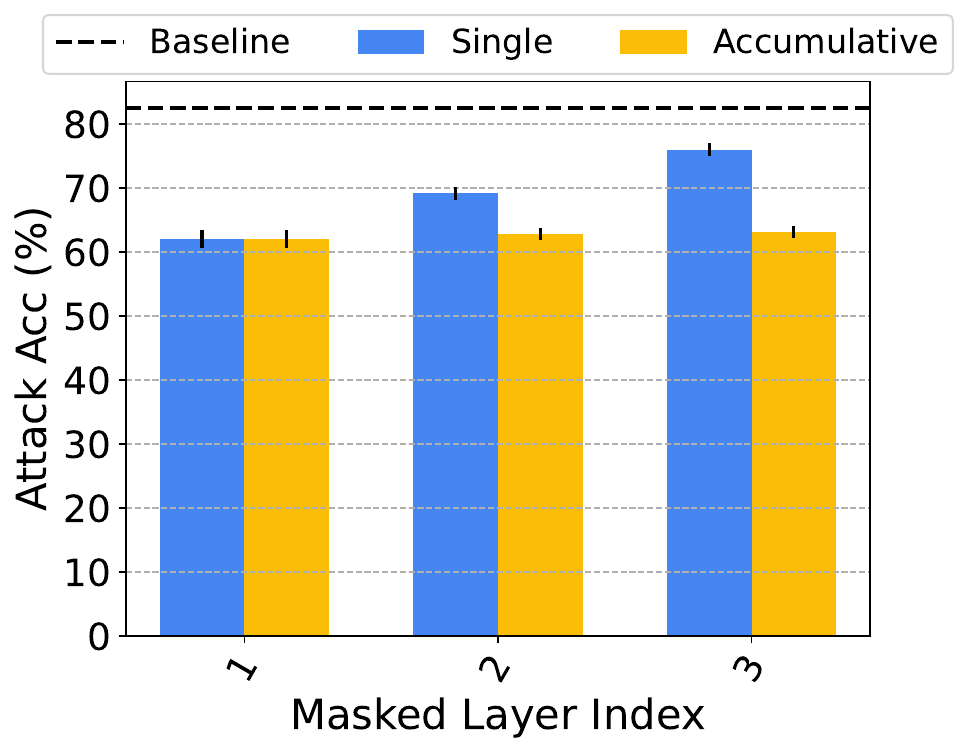}
		\caption{MLP3, TabMNIST}
		\label{mlp_tab_mnist_layers}
	\end{subfigure}\hfill
	\begin{subfigure}[b]{0.23\textwidth}
		\includegraphics[width=\textwidth]{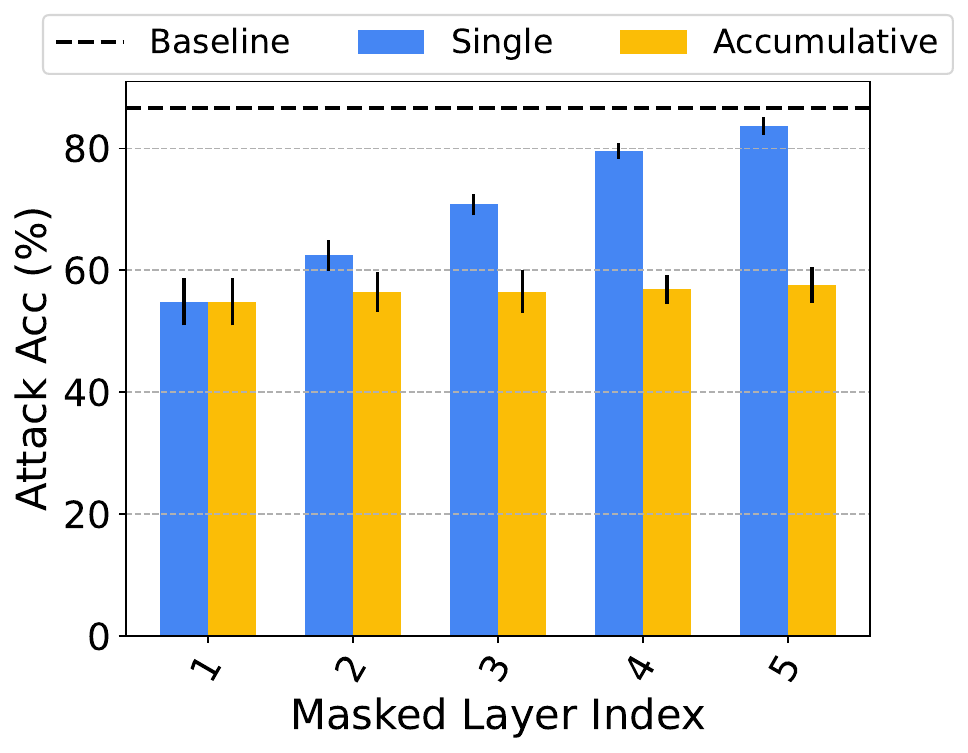}
		\caption{LeNet5, MNIST}
		\label{lenet_mnist_layers}
	\end{subfigure}\hfill
	\begin{subfigure}[b]{0.23\textwidth}
		\includegraphics[width=\textwidth]{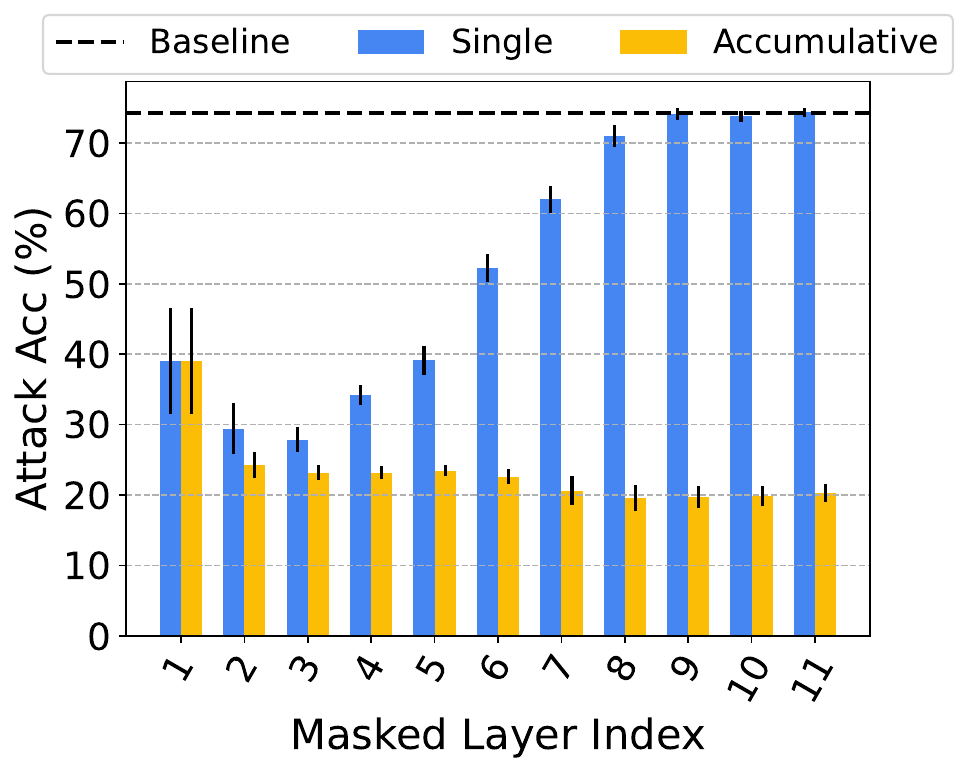}
		\caption{VGG13, CIFAR10}
		\label{vgg13_cifar10_layers}
	\end{subfigure}\hfill
	\begin{subfigure}[b]{0.23\textwidth}
		\includegraphics[width=\textwidth]{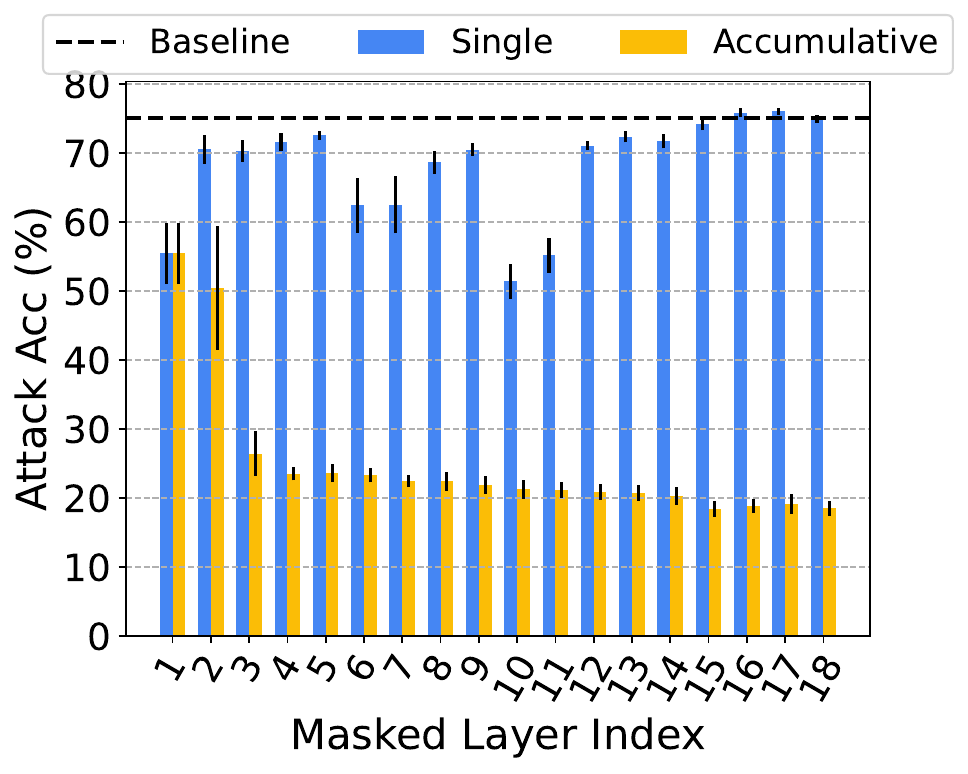}
		\caption{ResNet18, CIFAR10}
		\label{resnet18_cifar10_layers}
	\end{subfigure}\hfill
	\caption{
		The impact of single layer masking and accumulative layer masking on MC attack accuracy.
		The ``Baseline'' denotes MC attack accuracy obtained from the intact bottom model (without masking any layers) after VFL training.
    }
	\label{fig:motivation_layer_masking}
    \vspace{-8pt}
\end{figure*}

\subsection{Key Insight}

We begin by analyzing the cause of label leakage in MC attack.
From the VFL setting described in Section \ref{sec:vfl_setting}, we observe that the passive party (i.e., potential attacker) possesses the entire bottom model.
This bottom model is gradually optimized during the training process, enabling it to transform the input data into more representative feature embeddings.
In VFL, the passive party's feature embeddings typically contribute to the global classification task, giving them strong predictive capability for the private labels.
As a result, the attacker can classify these feature embeddings with high accuracy to infer labels by simply attaching an inference head model to the bottom model and fine-tuning it, even with limited samples.
Consequently, for the active party (label owner), this leads to a higher extent of label leakage compared to what would result from a random bottom model.

\begin{figure}
	\centering
	\begin{subfigure}[b]{0.23\textwidth} 
		\includegraphics[width=\textwidth]{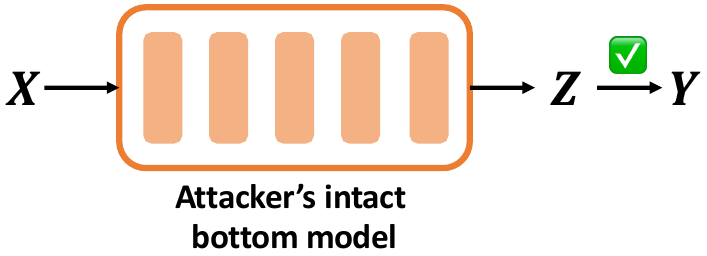}
		\caption{Intact bottom model}
		\label{fig:insight_intact}
	\end{subfigure}\hfill
	\begin{subfigure}[b]{0.23\textwidth}  
		\includegraphics[width=\textwidth]{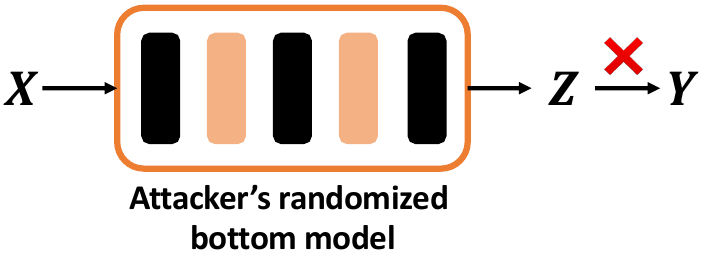}
		\caption{Randomized bottom model}
		\label{fig:insight_randomized}
	\end{subfigure}\hfill
	\caption{Our insight is to randomize certain layer parameters of the attacker's bottom model to disrupt the correlation between input data $X$ and feature embedding $Z$, thereby reducing the strong predictive capability of $Z$ for the label $Y$.} 
	\label{fig:insight}
    \vspace{-2pt}
\end{figure}

To mitigate the predictive capability of the feature embeddings, \emph{our key insight is to disrupt the strong correlation between the input data and the feature embeddings by randomizing certain layer parameters of the attacker's bottom model.}
For example, as shown in Figure \ref{fig:insight}, by randomizing certain layers, the original intact bottom model (Figure \ref{fig:insight_intact}) is segmented into multiple sub-models (Figure \ref{fig:insight_randomized}). 
Some of these sub-models retain the same parameters as the original model (layers 2 and 4), while the parameters of other sub-models are entirely random (layers 1, 3, and 5).
This disruption alters the original continuous mapping from input data to feature embeddings. 
We hypothesize that this makes it difficult for the attacker to restore the mapping through fine-tuning with limited labeled samples, thereby preventing them from obtaining representative feature embeddings and further decreasing the MC attack accuracy.

\subsection{Feasibility of Layer Masking} \label{sec:layer_masking}
We define \emph{layer masking} as the process of randomizing specific layer parameters within the attacker's bottom model.
To explore the feasibility of this approach in reducing MC attack accuracy, we first mask certain layer parameters in the attacker's trained bottom model and then allow the attacker to conduct MC attack on this masked model.
Below we conduct experiments with three different layer masking strategies, training a two-party VFL classification model using four different model architectures: MLP3, LeNet5 \cite{lecun1998lenet}, VGG13 \cite{simonyan2014vgg}, and ResNet18 \cite{he2016resnet}, alongside three datasets: TabMNIST, MNIST \cite{mnist_dataset}, and CIFAR10 \cite{cifar10_dataset}, each comprising 10 classes.
For each class, we assign 20 labels for the attacker to conduct MC attack.
Here we have three main observations.

\textbf{Observation 1: Masking a single low-level layer can effectively decrease attack accuracy.}
In this layer masking strategy, we mask only one layer, $\theta_k^j$ ($1\le j\le L$ and $L$ is the layer counts), in the trained bottom model $f(\theta_k)$, while keeping other layers unchanged. 
Figure \ref{fig:motivation_layer_masking} shows that masking low-level layers (closer to input) results in a greater decrease in attack accuracy compared to masking high-level layers
For instance, in (LeNet5, MNIST), masking layer $j=1$ results in an attack accuracy of 0.5481$\pm$0.0384, whereas masking layer $j=5$ raises it to 0.8373$\pm$0.0148. 
This is likely because low-level layers capture basic features.
Masking these layers forces the model to adjust parameters across all layers to compensate for the absence of these features, making fine-tuning significantly harder for attacker.

\textbf{Observation 2: Accumulatively masking more layers can gradually decrease attack accuracy.}
In this strategy, layers from $\theta_k^1$ to $\theta_k^j$ are masked, while layers $\theta_k^{j+1}$ to $\theta_k^L$ remain unmodified. 
As shown in Figure \ref{fig:motivation_layer_masking}, attack accuracy decreases progressively as more layers are masked. 
For instance, in (VGG13, CIFAR10), masking only the first layer results in an attack accuracy of 0.3902$\pm$0.0748, which drops to 0.2026$\pm$0.0129 when all layers are masked.
This trend indicates that masking more layers forces attacker to fine-tune more layers from scratch, thereby reducing attack accuracy.

 \begin{table}[t!]
    \footnotesize
	\caption{Comparison of MC attack accuracy(\%) under different combinations of masked layers. 
    }
	\centering
	\begin{tabular}{c|c|c|c}
	  \toprule
	\diagbox{Model}{Layers} & $(\theta_k^1, \theta_k^2)$ & $(\theta_k^1, \theta_k^3)$ & $(\theta_k^2, \theta_k^3)$ \\
	  \midrule
	  MLP3       & 62.76 \textpm 0.89 & \textbf{62.52 \textpm 1.69} & 67.90 \textpm 0.74 \\
	  LeNet5     & \textbf{56.45 \textpm 3.30} & 57.26 \textpm 3.98 & 62.07 \textpm 2.17 \\
	  VGG13     & 24.28 \textpm 1.85 & \textbf{23.92 \textpm 1.04} & 24.66 \textpm 0.91 \\
	  ResNet18  & 50.43 \textpm 8.95 & \textbf{27.71 \textpm 4.19} & 68.15 \textpm 1.58 \\
	  \bottomrule
	\end{tabular}
	\label{tab:crossed_layer_masking}
\end{table}

\textbf{Observation 3: Masking non-adjacent layers can decrease attack accuracy more effectively.}
In this strategy, non-adjacent layers are masked instead of consecutive ones. 
Table \ref{tab:crossed_layer_masking} shows that masking non-adjacent layers ($\theta_k^1$ and $\theta_k^3$) leads to a greater drop in attack accuracy compared to adjacent layers. 
For example, in (ResNet18, CIFAR10), masking $(\theta_k^1, \theta_k^3)$ results in an attack accuracy of only 0.2771$\pm$0.0419, whereas masking $(\theta_k^1, \theta_k^2)$ yields 0.5043$\pm$0.0895. 
This indicates that layer $\theta_k^3$ has a greater impact on reducing attack accuracy than $\theta_k^2$, highlighting the potential of targeting critical layers for masking to achieve a more effective reduction in attack accuracy.

These results demonstrate that layer masking is a feasible and effective approach for decreasing MC attack accuracy.

\subsection{Main Ideas of VMask}
In the previous section, we demonstrate that layer masking strategies effectively reduce MC attack accuracy.
However, applying these strategies in a VFL system raises two key questions: 
(1) How to maintain main task accuracy when some layers are masked?
(2) How can the defender automatically identify which layers to mask during training?
Below, we outline our main ideas for addressing these questions, which guide the design of VMask framework.

\textbf{Utilizing SS for layer masking.}
To maintain main task accuracy, we employ SS to share the layer parameters of the attacker's bottom model. 
This method is equivalent to layer masking because each shared layer parameter is completely random. 
Additionally, the masked layers are trained using SS, which preserves the integrity of linear computations (i.e., $\mathtt{Add}$ and $\mathtt{Mul}$), as detailed in Section \ref{sec:secret_sharing}.
This ensures that the main task accuracy is retained.

To answer the second question, we develop a strategy to identify critical layers, a subset of all layers, to mask.
Masking only these layers reduces the overhead of naively applying SS to the entire model and more effectively decreases attack accuracy, as shown in Observation 3. 
Our selection strategy is based on three main ideas.

\textbf{Adopting accumulative gradient norm as layer selection criterion.}
We define the accumulated gradient norm for the $j$-th layer as \( G^j_k = \sum_{t=1}^{T} |\nabla \theta^j_{k,t}| \), where \( T \) is the total number of training epochs. 
We prove that layers with larger accumulated gradient norms significantly impact the model's output and more effectively reduce MC attack accuracy when masked. 
Empirical results show that the largest norms often belong to non-adjacent layers, which is consistent with Observation 3. 
Thus, accumulated gradient norm serves as a reasonable criterion for selecting critical layers. 
Detailed proofs and empirical results are provided in Appendix \ref{appendix:vis_accumu_grad_norm}.

\textbf{Incorporating shadow model to estimate ground-truth gradient.}
In VFL, the active party cannot access the ground-truth gradient of the attacker's bottom model. 
To address this, we introduce a shadow model that has the same architecture (excluding parameters) as the attacker's model. 
Inspired by Gao et al. (2023) \cite{gao2023pcat}, the active party locally trains this shadow model synchronously with the VFL model using a small auxiliary dataset, enabling it to estimate the ground-truth gradient.

\textbf{Providing a tunable privacy budget to flexibly control label privacy level.}
Using the shadow model and auxiliary dataset, the active party can locally simulate the MC attack by constructing an attack dataset with ground-truth labels to estimate label leakage. 
This approach allows the active party to set a tunable label privacy budget, which defines an upper limit on the extent of label leakage.
For instance, if the estimated leakage exceeds the privacy budget, additional layers will be selected for masking until the leakage is within the budget.

Combining these ideas, we design the VMask framework, which will be elaborated upon in the next section.

\section{VMask Framework Design} \label{sec:framework}

\subsection{Threat Model} \label{sec:threat_model}


\textbf{Attacker.} 
We assume that \(K-1\) passive parties are semi-honest (honest-but-curious) and non-colluding, meaning they strictly follow the training protocol specified in VMask framework and do not collude with each other. 
However, one of the passive parties may potentially be controlled by an attacker.
The attacker is assumed to have full knowledge of VMask design, and its adversarial goal is first to reconstruct the masked layer parameters in its bottom model, and subsequently to conduct MC attack to infer the active party's private labels.



\discuss{
\textbf{Defender.}
For the active party in VMask, we assume it is also semi-honest.
It is aware of the bottom model architectures (excluding parameters) of other passive parties.
This knowledge is used to construct shadow models $f(\theta_{s,1})$, $f(\theta_{s,2})$, $\cdots$, $f(\theta_{s,K-1})$.
Additionally, we assume the active party has access to small auxiliary datasets, $D_{1}^{aux}$, $D_{2}^{aux}$, $\cdots$, $D_{K-1}^{aux}$, corresponding to other passive parties.
These auxiliary datasets, used for training shadow models, do not overlap with the original datasets of the passive parties. 
They may be either in-distribution or out-of-distribution relative to the original datasets, and are typically only 1\% to 5\% the size of the originals, as shown in our experiments (see Sections \ref{sec:effectiveness} and \ref{sec:ablation}).
This assumption is both reasonable and practical.
For example, the active party may purchase samples from online data marketplaces (e.g., public image or text datasets) to construct an out-of-distribution auxiliary dataset \cite{pasquini2021unleashing,chen2023practical}.
Alternatively, it may obtain unaligned samples (i.e., unused samples after PSI) from passive party to form an in-distribution auxiliary dataset.
Such samples may arise from historical logistics records with incompatible ID formats in e-commerce logistics collaborations \cite{cacitwhitepaper}, where PSI failures naturally produce non-sensitive, non-overlapping feature sets (e.g., anonymized shipping timestamps without customer identifiers).
Another example is time-window-mismatched financial transactions (e.g., 8-hour latency gaps in daily PSI synchronization) in collaborations between banks and e-commerce platforms \cite{ye2025vertical}, where the e-commerce party—acting as passive party—legally retains ownership of unaligned data and contractually permits limited sharing of de-identified transaction patterns. 
These practices are common in previous studies within the VFL community \cite{fu2022label,naseri2024badvfl,pang2023adi,he2023backdoor}, and are typically conducted under the legal doctrine of ``legitimate interest'' (GDPR Article 6(1)(f) \cite{gdpr2018}), which permits the processing of non-personal/non-identifiable data for business partnership purposes without requiring individual consent.


}

\begin{figure}[t!]
    \centering
    \includegraphics[width=0.9\linewidth]{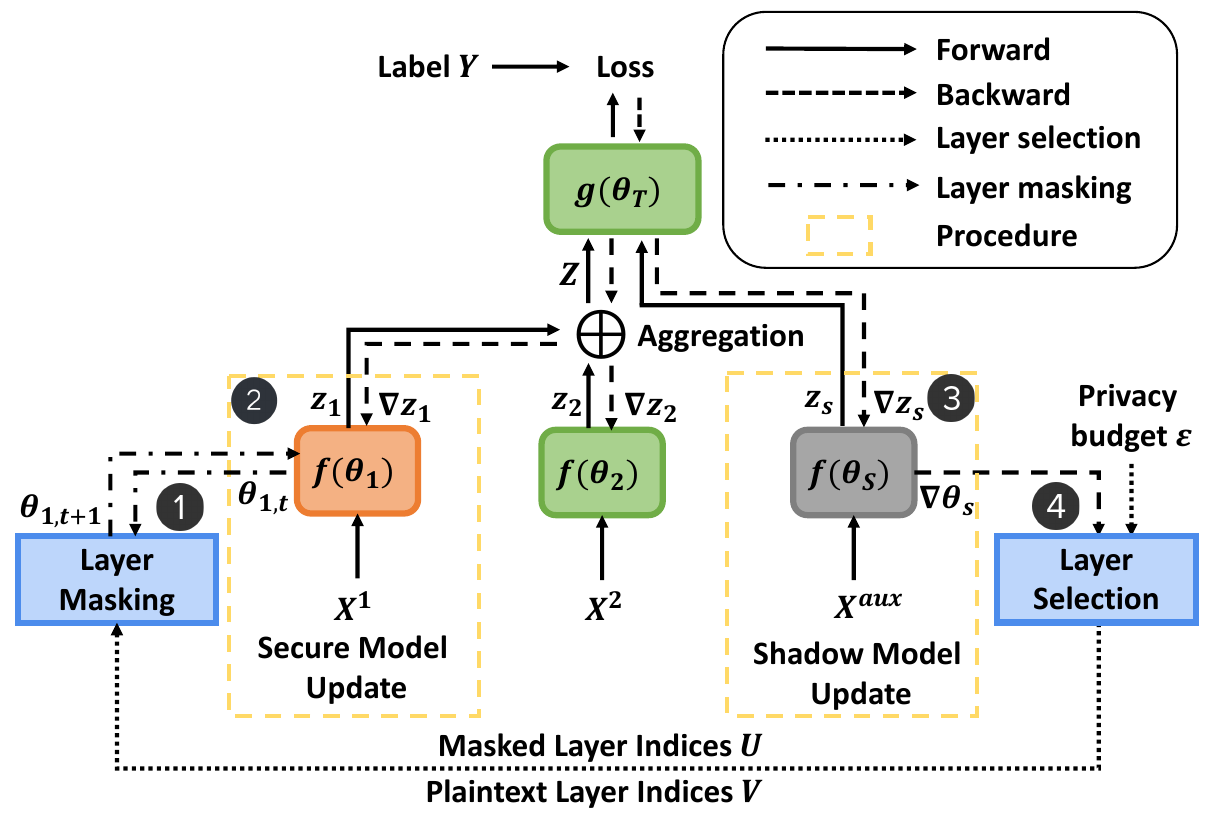}
    \caption{Illustration of VMask framework.}
    \label{fig:VMask_framework}
    \vspace{-6pt}
\end{figure}

\subsection{Overview} \label{sec:overview}

Figure \ref{fig:VMask_framework} provides an overview of the proposed VMask framework (for simplicity, we illustrate the framework in a two-party setting). 
In addition to the standard VFL setting (as shown in Figure \ref{fig:vfl_setting}), VMask incorporates two additional modules and two additional procedures.
\textcircled{1} \textbf{Layer masking module:} This module leverages SS primitives to mask and reconstruct passive party's bottom model parameters based on the masked and plaintext layer indices.
\textcircled{2} \textbf{Secure model update procedure:} This procedure performs secure forward and secure backward computations based on the bottom model after layer masking. 
For masked layers, the passive party computes the forward embedding and backward gradient in a secret-shared manner. 
For plaintext layers, the computations remain the same as in standard VFL.
Note that the active party's bottom model and top model are still trained in plaintext.
\textcircled{3} \textbf{Shadow model update procedure:} This procedure enables the active party to locally train the shadow model in plaintext using the auxiliary dataset, estimating the ground-truth gradients on the passive party's side.
These estimated gradients are sent to the layer selection module to determinate which layers need to be masked.
\textcircled{4} \textbf{Layer selection module:} This module determines the layers that need to be masked based on the pre-defined privacy budget and the accumulated gradient norm of the shadow model.
It then sends the masked and plaintext layer indices to the layer masking module for the next epoch.
These steps are iterated until the VFL model converges.
For a more detailed explanation of the relationships among these modules and procedures, please refer to Algorithm \ref{algo:framework} in Appendix ~\ref{appendix:vmask_framework}.

\subsection{Dive into VMask} \label{sec:vmask_details}
In this section, we dive into VMask framework, detailing each module and procedure.

\subsubsection{Initialization.} \label{sec:init}
The initialization stage prepares for the main VFL training loop. 
For each \(k \in [1,K-1]\), the passive party \(P_k\)'s bottom model parameter \(\theta_k\) and the active party's shadow model parameter \(\theta_{s,k}\) are initialized using the same random seed. 
The active party's bottom model parameter \(\theta_K\) and top model parameter \(\theta_T\) are also randomly initialized.
For each passive party \(P_k\), the masked layer indices (for first epoch) and plaintext layer indices are initialized as \(U_k^1 = \{1\}\) and \(V_k^1 = [L] - \{1\}\).
The layer masking module (described in Section \ref{sec:vmask_layer_masking}) is then invoked to mask the bottom model parameter \(\theta_k\) based on \(U_k^1\) and \(V_k^1\).
The decision to mask the first layer in the initial epoch is based on empirical observations. 
Our findings indicate that easier-to-train datasets, like MNIST, achieve high main accuracy after just one epoch. 
If the first layer is unmasked during this epoch, it poses a significant risk of label privacy breaches once the epoch finishes. 
This decision is further supported by the visualization of accumulated gradient norms (provided in Appendix \ref{appendix:vis_accumu_grad_norm}), which show that the first layer has a significantly larger gradient norm than other layers, making it a priority for selection by the layer selection module (described in Section \ref{sec:layer_selection}).
Additionally, for each passive party, the corresponding accumulated gradient norm (the layer selection criterion) is initialized to \(G_k=0\).

\subsubsection{Layer Masking via Secret Sharing.} \label{sec:vmask_layer_masking}
The layer masking module consists of two submodules: layer sharing and layer reconstruction. 
The former uses masked layer indices \(U\) for parameter sharing, while the latter uses plaintext layer indices \(V\) for reconstruction.

As shown in Algorithm \ref{algo:layer_masking}, each passive party \(P_k\) first computes the set difference \(\Delta U\) between the current epoch \(U_{curr}\) and the previous epoch \(U_{prev}\) (Line 1). 
Each element in \(\Delta U\) represents a newly added layer requiring masking for the upcoming epoch. 
For each \(u \in \Delta U\), passive party \(P_k\) masks the \(u\)-th layer parameter \(\theta^u\) by invoking the \(\mathtt{Share}(\cdot)\) primitive to share it into \(\llbracket \theta^u \rrbracket_k\) for itself and \(\llbracket \theta^u \rrbracket_K\) for active party \(P_K\). 
Then, the active party adds parameter noise \(\theta_r^u\) to its share \(\llbracket \theta^u \rrbracket_K\) (Lines 2-6). 
This step ensures that the passive party does not know the true layer parameter \(\theta^u\), with a detailed security analysis provided in Section \ref{sec:security_analysis}.

For layer reconstruction, passive party \(P_k\) calculates the set difference \(\Delta V\) between the current epoch \(V_{curr}\) and the previous epoch \(V_{prev}\) (Line 7). 
Each element in \(\Delta V\) represents a newly added layer to be reconstructed. 
For each \(v \in \Delta V\), active party \(P_K\) generates random layer noise \(\theta_r^v\) and adds it to its share \(\llbracket \theta^v \rrbracket_K\) (Lines 9-10), again to prevent passive party from knowing the true layer parameter \(\theta^v\). 
Finally, the passive party invokes the \(\mathtt{Reconstruct}(\cdot)\) primitive to reconstruct \(\theta^v\) for plaintext use in the next epoch (Line 11).

For each passive party \(P_k\), since each layer sharing or reconstruction operation only requires communication with the active party, the communication complexity of the layer masking module is linear \emph{w.r.t.}  the number of participants, ensuring good scalability.

\begin{algorithm}[t]
    \small
    \caption{Layer Masking via Secret Sharing}
    \label{algo:layer_masking}
    \SetKwInOut{Input}{Input}
    \SetKwInOut{Output}{Output}
    \Input{Masked layer index set of current epoch $U_{curr}$ and previous epoch $U_{prev}$, plaintext layer index set of current epoch $V_{curr}$ and previous epoch $V_{prev}$, bottom model parameter $\theta$, passive party rank $k$, noise standard variance $\sigma$.}
    \Output{Masked bottom model parameters $\theta$.}

    $\Delta U = U_{curr} - U_{prev}$ \;
    \For{each $u \in \Delta U$}{
        $\llbracket \theta^{u} \rrbracket_k, \llbracket \theta^{u} \rrbracket_K \gets \texttt{Share}(\theta^{u})$ \;
        Generate random layer parameters $\theta_r^u \gets \mathcal{N}(0, \sigma)$ \;
        $\llbracket \theta^{u} \rrbracket_K \gets \llbracket \theta^{u} \rrbracket_K + \theta_r^u$ \;
        $\theta^{u} \gets \llbracket \theta^u \rrbracket_k$\;
    }
    
    $\Delta V = V_{curr} - V_{prev}$ \;
    \For{each $v \in \Delta V$}{
        Generate random layer noise $\theta_r^v \gets \mathcal{N}(0, \sigma)$ \;
        $\llbracket \theta^{v} \rrbracket_K \gets \llbracket \theta^{v} \rrbracket_K + \theta_r^v$ \;
        $\theta^{v} \gets \texttt{Reconstruct}(\llbracket \theta^{v} \rrbracket_k, \llbracket \theta^{v} \rrbracket_K)$ \;
    }
    \Return{$\theta$}
\end{algorithm}

\subsubsection{Secure Model Update.}
After sharing and reconstructing the passive parties' layer parameters in the layer masking module, we proceed with the secure model update procedure. 
This includes secure forward and backward sub-procedures to train the VFL models (\(\theta_1, \ldots, \theta_K, \theta_T\)) using secret sharing.

In the secure forward phase, each passive party $P_k, k\in[1,K-1]$ checks whether the layer parameter $\theta_k^j, j\in[1,L]$ is masked.
If it is masked, model forwarding is performed in a secret-shared manner: the layer input is shared between  $P_k$ and the active party \(P_K\).
Each input share is fed into the shared layer parameter to obtain the output share, which is then reconstructed to serve as the input for the next layer. 
If the layer is not masked, the layer input is directly fed into the plaintext layer to derive the output. 
This process is repeated until each passive party produces its bottom model embedding. 
The active party then aggregates all embeddings and feeds them into the top model to calculate loss.

In the secure backward phase, the procedure follows a similar approach. 
The gradients of the active party's top and bottom models are first computed in plaintext. 
For each layer parameter of the passive party's bottom model, if the layer is masked, its gradient is calculated in a secret-shared manner; otherwise, the gradient is computed in plaintext. 
A detailed explanation of this procedure is provided in Algorithm \ref{algo:secure_model_update} in Appendix \ref{appendix:secure_model_update}.

\subsubsection{Shadow Model Update.}

The approach proposed by Gao et al. (2023) \cite{gao2023pcat} is adopted to train the shadow models.
Using the auxiliary datasets, the training process is similar to the standard VFL model training but is conducted solely by the active party in plaintext.
Unlike standard training, the top model parameter \(\theta_T\) is kept fixed, and the active party's bottom model is not involved in either the forward or backward phase.
This procedure is explained in more detail in Algorithm \ref{algo:shadow_model_update} in Appendix \ref{appendix:shadow_model_update}.

\subsubsection{Layer Selection According to Tunable Privacy Budget.} \label{sec:layer_selection}

In Algorithm \ref{algo:layer_selection}, the active party first adds the \(L_1\) gradient norm of the shadow model at the current epoch to its accumulated gradient norm \(G\) (Line 1), and then sorts \(G\) in descending order to create the layer index list \(Q\), with the first element corresponding to the layer with the largest accumulated gradient norm (Line 2).

We derive two variant frameworks based on the layer selection strategy: VMask (with replacement) and VMask-AS (without replacement). 
For VMask, the masked layer indices $U$ is initialized as an empty set, allowing any layer to be selected for masking (Lines 5-6). 
In contrast, VMask-AS excludes already-masked layer indices by subtracting $U$ from \(Q\), ensuring that layers are selected only from \(Q-U\), implying an accumulative selection of masked layers (Lines 3-4).

Once the candidate layers are identified, the active party $P_K$ greedily constructs the masked layer indices for the next epoch. 
First, $P_K$ constructs an attack dataset \(D_A\) using the auxiliary dataset and ground truth labels (Line 7). 
At each iteration (Lines 8-18), $P_K$ masks the shadow model by replacing each \(\theta_{s,mask}^u, u\in U\), with random values. 
Using this masked shadow model, $P_K$ simulates MC attack locally to estimate the attack accuracy (Line 13, see Algorithm \ref{algo:mc_attack} in Appendix \ref{appendix:mc_attack}).
If the estimated attack accuracy exceeds the privacy budget, the largest element \(q\) is popped from \(Q\) and added  to \(U\) for the next iteration (Lines 14-16). 
This loop continues until either the estimated attack accuracy falls below the privacy budget (Lines 17-18) or no more layers are left to select (Line 8). 
Finally, the plaintext layer indices $V$ are calculated as \( [L] - U\) (Line 19).

\begin{algorithm}[t]
    \small
    \caption{Layer Selection}
    \label{algo:layer_selection}
    \SetKwInOut{Input}{Input}
    \SetKwInOut{Output}{Output}
    \Input{Auxiliary dataset $D^{aux}$, shadow model parameters $\theta_s$ and its gradient $\nabla \theta_s$, masked layer index set $U$, privacy budget $\epsilon$, historical accumulated gradient norm $G$, number of bottom model layers $L$, number of labeled samples $M$, ground truth label $Y$,  MC attack training epochs $T'$, learning rate $\eta$.}
    \Output{Updated masked layer index set $U$, updated plaintext layer index set $V$, updated accumulated gradient norm $G$.}

    $G \gets G + L_1(\nabla \theta_s)$; 

    $Q \gets \text{argsort}(G, \text{reverse=True})$; 

    \eIf{selection without replacement}{
        $Q \gets Q - U, U \gets U$ ; \tcp{VMask-AS}
    }{
        $Q \gets Q, U \gets \emptyset$ ; \tcp{VMask}
    }
    Randomly select $M$ samples and labels from $(D^{aux}, Y)$ to construct attack dataset $D_A$ \;
    \While{$Q$ is not empty}{
        $\theta_{s, mask} \gets \theta_s$ \;
        \For{each $u \in U$}{
            Generate random layer parameters $\theta_r^u$ \;
            $\theta_{s, mask}^u \gets \theta_r^u$ \;
        }
        $\epsilon' \gets \text{\textbf{MCAttack}}(D_A, \theta_{s, mask}, T', \eta)$ \;
        \eIf{$\epsilon' > \epsilon$}{
            Pop first element $q$ from $Q$ \;
            $U \gets U + q$ \;
            
        }{
            \textbf{break}\;
        }
    }
    $V \gets [L] - U$ \;
    \Return{$U, V, G$}
\end{algorithm}


\subsection{Security Analysis} \label{sec:security_analysis}
\discuss{
    We investigate the potential for a semi-honest attacker to reconstruct the parameters of the masked layers within the layer masking module and secure model update procedure.
    If these masked parameters are successfully recovered, the attacker can subsequently launch MC attack to infer private labels.
    Therefore, if parameter reconstruction is infeasible, VMask can be considered secure against semi-honest attackers.
}

\textbf{Security of the layer masking module.}
Consider a training epoch \( t \), where \( t \in [1, T] \). 
Let \( \theta_t \) and \( \theta_{t+1} \) represent the bottom model parameters of the attacker before and after model updating, respectively. 
As specified in Algorithm \ref{algo:layer_masking}, each element \( u \) in \( \Delta U \) denotes a newly added layer of \( \theta_{t+1} \) that requires masking for epoch \( t+1 \).  
If \( \theta_{t+1}^u \) is shared directly, the attacker could reconstruct the ground truth layer parameters, as it already possesses its own share of the layer. 
Given that SS preserves the computational integrity of linear layers, the attacker could use the gradient from epoch \( t+1 \) to update \( \theta_{t+1}^u \) using stochastic gradient descent (SGD), thereby obtaining \( \theta_{t+2}^u \). 
To mitigate this risk, random noise is added to the active party's share of the layer parameters, preventing the attacker from inferring the ground truth layer parameters.  

Similarly, for each epoch \( t \), every element \( v \) in \( \Delta V \) represents a newly added layer of \( \theta_{t+1} \) that needs to be reconstructed for epoch \( t+1 \). 
If \( \theta_{t+1}^v \) is directly reconstructed for the attacker, it could use the gradients to infer the ground truth layer parameters \( \theta_t^v \) of epoch \( t \) with SGD, leading to leakage of the \( t \)-th epoch's parameters. 
To prevent this, random noise is introduced to the active party's share of the layer parameters before reconstruction. 
This addition ensures that the attacker cannot derive the ground truth layer parameters of epoch \( t \).

\textbf{Security of the secure model update procedure.}
For each masked layer, the attacker knows the layer input and output and could potentially launch a reconstruction attack to infer the masked layer parameters.

For fully connected (FC) layers, we denote the layer parameters as \( W \in \mathbb{R}^{n_2 \times n_1} \), the layer input as \( A \in \mathbb{R}^{B \times n_1} \), and the layer output as \( C \in \mathbb{R}^{B \times n_2} \), where \( B \) is the batch size, and \( n_1 \) and \( n_2 \) are the input and output dimensions, respectively. 
The output is computed as \( C = W A^{T} \), where \( W \) is unknown.  
For each training batch, the attacker can solve this linear system only if \( \text{rank}(A^T) = n_1 \), which requires \( B \geq n_1 \). 
If \( B < n_1 \), then \( \text{rank}(A^T) < n_1 \), making it impossible for the attacker to reconstruct the true layer parameters. 
Furthermore, since layer parameters are updated in each batch, the attacker remains unable to derive the parameters throughout the training process. 
Therefore, to secure masked parameters in linear layers, it is crucial to set a small batch size such that \( B < n_1 \).  

For convolutional (Conv) layers, assuming the input channel equals 1, we denote the input shape as \( h \times h \), the kernel size as \( n \times n \), the padding as \( p \), and the stride as \( s \). 
The output shape of a convolutional layer is computed as \( (h - n + 2p)/s + 1 \) \cite{conv_output_shape}.  
The attacker can construct a linear system using the input and output, but this system has a solution only if \( (h - n + 2p)/s + 1 \geq n \), which implies \( n < (h + 2p + s)/(s + 1) \). 
If this condition is not met, the attacker cannot reconstruct the layer parameters. Thus, to secure masked parameters in convolutional layers, it is crucial to use a large kernel size and stride to satisfy \( n > (h + 2p + s)/(s + 1) \).

\section{Evaluation} \label{sec:evaluation}

\vspace{-3pt}
\subsection{Experimental Setup} \label{sec:exp_setup}
\noindent \textbf{Models and Datasets.}
The bottom model employs five architectures: MLP3, LeNet5 \cite{lecun1998lenet}, VGG13 \cite{simonyan2014vgg}, ResNet18 \cite{he2016resnet}, and Transformer-based DistilBERT \cite{sanh2019distilbert}, for both active and passive parties. 
We use 13 datasets across different modalities. 
\emph{Image}: MNIST (abbreviated as M) \cite{mnist_dataset}, FMNIST (FM) \cite{fmnist_dataset}, SVHN \cite{svhn_dataset}, CIFAR10 (CF10), CIFAR100 (CF100) \cite{cifar10_dataset}, CINIC10 (CN10) \cite{cinic10_dataset}, and TinyImageNet (TI) \cite{tiny_imagenet_dataset}.
\emph{Tabular}: TabMNIST (TM), TabFMNIST (TFM), and CRITEO (CR) \cite{criteo_dataset}.
\emph{Text}: TREC \cite{trec_dataset}, AG's News (NEWS) \cite{ag_news_dataset}, and IMDB \cite{imdb_dataset}.
A detailed explanation of these datasets is provided in Appendix \ref{appendix:configs}.
For image/text datasets, data is evenly split into sub-images/sub-texts for each party. 
Similarly, for tabular datasets, feature vectors are evenly divided into sub-vectors.

\noindent \textbf{Metrics.}
For both the VFL main task and MC attack, we report \textit{Top-1} accuracy across all datasets, except for CR and IMDB, for which we report \textit{AUC}, and CF100 and TI, where we report \textit{Top-5} accuracy.
To evaluate the efficiency of different methods, we measure the average running time per training epoch and total running time required to achieve a targeted accuracy threshold (e.g., 90\%).

\noindent \textbf{Compared Methods.}
We compare VMask with 12 other defense methods, including perturbation-based methods NG, CG, DG \cite{fu2022label}, and FedPass \cite{gu2023fedpass}, confusion-based methods LabelDP \cite{ghazi2021labeldp} and KD$k$ \cite{arazzi2024kdk}, regularization-based methods MID \cite{zou2023mutual} and dCor \cite{sun2022label}, and cryptography-based methods ACML \cite{zhang2020additively}, SPNN \cite{zhou2022toward}, SFA \cite{cai2022secure}, and BlindFL \cite{fu2022blindfl}. Detailed descriptions of these methods are provided in Section \ref{sec:related_work}. To demonstrate the efficacy and efficiency of our layer selection strategy, we also compare VMask with 3 variants: VMask-AS, VMask-RS, and VMask-ALLS. These variants use the same training framework as VMask but differ in their layer selection strategies. Specifically, VMask-AS employs an accumulative layer selection strategy, VMask-RS selects the same number of layers as VMask in each epoch but does so randomly, and VMask-ALLS selects all layers for masking in every epoch. Further details on the hyperparameters of these methods are provided in Appendix ~\ref{appendix:configs}.


\noindent \textbf{Implementations.} 
We build a VFL system on a cluster of Ubuntu 22.04 servers, each equipped with a 64-core Intel Xeon CPU, 128GB RAM, and two NVIDIA RTX 3090 GPUs, connected via Gigabit Ethernet (836 Mbit/s).
All our implementations are based on this VFL system.
\emph{(1) VMask Implementation.} 
Using PyTorch \cite{paszke2019pytorch} and Transformer \cite{wolf-etal-2020-transformers} frameworks, we train plaintext layers and implement a custom additive SS scheme for training masked layers. 
Auxiliary datasets are generated by randomly flipping/cropping images or by inserting/deleting words in text datasets. 
Communication is managed via Python socket \cite{python-socket}.
\emph{(2) Other Defense Methods.}  
We use open-sourced code when available, and reproduce methods when not. 
For Paillier-based cryptography, we use the python-paillier library \cite{PythonPaillier} with a 1024-bit key.
\emph{(3) MC Attack Implementation.} 
We append an MLP2 inference head to the trained bottom model and train it with labeled samples in a supervised manner.
Details about the number of labeled samples used for each dataset are provided in Table \ref{tab:models_and_datasets} in Appendix \ref{appendix:configs}.

\begin{table*}
	\caption{
		Comparison of the effectiveness of various defense methods in defending against MC attack while maintaining main task accuracy, evaluated using the best-performing model from the VFL training history.
	}
	\label{tab:effectiveness_best_model}
	\centering
	\resizebox{\linewidth}{!}{%
	\begin{threeparttable}[t]
		\footnotesize
		\begin{tabular}{c||c||ccc|c||ccc|c||ccc|c||ccc|c||ccc|c}
			\toprule

			\multirow{2}{*}{Metric} & \multirow{2}{*}{Method}  & \multicolumn{3}{c|}{MLP3} & \multirow{2}{*}{\makecell{Avg.\\Diff.}} & \multicolumn{3}{c|}{LeNet5} & \multirow{2}{*}{\makecell{Avg.\\Diff.}} & \multicolumn{3}{c|}{VGG13} & \multirow{2}{*}{\makecell{Avg.\\Diff.}} & \multicolumn{3}{c|}{ResNet18} & \multirow{2}{*}{\makecell{Avg.\\Diff.}} & \multicolumn{3}{c|}{\add{DistilBERT}} & \multirow{2}{*}{\makecell{Avg.\\Diff.}}\\
		  	& & TM & TFM & CR &  & M & FM & SVHN &  & CF10 & CF100 & \add{TI} &  & CF10 & CF100 & \add{TI} &  & \add{TREC} & \add{NEWS} & \add{IMDB} & \\
			\midrule

			\multirow{14}{*}{\makecell{Main\\Accuracy(\%)}} & Alone  & 91.87 & 85.61 & 64.69 & -6.99& 93.92 & 86.33 & 73.44 & -7.76& 84.93 & 80.39 & \add{62.40} &-7.07 & 87.64 & 86.06  & \add{63.81} & -6.75 & \add{25.20} & \add{33.43} & \add{52.03} & -58.42 \\
			& Vanilla  & 99.02 & 90.59 & 73.52 & 0& 98.72 & 90.99 & 87.26 & 0& 90.74 & 85.24  & \add{72.95} & 0& 92.11 & 91.14  & \add{74.50} & 0& \add{95.60} & \add{92.57} & \add{97.75} & 0\\
			
			\cmidrule(r){2-22}
			& NG \cite{fu2022label}       & 98.97 & 90.44 & 73.31 & -0.14 & 98.55 & 91.11 & 87.88 &  +0.19 & 90.45 & 84.72  & \add{64.02} &  -3.25 & 91.95 & 88.01  & \add{66.17} &-3.87  & \add{95.40} & \add{92.36} & \add{97.71} & -0.15 \\
			  
			& CG \cite{fu2022label}       & 99.05 & 90.72 & 73.46 & +0.03& 98.71 & 91.03&86.22 & -0.34&87.78 &85.19 & \add{70.27} & -1.90 & 89.42&90.00  & \add{72.28} &-2.02 &\add{96.40} &\add{92.34} &\add{97.70} & +0.17  \\
			
			& DG \cite{fu2022label}        &98.91 &90.25 &70.71 & -1.09&98.45 &90.16 &84.61 & -1.25&90.63 &80.60  & \add{61.37} & -5.44& 92.09 &84.90 & \add{62.30} & -6.15& \add{94.80}& \add{92.43}&\add{97.76} & -0.31 \\
			
			& FedPass \cite{gu2023fedpass}   & 98.98&90.21 &73.43 &-0.17 &98.29 &90.89 & 87.46 & -0.11& 90.77&86.31& \add{67.94} &-1.30 & 91.84 &89.75  & \add{68.08} & -2.69&\add{94.00} & \add{91.05}&\add{97.73}&-1.05  \\

			& LabelDP \cite{ghazi2021labeldp}     & 98.98&90.34 &73.39 & -0.14&97.49 &89.97 &85.18 &-1.44 &87.24 &29.19  &\add{4.23} &-42.76 & 89.28 & 38.25 & \add{5.45} & -41.59&\add{94.00} &\add{91.37} &\add{97.64} & -0.97 \\

			& \add{KD$k$} \cite{arazzi2024kdk}     & \add{97.36} & \add{88.68}& \add{62.53} & -4.85& \add{94.70}& \add{86.74}& \add{76.49} & -6.35& \add{85.64}& \add{82.24} & \add{68.77} & -4.09& \add{88.18}& \add{88.74}& \add{70.98} & -3.28 & \add{20.20}& \add{33.47}& \add{51.11}&-60.38  \\
			
			& dCor \cite{sun2022label}     & 99.01 & 90.35 & 71.10 &-0.89 & 98.72 & 91.05 & 81.75 &-1.82 & 89.96 & 85.56 & \add{63.56}& -3.28& 91.06 & 91.37 &\add{64.23} & -3.70 & \add{94.20}& \add{25.26}&\add{50.13}& -38.78 \\

			& MID \cite{zou2023mutual}    & 98.46& 89.67&71.68 & -1.11&97.82 &90.58 &75.72 &-4.28 &85.95 &80.79  & \add{62.08} & -6.70 & 89.80 &85.57  &\add{63.83} &-6.18 & \add{96.40}& \add{91.37}&\add{97.71}&-0.15  \\

			\cmidrule(r){2-22}
			
			& VMask-RS  & 98.95 & 90.45 & 73.44 & -0.10& 98.70 & 90.73 & 87.15 & -0.13& 90.08 & 85.10  & \add{71.93} &-0.61 & 91.40 & 91.08  & \add{73.58} & -0.56& \add{95.20}& \add{92.25}&\add{97.21} & -0.42 \\
            & \textbf{VMask-AS} & 99.01 & 90.53 & 73.47 & \textbf{-0.04} & 98.72 & 90.83 & 87.11 & \textbf{-0.10} & 90.16 & 85.22 & \add{72.22} & \textbf{-0.44} & 92.03 & 90.78  & \add{73.48} & \textbf{-0.49} & \add{96.00}& \add{92.37}& \add{97.72} & \textbf{+0.06} \\
            & \textbf{VMask-ALLS}  & 99.02 & 90.52 & 73.46 & \textbf{-0.04} & 98.69 & 90.85 & 87.09 & \textbf{-0.11} & 90.24 & 85.11  & \add{72.08}& \textbf{-0.50}&  92.00 & 90.79 &\add{73.71} & \textbf{-0.42} & \add{95.30}& \add{91.78}&\add{97.32} & \textbf{-0.51} \\
			& \textbf{VMask} & 99.00 & 90.49 & 73.46 & \textbf{-0.06} & 98.66 & 90.64 & 87.11 & \textbf{-0.19} & 90.28 & 85.22  & \add{73.19} &  \textbf{-0.08} & 92.01 & 90.85 &    \add{73.88} & \textbf{-0.34} & \add{95.80}& \add{92.13}&\add{97.72} & \textbf{-0.09} \\
			
			\hhline{======================}
			    
			\multirow{14}{*}{\makecell{Attack \\Accuracy(\%)}} & Scratch  &41.35 & 50.25& 52.42 & 0&37.67 & 45.37& 15.59 &0 & 14.46&21.56  & \add{9.88} & 0& 16.87 &20.50  & \add{9.12} &0 & \add{19.55} & \add{28.66} & \add{51.75} & 0\\  
			& Vanilla  &80.72 &67.34  & 58.64 & +20.89&79.44 &  75.21& 53.72 & +36.58&73.62 &63.44 &  \add{46.95} &+46.04 & 72.87& 63.49 & \add{43.41} &+44.43 & \add{40.40} & \add{51.61} & \add{93.45} &+28.50 \\  
			
			\cmidrule(r){2-22}
			& NG \cite{fu2022label}      & 77.26&65.83 &57.60 & +18.89&77.90 &75.62 &55.10 &+36.66&73.50 &39.88 &\add{55.40} & +40.96&71.29 &29.15  &\add{55.97} & +36.64 & \add{49.15}& \add{44.16}&\add{76.28} & +23.21 \\
			& CG \cite{fu2022label}     &81.32 &67.80 &56.85 &+20.65 &78.70 &75.14 &50.33 & +35.18&49.19 &62.82 & \add{48.20} &  +38.1&49.44 &65.46 &\add{46.85} & +38.42& \add{42.35}&\add{53.30} &\add{92.64}& +29.44 \\
			   
			& DG \cite{fu2022label}     & 78.25& 68.67&51.98 &+18.29 &77.99 &67.02 &51.57 &+32.65 &67.12 &7.28 & \add{56.09} &+28.2  &70.79 &10.41  &\add{56.31} & +30.34&\add{33.25} & \add{49.98}&\add{79.37} & +20.88 \\
			   
			& FedPass \cite{gu2023fedpass}  &80.61 & 67.34& 58.18 & +20.70&76.40 &75.32 &55.41 & +36.17&74.51 &63.70  & \add{42.73} &  +45.01 &74.75&66.52  &\add{47.23} &  +47.34&\add{38.75} &\add{51.19}&\add{59.73} & +16.57 \\
			   
			& LabelDP \cite{ghazi2021labeldp}  & 82.54& 71.11& 57.63 &+22.42 &76.83 &74.86 &48.03 &  +33.70&55.91 &18.09 & \add{2.50} & +10.20 &57.56 &19.28  &\add{5.87} & +12.07&\add{25.05} &\add{51.06} &\add{59.37} &+11.84 \\
			   
			& \add{KD$k$} \cite{arazzi2024kdk}     &\add{81.93} &\add{72.85 }& \add{51.57} & +20.78&\add{73.09} &\add{69.84} &\add{49.88} & +31.39&\add{61.91} &\add{45.91}  &\add{36.06} & +32.66 &\add{59.71} &\add{56.65}  &\add{38.16} & +36.01&\add{20.45} &\add{31.05} &\add{60.72} &+4.09 \\
			    
			& dCor \cite{sun2022label}    & 75.13 & 61.81 & 55.20 & +16.04& 76.97 & 73.03 & 33.72 & +28.36& 62.06 & 57.25  &\add{42.54} & +38.65& 58.00 & 60.94 &\add{40.29} &+37.58 & \add{46.80}& \add{25.00}& \add{51.23} &+7.69 \\
			   
			& MID \cite{zou2023mutual}     &72.01 &57.26 &50.00 & +11.75&59.79 &60.31 &14.42 &+11.96 &26.38 & 5.00 &\add{56.14} &  +13.87&45.71 &15.96  &\add{55.98} & +23.72 &\add{60.90} &\add{45.56} &\add{89.30} &+31.93 \\

			\cmidrule(r){2-22}
			& VMask-RS  & 43.81 & 61.12 & 56.44 &+5.78 & 41.84 & 40.62 & 16.42 & +0.08& 59.20 & 55.00 & \add{14.31} &+27.54  & 65.30 & 58.34  & \add{39.74} & +38.96& \add{29.33} & \add{38.34}&\add{64.39} &+10.7 \\
            & \textbf{VMask-AS}  & 43.44 & 50.92 & 52.89 & \textbf{+1.08} & 37.95 & 43.97 & 14.72 & \textbf{-0.66} & 15.94 & 25.67  & \add{12.23} & \textbf{+2.65} & 18.04 & 42.92  & \add{18.51} &\textbf{+10.99} & \add{27.00}& \add{38.42}&\add{67.98} & \textbf{+11.15} \\
            & \textbf{VMask-ALLS}  & 42.60 & 48.61 & 52.22 & \textbf{-0.20} & 38.27 & 45.09 & 13.05 &  \textbf{-0.74} & 13.14 & 11.81 &\add{5.43} &  \textbf{-5.17} & 13.42 & 9.91 & \add{5.60} & \textbf{-5.85} & \add{19.34} &\add{27.78} &\add{51.66} & \textbf{-0.39} \\
			& \textbf{VMask}  & 42.85 & 50.76 & 52.78 & \textbf{+0.79} & 38.64 & 44.14 & 14.33 &  \textbf{-0.51} & 15.29 & 22.75 & \add{10.31} & \textbf{+0.82} & 17.52 & 25.98 & \add{13.22} &  \textbf{+3.41} & \add{28.47} &\add{38.21} &\add{57.65}& \textbf{+8.12} \\

			\bottomrule
		\end{tabular}
		\begin{tablenotes}[flushleft]
			\footnotesize
			\item[1] ``Vanilla'' means standard VFL model training without applying any defense methods. 
			``Alone'' means the active party trains the model independently with its own features and labels without the collaboration with any passive party. 
			``Scratch'' means MC attack based on randomly initialized bottom model. 
			The Avg. Diff. column shows the difference in main/attack accuracy between each method and the baseline  (``Vanilla'' for Main Accuracy or ``Scratch'' for Attack Accuracy) for each model architecture, averaged across three datasets. 
			For Main Accuracy, a negative value in Avg. Diff. column indicates a drop in model accuracy compared to vanilla VFL. For Attack Accuracy, a positive value in Avg. Diff. column indicates increased label leakage compared to a random bottom model.
		\end{tablenotes}
	\end{threeparttable}
	}
	\vspace{-11pt}
\end{table*}

\begin{figure*}
	\captionsetup[subfigure]{aboveskip=-0.5pt,belowskip=-0.5pt}
	\centering
	\begin{minipage}{1.0\textwidth}
		\centering
		\includegraphics[width=0.95\linewidth, trim=50 450 50 0, clip]{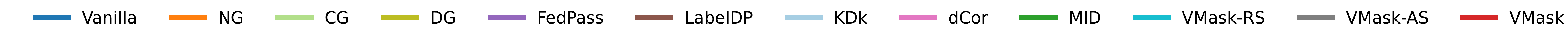}
	\end{minipage}

	\vspace{-2pt}

	\begin{minipage}{1.0\textwidth}
		\centering
		\begin{subfigure}[b]{0.19\textwidth} 
			\includegraphics[width=\textwidth]{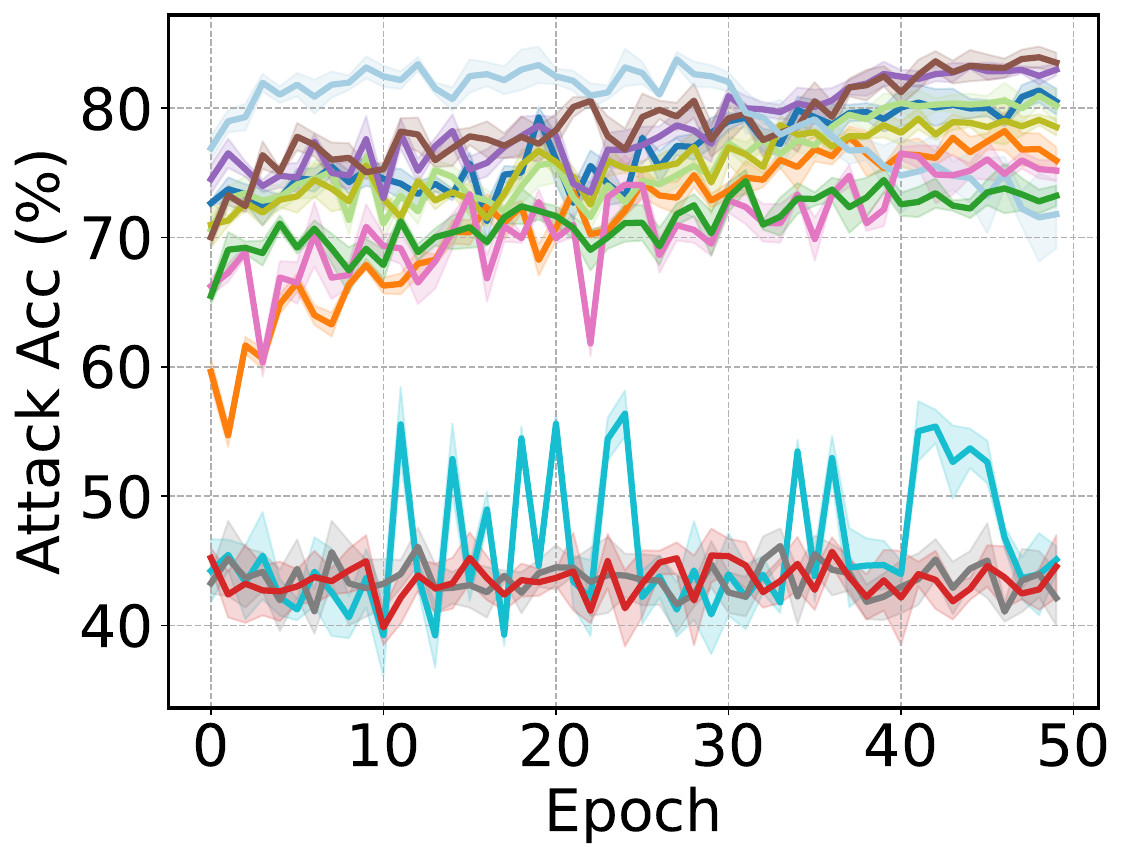}
			\caption{MLP3, TM}
			\label{mlp_tabmnist_every_epoch_attack}
		\end{subfigure}\hfill
		\begin{subfigure}[b]{0.19\textwidth}
			\includegraphics[width=\textwidth]{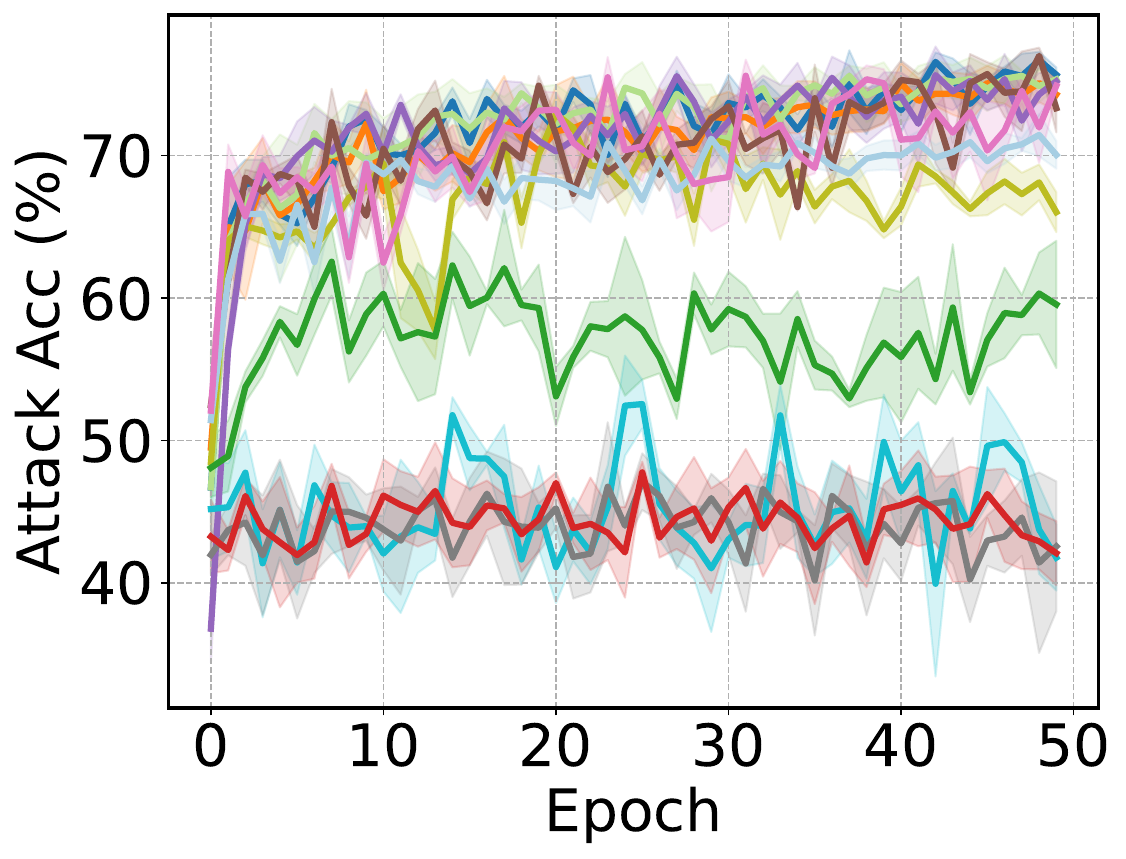}
			\caption{LeNet5, FM}
			\label{lenet_fmnist_every_epoch_attacl}
		\end{subfigure}\hfill
		\begin{subfigure}[b]{0.19\textwidth}
			\includegraphics[width=\textwidth]{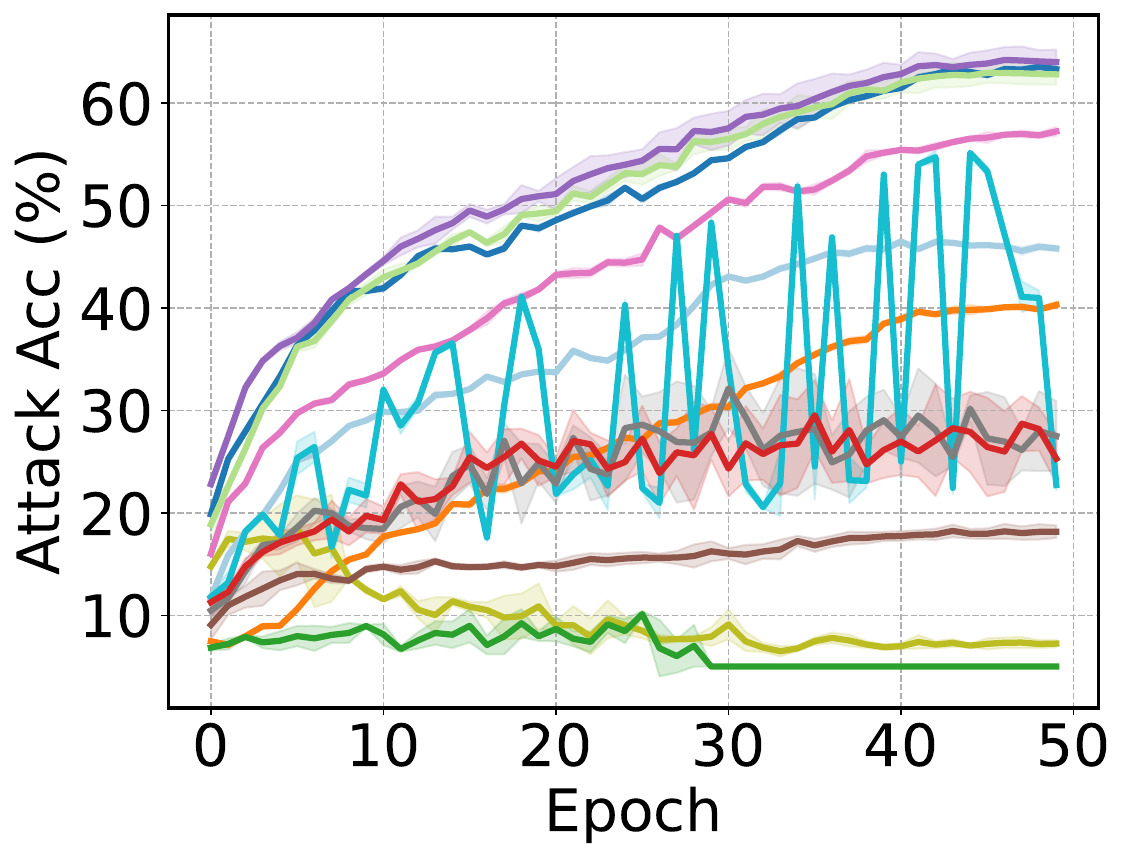}
			\caption{VGG13, CF100}
			\label{vgg13_cifar100_every_epoch_attack}
		\end{subfigure}\hfill
		\begin{subfigure}[b]{0.19\textwidth}
			\includegraphics[width=\textwidth]{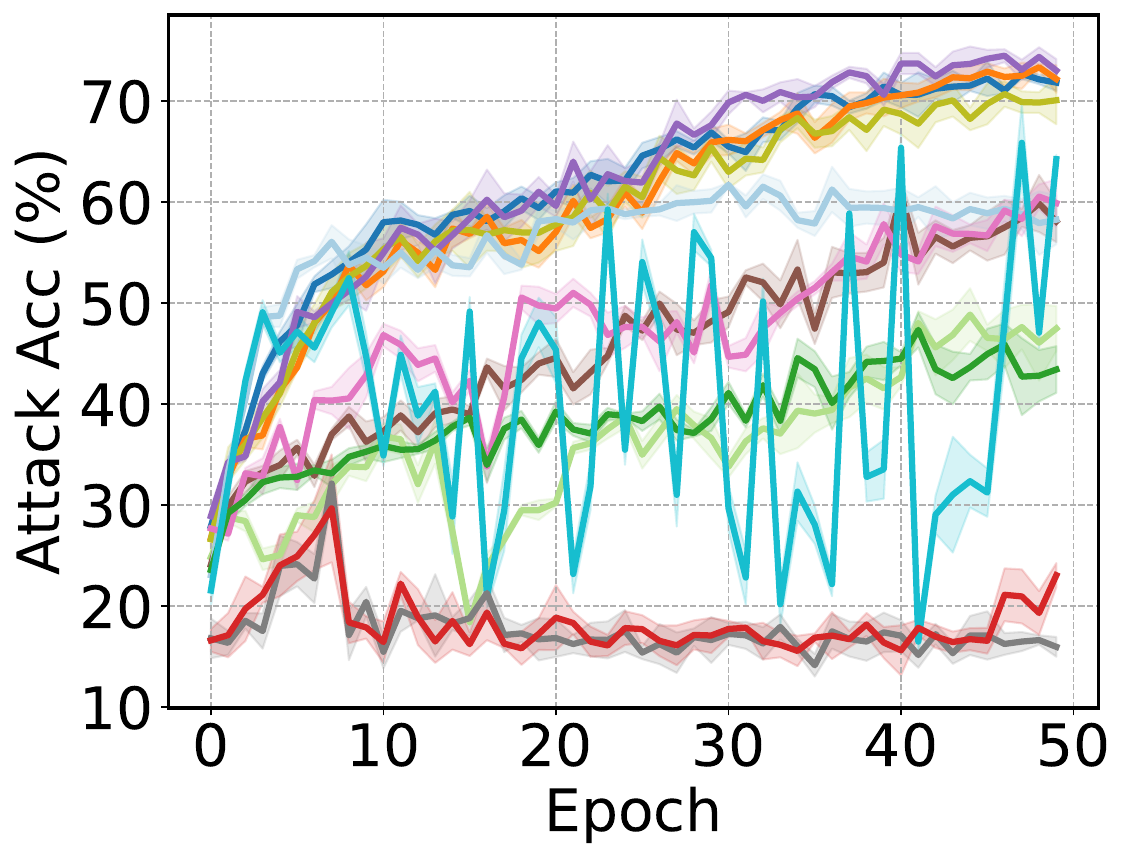}
			\caption{ResNet18, CF10}
			\label{resnet18_cifar10_every_epoch_attack}
		\end{subfigure}\hfill
		\begin{subfigure}[b]{0.19\textwidth}
			\includegraphics[width=\textwidth]{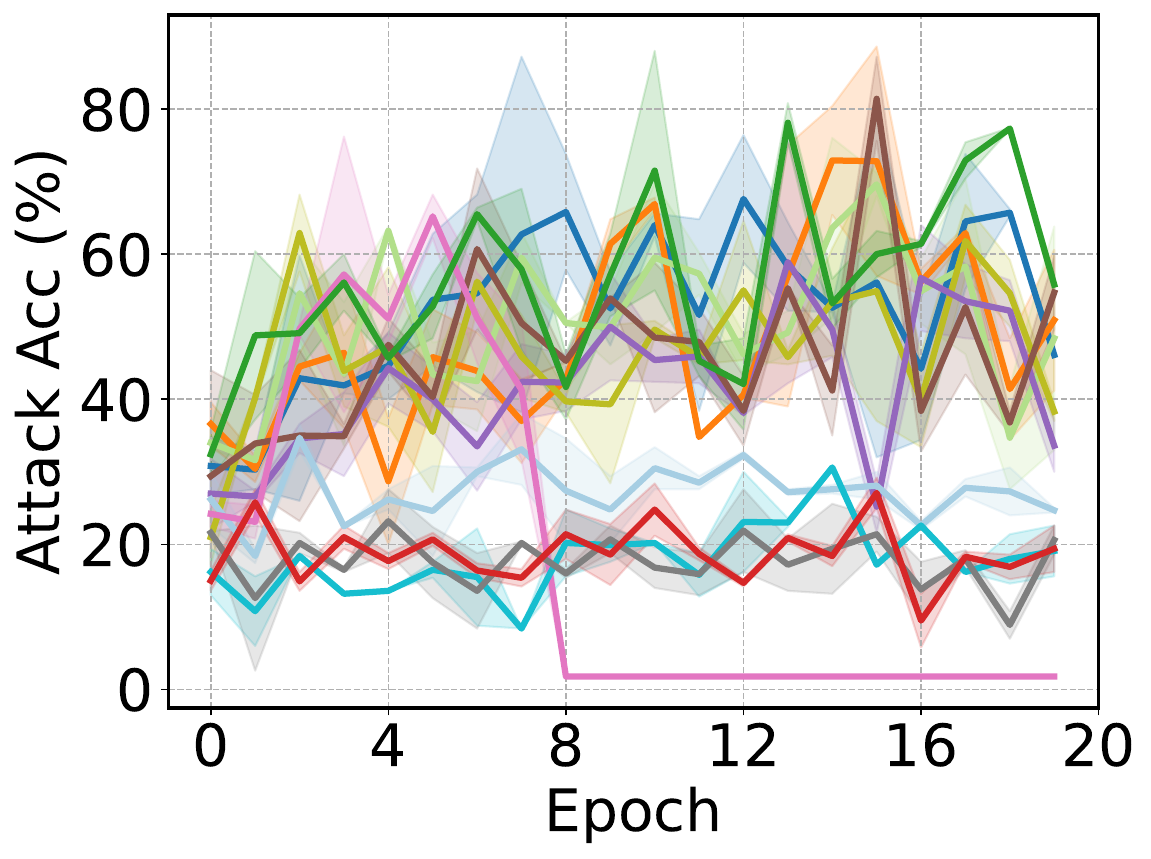}
			\caption{DistilBERT, TREC}
			\label{distilbert_trec_every_epoch_attack}
		\end{subfigure}\hfill
	\end{minipage}
	
	\caption{Comparison of defense methods' effectiveness against MC attack, evaluated using models trained at each epoch.}
	\label{fig:every_epoch_attack}
    \vspace{-12pt}
\end{figure*}

\begin{figure*}[t]
	\captionsetup[subfigure]{aboveskip=-0.5pt,belowskip=-0.5pt}
	\centering
	\begin{minipage}{1.0\textwidth}
		\centering
		\includegraphics[width=0.95\linewidth, trim=50 454 50 0, clip]{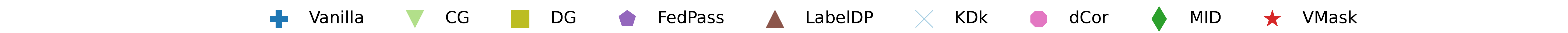}
	\end{minipage}

	\vspace{-2pt}

	\begin{minipage}{1.0\textwidth}
		\centering
		\begin{subfigure}[b]{0.19\textwidth} 
			\includegraphics[width=\textwidth]{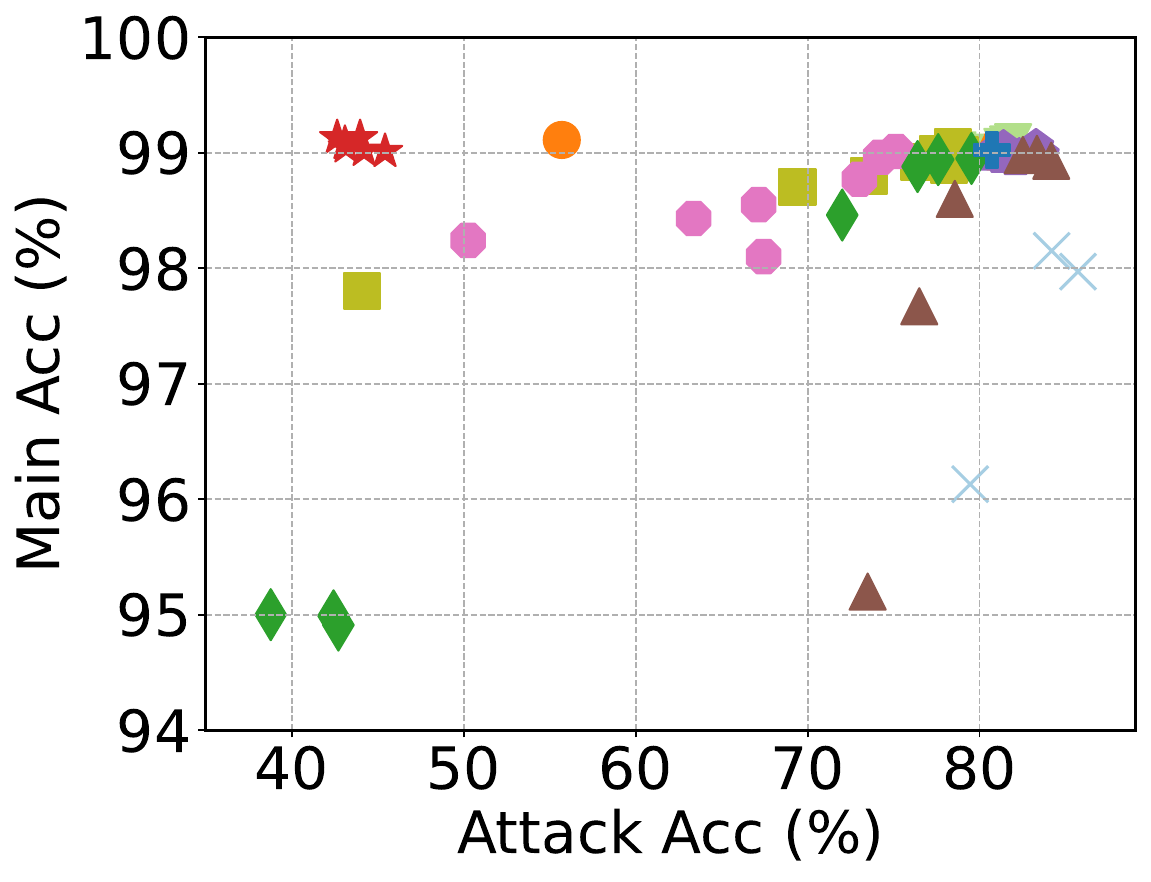}
			\caption{MLP3, TM}
			\label{tradeoff_mlp_tab_mnist}
		\end{subfigure}\hfill
		\begin{subfigure}[b]{0.19\textwidth}
			\includegraphics[width=\textwidth]{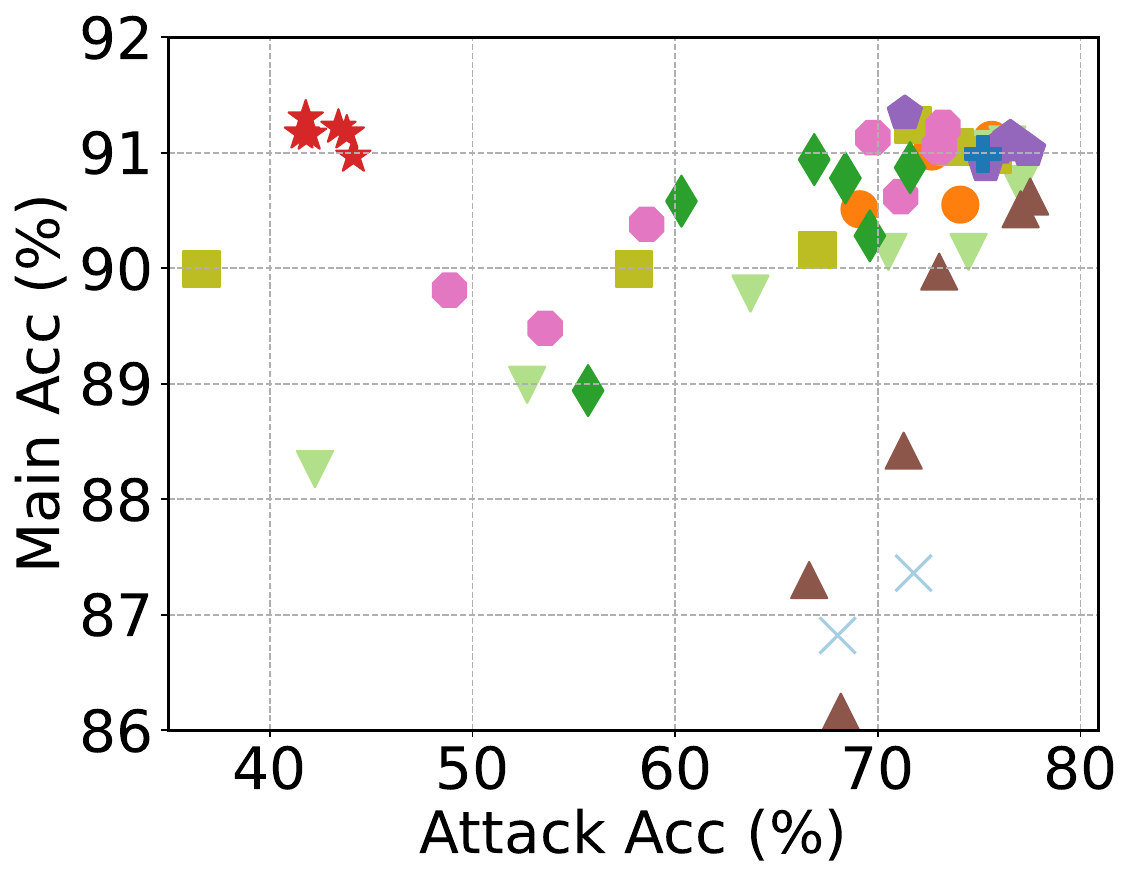}
			\caption{LeNet5, FM}
			\label{tradeoff_lenet_fashion_mnist}
		\end{subfigure}\hfill
		\begin{subfigure}[b]{0.19\textwidth}
			\includegraphics[width=\textwidth]{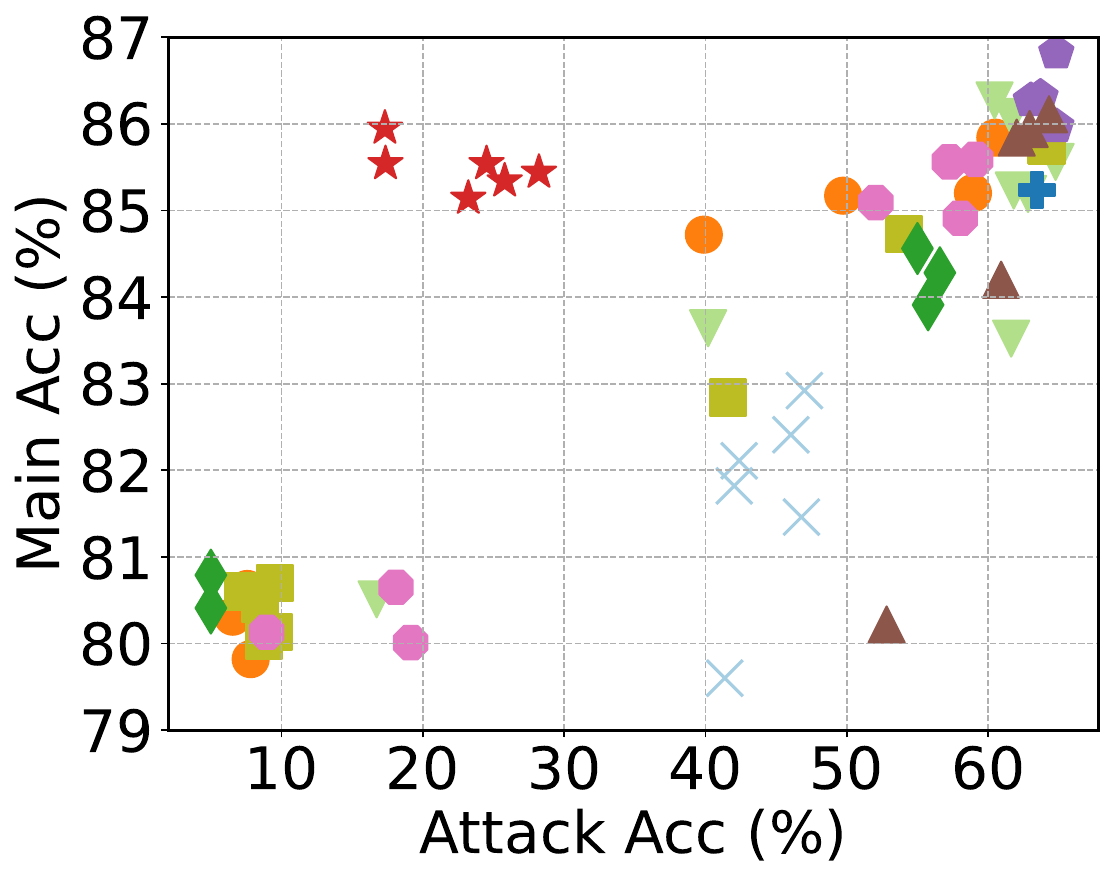}
			\caption{VGG13, CF100}
			\label{tradeoff_vgg13_cifar100}
		\end{subfigure}\hfill
		\begin{subfigure}[b]{0.19\textwidth}
			\includegraphics[width=\textwidth]{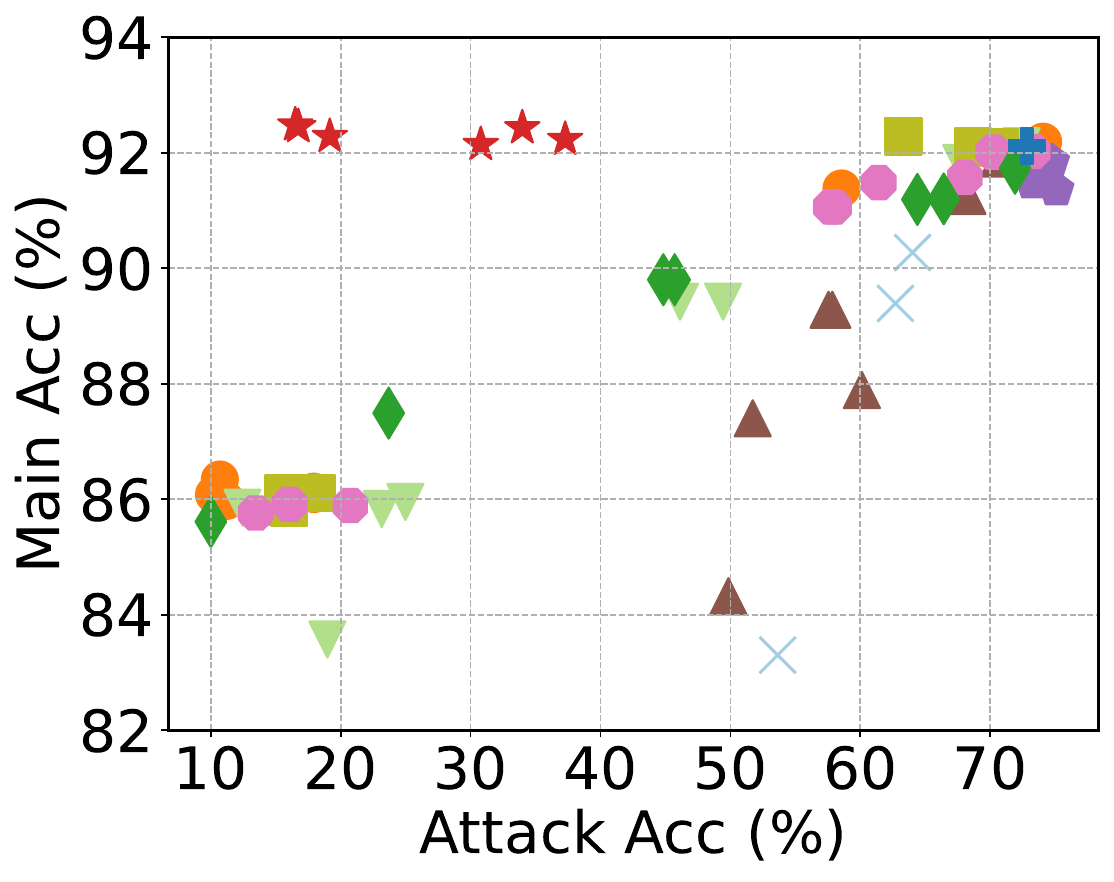}
			\caption{ResNet18, CF10}
			\label{tradeoff_resnet18_cifar10}
		\end{subfigure}\hfill
		\begin{subfigure}[b]{0.19\textwidth}
			\includegraphics[width=\textwidth]{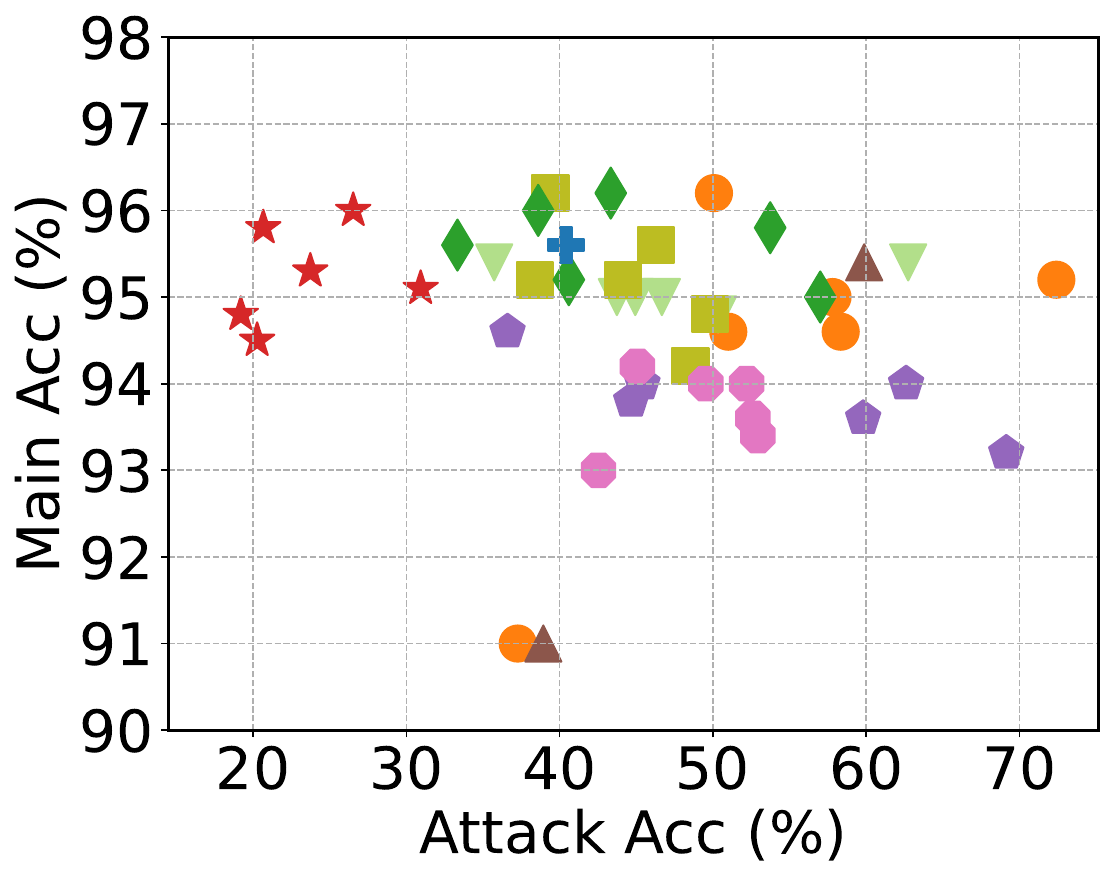}
			\caption{DistilBERT, TREC}
			\label{tradeoff_distilbert_trec}
		\end{subfigure}\hfill
	\end{minipage}
	
	\caption{Evaluation of the trade-off between label privacy and model utility across different defense methods.}
	\label{fig:tradeoff}
    \vspace{-14pt}
\end{figure*}

\subsection{Effectiveness of VMask} \label{sec:effectiveness}

We evaluate VMask's effectiveness in protecting label privacy in a two-party VFL setting, comparing it with other defense methods. 
Details about the privacy budget and auxiliary dataset size (typically only 1\% to 5\% of passive party's original dataset size) are shown in Table \ref{tab:models_and_datasets} in Appendix \ref{appendix:configs}. 
Each defense method is tested with three VFL model trainings and five MC attack runs, resulting a total of 15 runs. 
The reported accuracies are averages from these experiments.

First, we assess VMask's effectiveness on the best-performing bottom model from the VFL training history. 
As shown in Table \ref{tab:effectiveness_best_model}, VMask maintains nearly identical main task accuracy compared to vanilla VFL, with an averaged drop of at most 0.34\% in all cases and only 0.09\% in DistilBERT, 
while significantly reducing MC attack accuracy to levels comparable to the ``Scratch'' case. 
Here, ``Scratch'' refers to the attack accuracy obtained from a randomized bottom model, an inevitable leakage in any VFL system.
For instance, in the VGG13 model, the averaged attack accuracy increases by only 0.82\% compared to ``Scratch'', and in LeNet5, it is even 0.51\% lower than ``Scratch''. 
This suggests that the masked bottom model's mixture of randomized and well-trained layers complicates the attack. 
In contrast, other methods either fail to maintain high main accuracy or prevent label leakage. 
For instance, NG and FedPass preserve good main accuracy but exhibit high MC attack accuracy, nearing ``Scratch'' levels. 
Specifically, in the case of (DistilBERT, TREC), NG's leaked label privacy reaches 49.15\%, even higher than ``Scratch'' at 40.40\%.
On the other hand, Methods like MID and dCor reduce attack accuracy but significantly sacrifice main accuracy. 
For example, in (DistilBERT, NEWS), dCor achieves only 25.26\% main accuracy, which is even lower than ``Alone'' at 33.43\%. 
Similar results are observed in the CN10 dataset (see Table \ref{tab:appendix_effectiveness_cinic10} in Appendix \ref{appendix:effectiveness_cinic10}).

We also evaluate VMask's effectiveness on bottom models trained at each epoch, an aspect often overlooked in prior studies. 
As shown in Figure \ref{fig:every_epoch_attack}, for vanilla VFL, MC attack accuracy increases with each epoch due to improved feature embeddings extraction. 
Other defense methods, such as dCor, LabelDP, and MID, reduce attack accuracy but still exhibit a rising trend as training progresses. 
In contrast, VMask consistently achieves much lower attack accuracy compared to other methods, controlling leakage levels close to ``Scratch'' across all training epochs.
For example, in (DistilBERT, TREC) case, despite the instability of attack accuracy due to a limited number of labeled samples, a clear gap remains between VMask and other methods, with attack accuracy from VMask being relatively stable and close to the 19.55\% from ``Scratch''.


\begin{figure*}[t]
	\captionsetup[subfigure]{aboveskip=-0.5pt,belowskip=-0.5pt}
	\centering
	\begin{subfigure}[b]{0.085\textwidth} 
		\includegraphics[width=\textwidth]{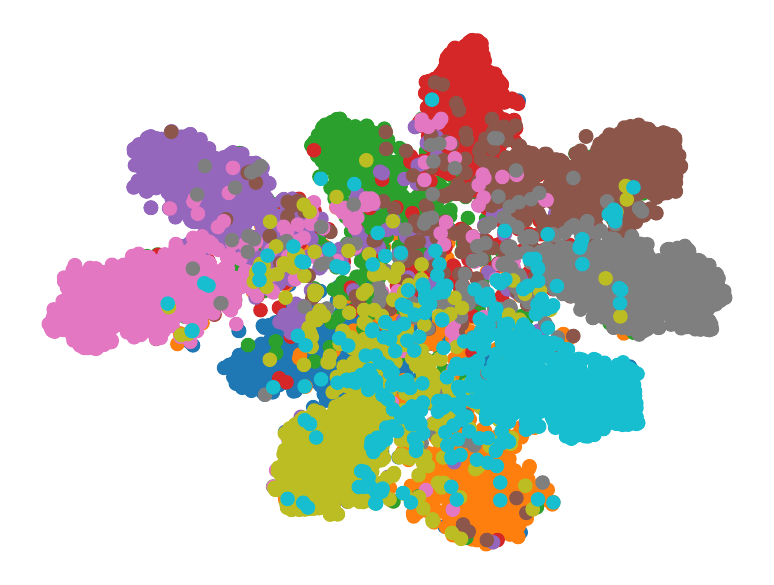}  
		\caption{Vanilla}
		\label{embedding_vanilla}
	\end{subfigure}\hfill
	\begin{subfigure}[b]{0.085\textwidth} 
		\includegraphics[width=\textwidth]{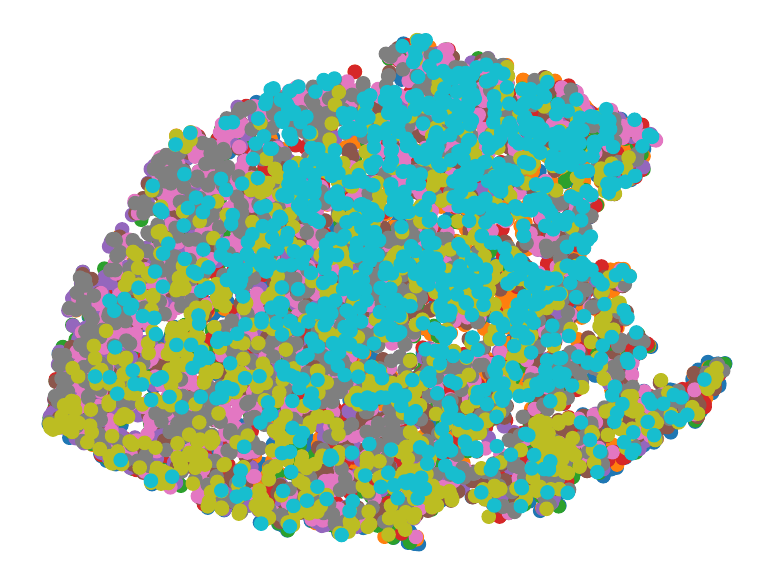}   
		\caption{Scratch}
		\label{embedding_random}
	\end{subfigure}\hfill
	\begin{subfigure}[b]{0.085\textwidth} 
		\includegraphics[width=\textwidth]{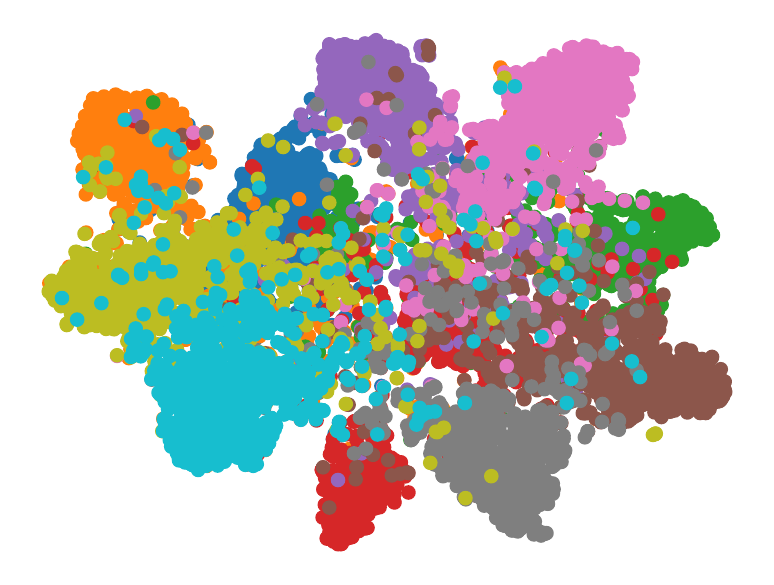}
		\caption{NG}
		\label{embedding_ng}
	\end{subfigure}\hfill
    \begin{subfigure}[b]{0.085\textwidth} 
		\includegraphics[width=\textwidth]{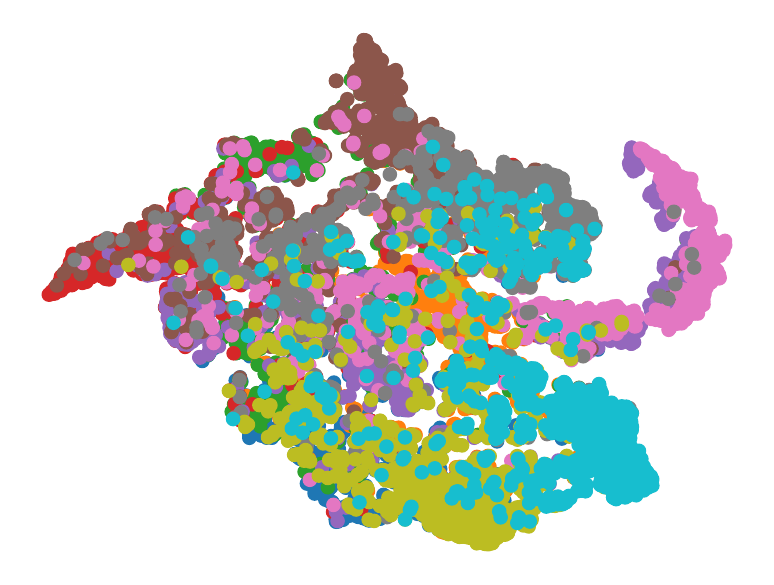}
		\caption{CG}
		\label{embedding_cg}
	\end{subfigure}\hfill
	\begin{subfigure}[b]{0.085\textwidth} 
		\includegraphics[width=\textwidth]{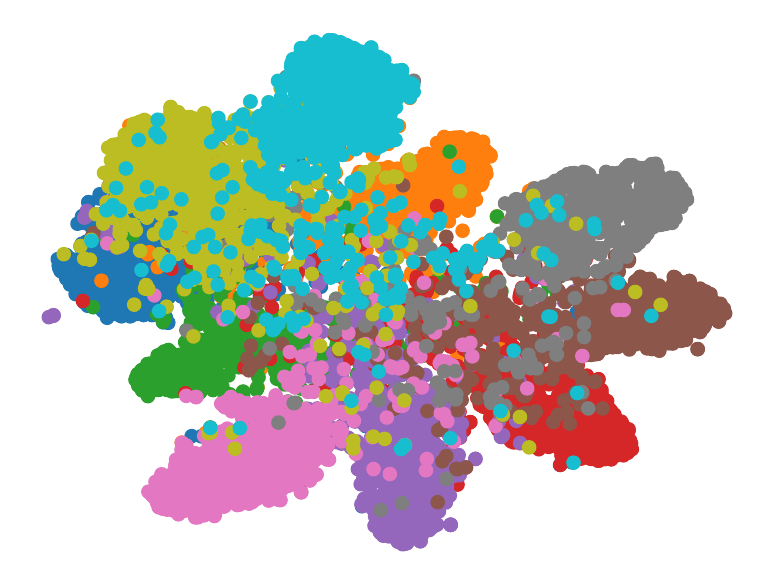}
		\caption{DG}
		\label{embedding_dg}
	\end{subfigure}\hfill
	\begin{subfigure}[b]{0.085\textwidth} 
		\includegraphics[width=\textwidth]{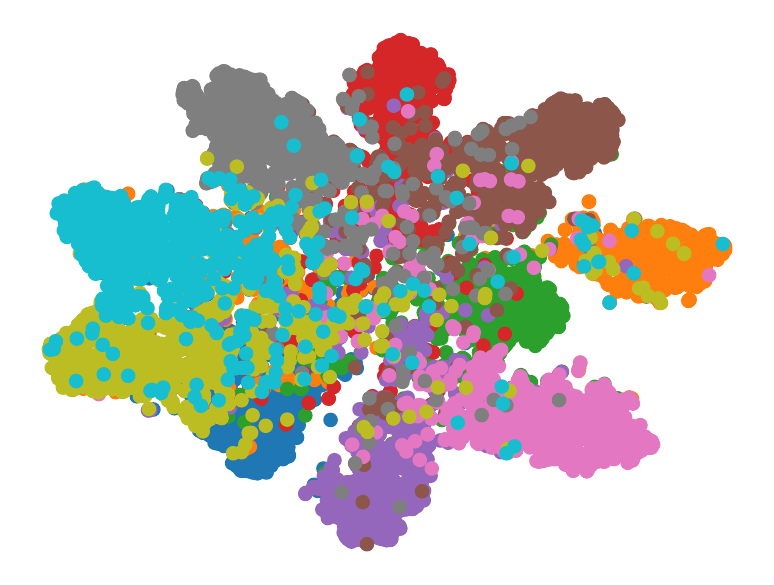}
		\caption{FedPass}
		\label{embedding_fedpass}
	\end{subfigure}\hfill
	\begin{subfigure}[b]{0.085\textwidth} 
		\includegraphics[width=\textwidth]{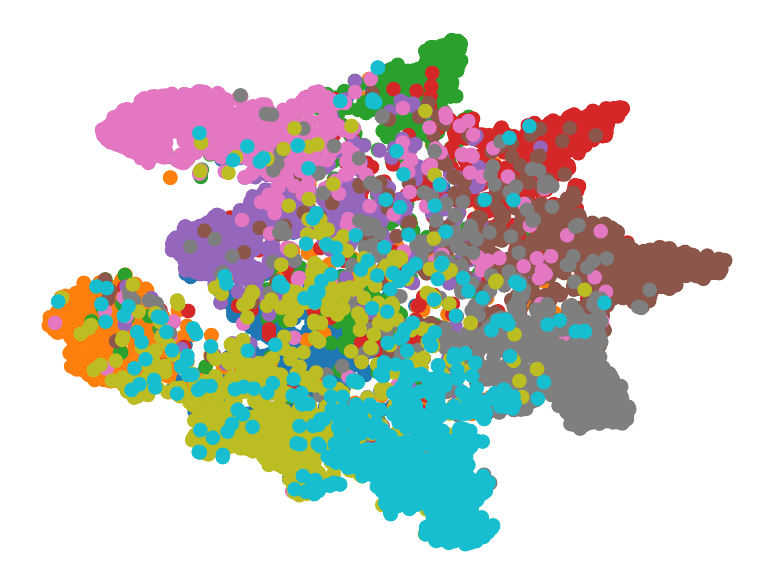}
		\caption{LabelDP}
		\label{embedding_labeldp}
	\end{subfigure}\hfill
	\begin{subfigure}[b]{0.085\textwidth} 
		\includegraphics[width=\textwidth]{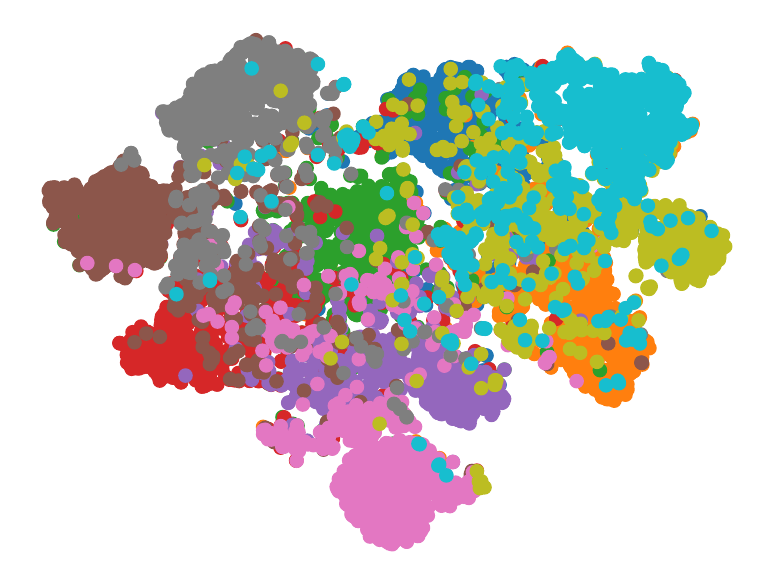}  
		\caption{KD$k$}
		\label{embedding_kdk}
	\end{subfigure}\hfill
	\begin{subfigure}[b]{0.085\textwidth} 
		\includegraphics[width=\textwidth]{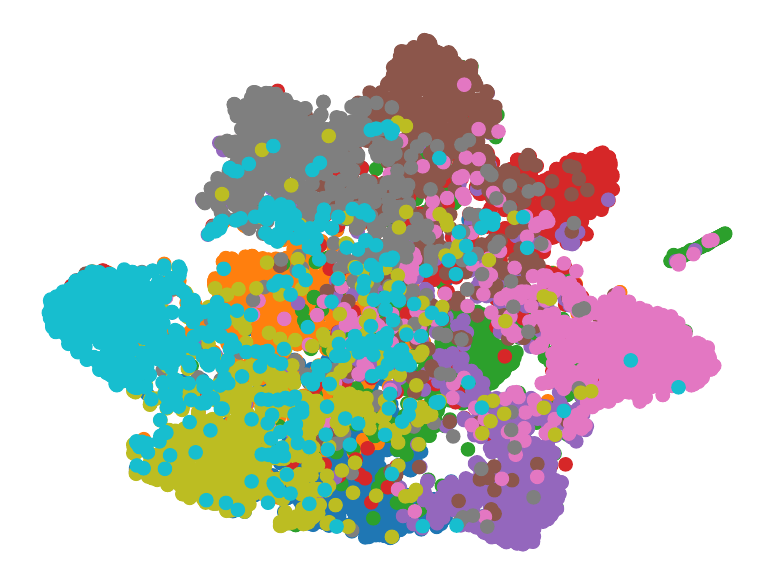}
		\caption{dCor}
		\label{embedding_dCor}
	\end{subfigure}\hfill
	\begin{subfigure}[b]{0.085\textwidth} 
		\includegraphics[width=\textwidth]{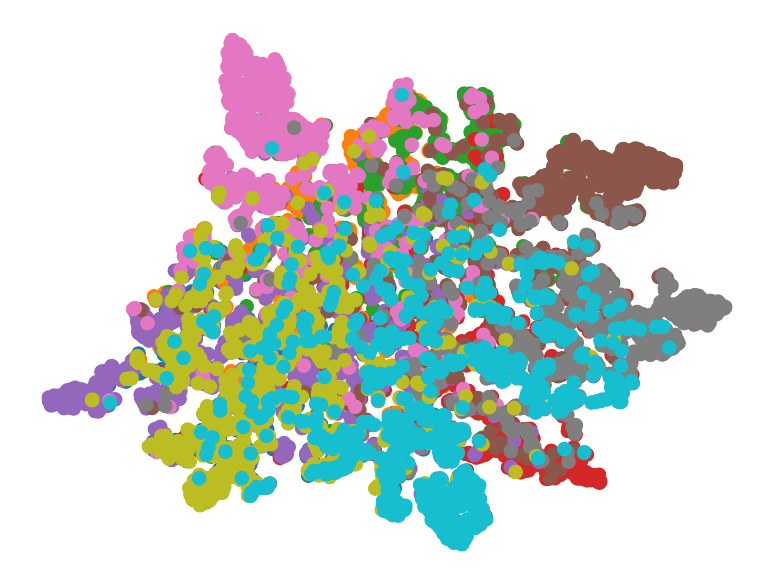}
		\caption{MID}
		\label{embedding_mid}
	\end{subfigure}\hfill
	\begin{subfigure}[b]{0.085\textwidth} 
		\includegraphics[width=\textwidth]{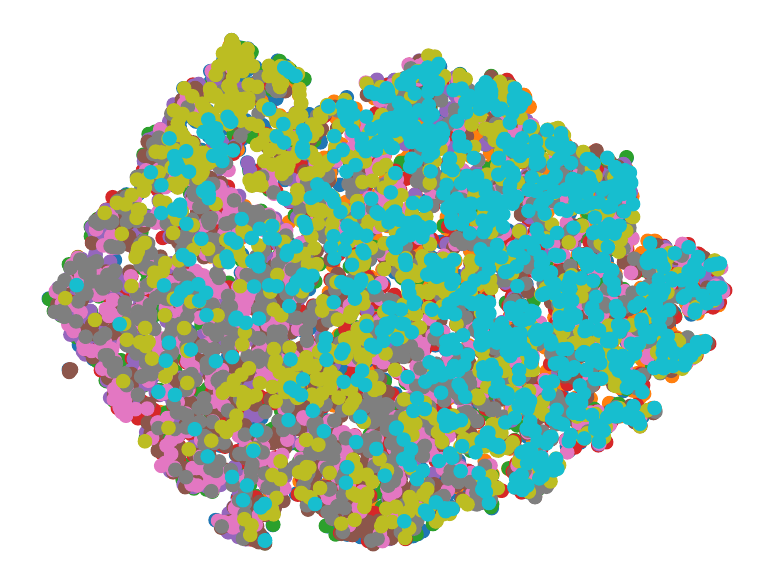}  
		\caption{VMask}
		\label{embedding_vmask}
	\end{subfigure}\hfill

    \caption{Comparison of feature embedding distributions across various defense methods in the (ResNet18, CF10) case.}
	\label{fig:embedding_comparision}
    \vspace{-18pt}
\end{figure*}

\subsection{Privacy-Utility Trade-off} \label{sec:tradeoff_evaluation}
In this section, we evaluate the trade-off between model utility and label privacy achieved by VMask and other defense methods. 
These methods typically involve a hyperparameter that controls the strength of label protection, such as privacy budget in LabelDP or mutual information weight in MID. 
For each method, we train VFL model with various hyperparameters, execute MC attack, and plot each (Attack Accuracy, Main Accuracy) pair, as depicted in Figure \ref{fig:tradeoff}.
We observe that the scatter points from other defense methods cluster in the top-right and bottom-left corners of the figure. 
The former indicates high main task accuracy but poor defense against MC attack, while the latter reflects effective prevention of MC attack at the cost of reduced main task accuracy, often dropping to the level achievable by the active party alone.
None of these methods are able to reduce attack accuracy while maintaining high main task accuracy, highlighting the inherent privacy-utility trade-off they cannot overcome.
In contrast, VMask's scatter points are concentrated in the top-left corner, a region other methods fail to reach, particularly evident for (ResNet18, CF10) and (DistilBERT, TREC). 
Notably, as the label protection strength increases,
VMask still maintains main task accuracy comparable to vanilla VFL. 
These results demonstrate that VMask is the only defense method capable of effectively breaking the privacy-utility trade-off.

\begin{figure}[t]
	\captionsetup[subfigure]{aboveskip=-0.5pt,belowskip=-0.5pt}
	\centering
	\begin{minipage}{1.0\linewidth}
		\centering
		\includegraphics[width=0.9\linewidth, trim=50 500 50 0, clip]{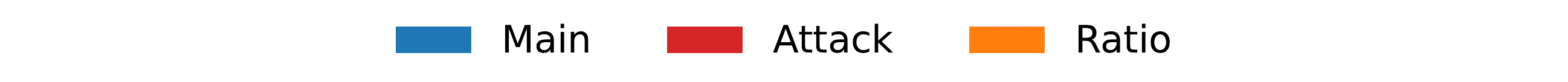}
	\end{minipage}

	\vspace{-2pt}
	
	\begin{minipage}{1.0\linewidth}
		\centering
		\begin{subfigure}[b]{0.49\linewidth}  
			\includegraphics[width=\linewidth]{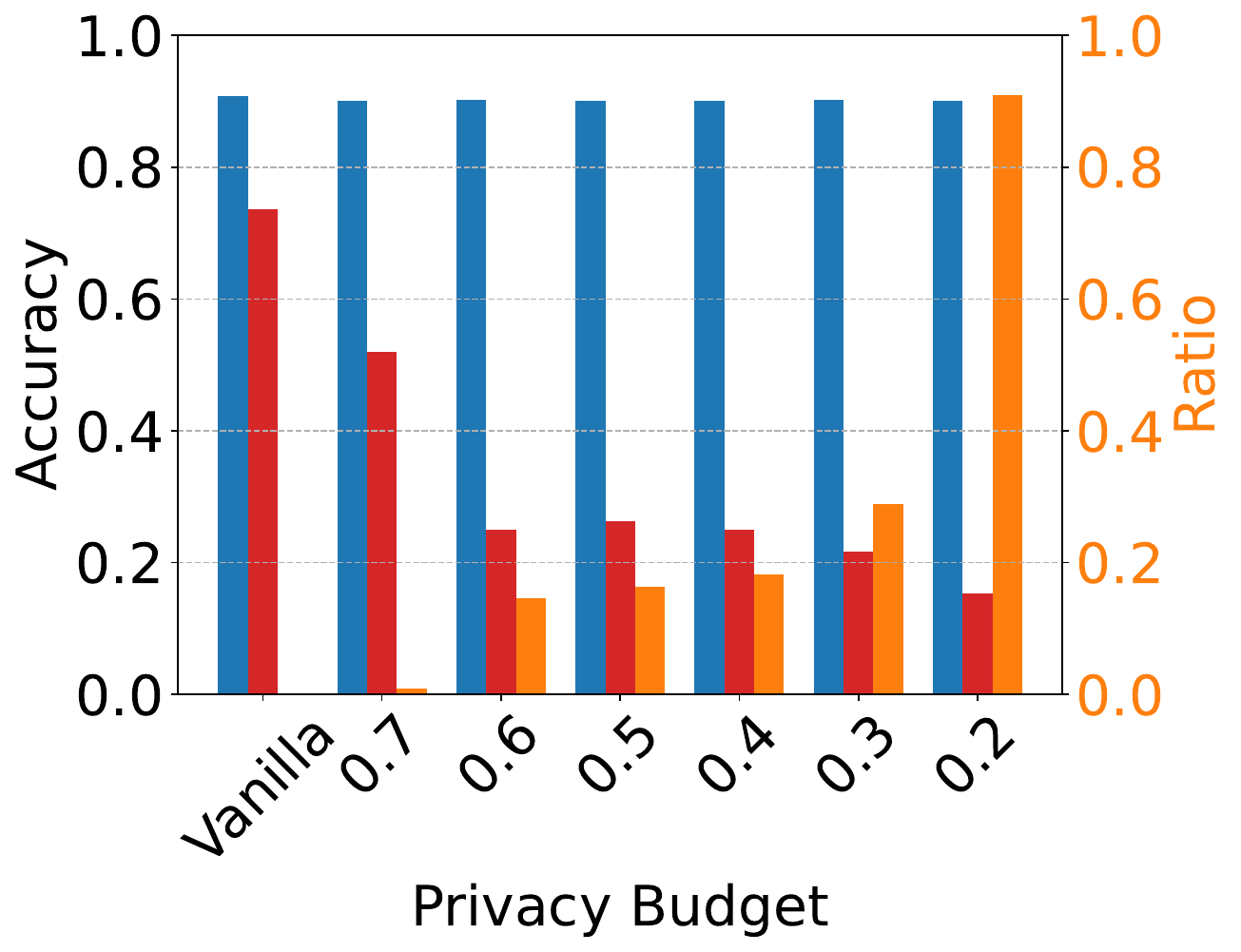}
			\caption{VGG13, CF10}
			\label{ablation_privacy_budget_vgg13}
		\end{subfigure}\hfill
		\begin{subfigure}[b]{0.49\linewidth} 
			\includegraphics[width=\linewidth]{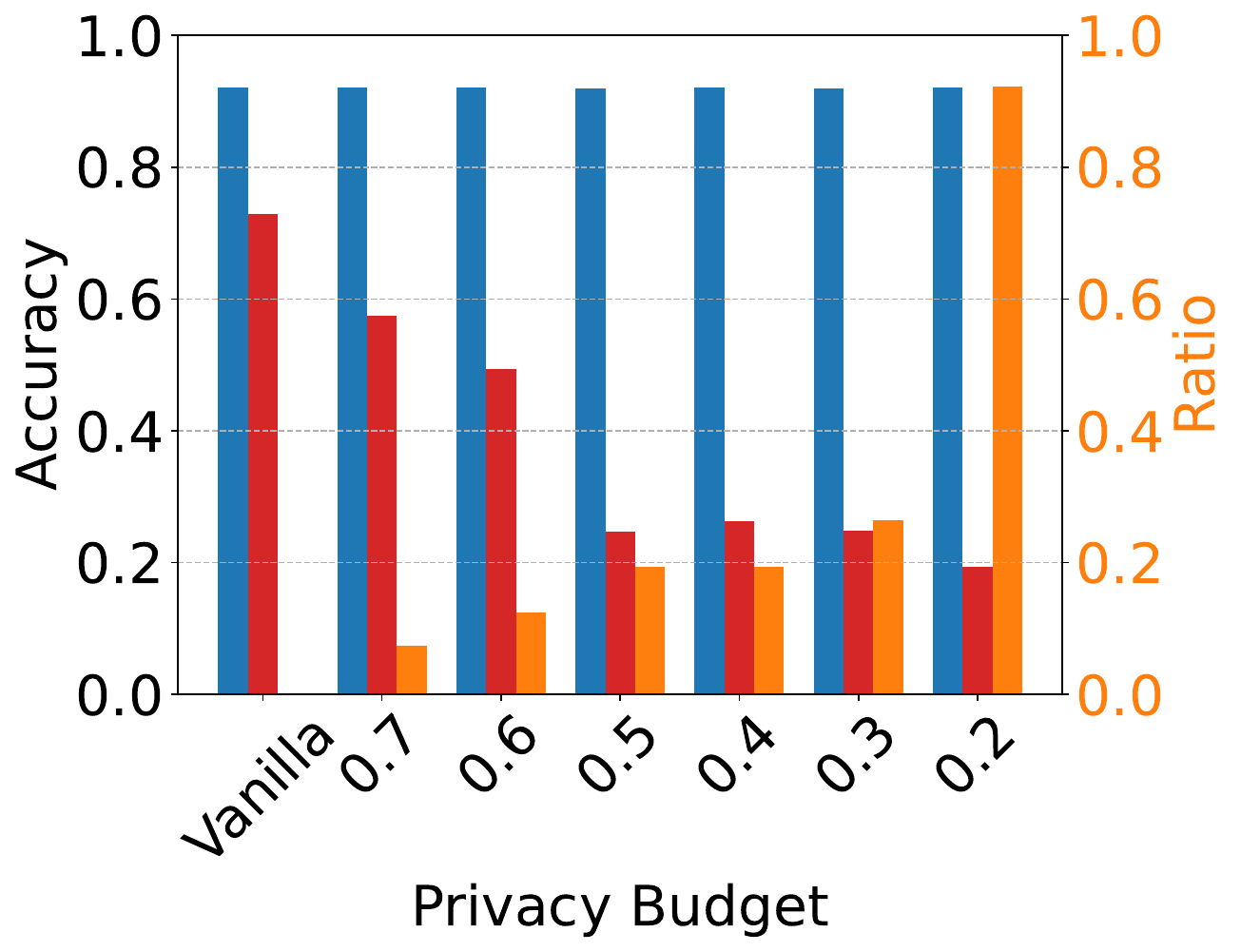}
			\caption{ResNet18, CF10}
			\label{ablation_privacy_budget_resnet18}
		\end{subfigure}\hfill
		
	\end{minipage}
	
	\caption{Impact of the tunable label privacy budget.}
	\label{fig:tunable_privacy_budget}
	\vspace{-4pt}
\end{figure}

\subsection{Tunable Label Privacy Protection} \label{sec:tunable_privacy_protection}
Experiments are conducted using the (VGG13, CF10) and (ResNet18, CF10) configurations.
As depicted in Figure \ref{fig:tunable_privacy_budget}, a decrease in the privacy budget corresponds to a gradual decline in attack accuracy. 
For every privacy budget, the attack accuracy remains lower than the budget itself, suggesting that VMask  effectively limits the extent of label leakage to not exceed the pre-set privacy budget.
Meanwhile, the main task accuracy remains relatively stable and nearly identical to that of vanilla VFL, showing no impact from the reduction of the privacy budget. 
Additionally, as the privacy budget decreases, the \emph{mask ratio}, defined as \(\sum_{t=1}^T L_t/(T\times L)\) (\(L_t\) represents the number of masked layers in epoch \(t\)), gradually increases. 
This trend can be attributed to the stringent privacy requirements necessitating more number of layers to be masked.
A larger mask ratio corresponds to increased computation and communication overhead, as more layers are trained using SS.
Therefore, the defender can adjust the tunable privacy budget to achieve an acceptable level of label privacy based on the available computational and communication resources.

\subsection{Visualization of Feature Embedding} \label{sec:visualization}

To understand why VMask is effective at defending against MC attack, we employ t-SNE \cite{tsne}, a popular unsupervised dimensionality reduction method often used for two-dimensional visualization, to analyze the feature embedding distribution in the (ResNet18, CIFAR10) case. 
We compare the feature embeddings generated by VMask with those generated by other defense methods.
As depicted in Figure \ref{fig:embedding_comparision}, for other defense methods, such as NG, DG, and FedPass, the clusters corresponding to each class are well-separated, much like the vanilla VFL shown in Figure \ref{embedding_vanilla}.
In contrast, for VMask, the clusters of each class overlap significantly and are difficult to distinguish, similar to the ``Scratch'' case shown in Figure \ref{embedding_random}, where the bottom model parameters are entirely random. 
These visualizations indicate that it is more challenging for an attacker to fine-tune an attack model based on the bottom model trained from VMask than from other defense methods.
This explains why VMask can effectively thwart the MC attack while other methods cannot.
For VMask, we also visualize the feature embedding distribution under different privacy budgets.
Due to space constraints, the details of this analysis are included in Appendix \ref{appendix:vis_embed}.
\subsection{Running Time} \label{sec:runtime}
In this section, we assess the running time of VMask and compare it with other defense methods and its variants. 
Specifically, we evaluate the running time from two perspectives: the average running time per training epoch and the total running time required to achieve a targeted main task accuracy threshold.
We set these thresholds at 99\%, 90\%, 85\%, 90\%, and 94\%, for the TM, FM, CF100, CF10, and TREC datasets, respectively, with each threshold chosen to be slightly lower than that of vanilla VFL.
Table \ref{tab:runtime} clearly shows that VMask's running time is significantly lower than that of methods based on cryptography across all cases. 
For instance, in the case of (VGG13, CF100), VMask achieves a total running time that is 531, 60846, 251, and 506 times faster than ACML, SPNN, SFA, and BlindFL, respectively.
Additionally, when compared to its three variants, VMask demonstrates higher efficiency. 
Specifically, in the case of (ResNet18, CF10), the running time per training epoch for VMask-AS, VMask-RS, and VMask-ALLS are 377.1s, 329.3s, and 536.7s, respectively, whereas VMask itself records a time of only 318.6s.
This highlights the effectiveness of VMask's layer selection strategy.
Although VMask incurs a higher running time compared to other ML-based defense methods (e.g., NG, LabelDP) and vanilla VFL, the increase is marginal and generally considered acceptable within the field. 
For example, even for the large Transformer-based model DistilBERT (67M parameters), the total running time of VMask is only 1.8 times greater than that of vanilla VFL.
This is because, in large Transformer-based models, the training time primarily results from computational overhead, while the communication overhead associated with layer sharing and reconstruction is relatively small. 
Thus, the overall running time of VMask remains only slightly larger than that of vanilla VFL.

\begin{table}[t!]
    \centering
    \footnotesize
    
	\caption{Comparison of the running time per training epoch and the total running time of different defense methods.}
    \label{tab:runtime}
    \resizebox{\linewidth}{!}{%
    \begin{tabular}{c||cc|cc|cc|cc|cc}
		\toprule
		\multirow{3}{*}{Method} &  \multicolumn{2}{c|}{\makecell{MLP3\\TM}}  & \multicolumn{2}{c|}{\makecell{LeNet5\\FM}}  & \multicolumn{2}{c|}{\makecell{VGG13\\CF100}}  & \multicolumn{2}{c|}{\makecell{ResNet18\\CF10}} & \multicolumn{2}{c}{\makecell{\add{DistilBERT}\\\add{TREC}}} \\
		\cline{2-11}
		& \makecell{Per\\Epoch} & \add{Total} & \makecell{Per\\Epoch} & \add{Total}  & \makecell{Per\\Epoch} & \add{Total}  & \makecell{Per\\Epoch} & \add{Total}  & \makecell{\add{Per}\\\add{Epoch}} & \add{Total}  \\
		\midrule
		Vanilla       & 1.9e0 & \add{8.9e1} & 4.1e0 & \add{1.6e2} & 8.5e0 & \add{3.8e2} & 1.4e1 & \add{5.6e2} & \add{8.0e1} & \add{8.0e2} \\
		\cmidrule(lr){1-11}
		
		NG/CG/DG \cite{fu2022label}      & 2.0e0 & \add{9.6e1} & 4.2e0 & \add{1.8e2} & 8.6e0 & \add{4.1e2} & 1.4e1 & \add{6.4e2} & \add{8.3e1} & \add{8.3e2} \\
		FedPass \cite{gu2023fedpass}      & 2.6e0 & \add{1.2e2} & 4.4e0 & \add{1.8e2} & 9.0e0 & \add{4.1e2} & 1.4e1 & \add{5.7e2} & \add{8.5e1} & \add{1.7e3} \\
		LabelDP \cite{ghazi2021labeldp}      & 4.8e0 & \add{2.3e2} & 5.1e0 & \add{2.5e2} & 2.1e1 & \add{1.1e3} & 1.7e1 & \add{8.3e2} & \add{8.0e1} & \add{1.6e3} \\
		\add{KD$k$} \cite{arazzi2024kdk}        & \add{2.0e0} & \add{1.0e2} & \add{4.5e0} & \add{2.3e2} & \add{9.8e0} & \add{4.9e2} & \add{1.5e1} & \add{7.5e2} & \add{8.1e1} & \add{1.6e3} \\
		dCor \cite{sun2022label}         & 8.4e2 & \add{4.1e4} & 8.6e2 & \add{3.8e4} & 7.6e2 & \add{3.4e4} & 7.7e2 & \add{3.1e4} & \add{1.2e2} & \add{1.7e3} \\
		MID \cite{zou2023mutual}          & 2.6e0 & \add{1.3e2} & 4.4e0 & \add{1.8e2} & 1.1e1 & \add{5.4e2} & 1.8e1 & \add{9.2e2} & \add{1.1e2} & \add{1.1e3} \\
		ACML \cite{zhang2020additively}         & 2.6e3 & \add{1.2e5} & 2.6e3 & \add{1.1e5} & 8.5e4 & \add{3.8e6} & 2.2e3 & \add{8.9e4} & \add{4.2e3} & \add{4.2e4} \\
		SPNN \cite{zhou2022toward}         & 4.5e2 & \add{2.1e4} & 1.1e6 & \add{4.3e7} & 9.7e6 & \add{4.4e8} & 9.6e6 & \add{3.9e8} & \add{1.6e4} & \add{1.6e5} \\
		SFA \cite{cai2022secure}          & 1.6e3 & \add{7.5e4} & 1.5e3 & \add{6.2e4} & 4.0e4 & \add{1.8e6} & 9.5e2 & \add{3.8e4} & \add{2.2e3} & \add{2.2e4} \\
		BlindFL \cite{fu2022blindfl}       & 2.6e6 & \add{1.2e8} & 9.0e3 & \add{3.6e5} & 8.0e4 & \add{3.6e6} & 8.1e4 & \add{3.2e6} & \add{1.4e6} & \add{1.4e7} \\
		
		\cmidrule(lr){1-11}
		VMask-RS      & 6.5e0 & \add{3.3e2} & 2.3e1 & \add{9.9e2} & 1.5e2 & \add{6.7e3} & 3.3e2 & \add{1.5e4} & \add{1.2e2} & \add{1.4e3} \\
		VMask-AS      & 6.3e0 & \add{2.9e2} & 2.3e1 & \add{9.4e2} & 1.9e2 & \add{8.6e3} & 3.8e2 & \add{1.5e4} & \add{1.5e2} & \add{1.8e3} \\
		VMask-ALLS    & 9.3e0 & \add{4.4e2} & 3.1e1 & \add{1.3e3} & 2.8e2 & \add{1.3e4} & 5.4e2 & \add{2.3e4} & \add{1.8e2} & \add{2.2e3} \\
		
		\cmidrule(lr){1-11}
		\textbf{VMask}         & 6.2e0 & \add{2.9e2} & 2.3e1 & \add{9.4e2} & 1.5e2 & \add{7.1e3} & 3.2e2 & \add{1.2e4} & \add{1.2e2} & \add{1.4e3} \\
		\bottomrule
	\end{tabular}
	}
	\vspace{-14pt}
\end{table}

\subsection{Ablation Study} \label{sec:ablation}

\begin{figure}[t]
	\captionsetup[subfigure]{aboveskip=-0.5pt,belowskip=-0.5pt}
	\centering
	\begin{minipage}{1.0\linewidth}
		\centering
		\includegraphics[width=0.9\linewidth, trim=50 500 50 0, clip]{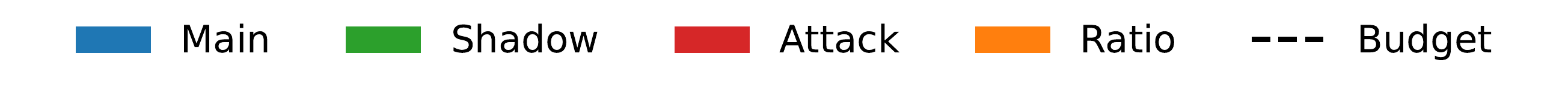}
	\end{minipage}

	\vspace{-2pt}

	\begin{minipage}{1.0\linewidth}
		\centering
		\begin{subfigure}[b]{0.49\linewidth} 
			\includegraphics[width=\linewidth]{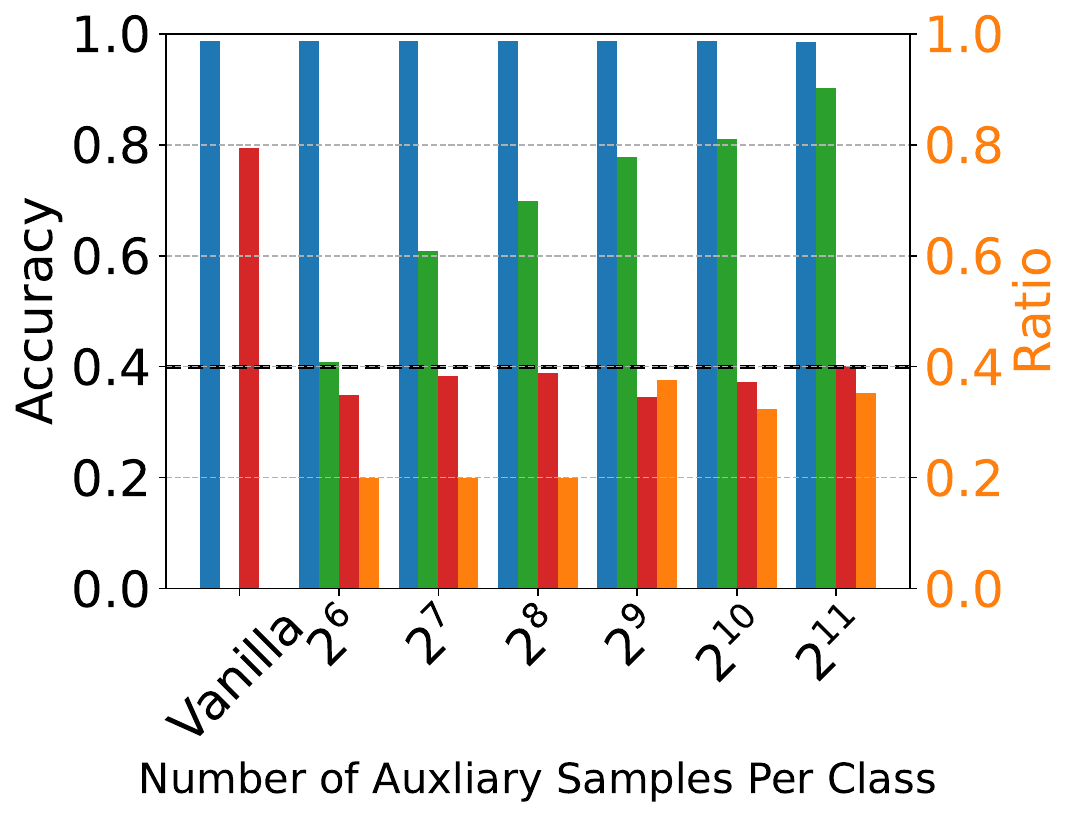}
			\caption{LeNet5, M}
			\label{ablation_aux_samples_lenet5}
		\end{subfigure}\hfill
		\begin{subfigure}[b]{0.49\linewidth} 
			\includegraphics[width=\linewidth]{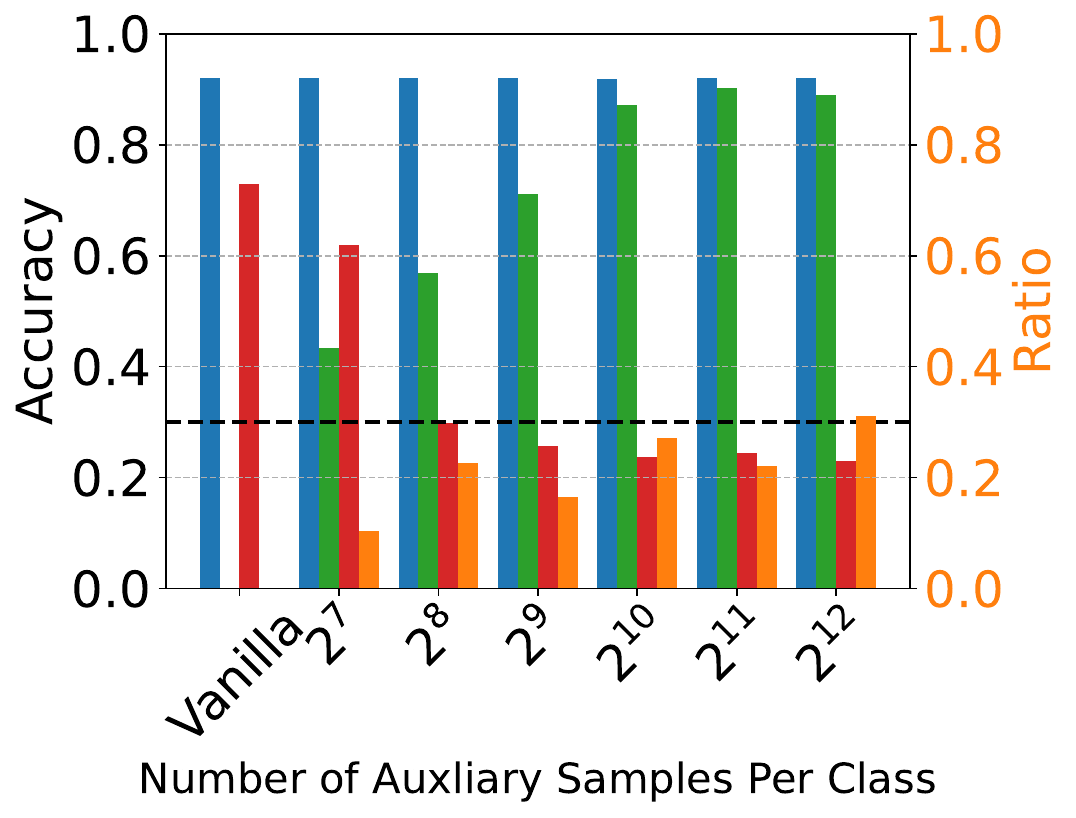}
			\caption{ResNet18, CF10}
			\label{ablation_aux_samples_resnet18}
		\end{subfigure}\hfill
	\end{minipage}
	
	\caption{Impact of auxiliary dataset size.}
	\label{fig:ablation_aux_samples}
	\vspace{-4pt}
\end{figure}

\noindent \textbf{Auxiliary Dataset Size.}
Experiments are conducted under the (LeNet5, M) and (ResNet18, CF10) cases, with the privacy budget fixed at 0.4 for the former and 0.3 for the latter.
As demonstrated in Figure \ref{fig:ablation_aux_samples}, an increase in the number of auxiliary samples leads to a gradual improvement in the shadow model's accuracy. 
An enhanced shadow model yields more precise estimated gradients and more accurate estimated label leakage. 
Consequently, the defender must mask more layers to meet privacy requirements, which explains the observed gradual decrease in attack accuracy and the corresponding increase in the mask ratio.
In addition, the main task accuracy remains as stable as that of the vanilla VFL, demonstrating VMask's robust ability to maintain model performance.
Notably, when the number of auxiliary samples per class reaches $2^6$ for M dataset and $2^8$ for CF10 dataset, the auxiliary dataset size constitutes merely 1\% ($2^6/6000$) and 5\% ($2^8/5000$) of the passive party's original dataset, respectively.
At this point, the attack accuracy is reduced to 0.35 and 0.29, both within the privacy budget.
Experimental results for (VGG13, CF10) are similar and are provided in Appendix \ref{appendix:ablation}. 

\noindent \textbf{Auxiliary Dataset Distribution.}
We evaluate VMask's effectiveness when the auxiliary dataset and the passive party's original dataset are either in-distribution (but non-independently and identically distributed, i.e., non-IID) or out-of-distribution.

\begin{figure}[t]
	\captionsetup[subfigure]{aboveskip=-0.5pt,belowskip=-0.5pt}
	\centering
	\begin{subfigure}[b]{0.49\linewidth} 
		\includegraphics[width=\linewidth]{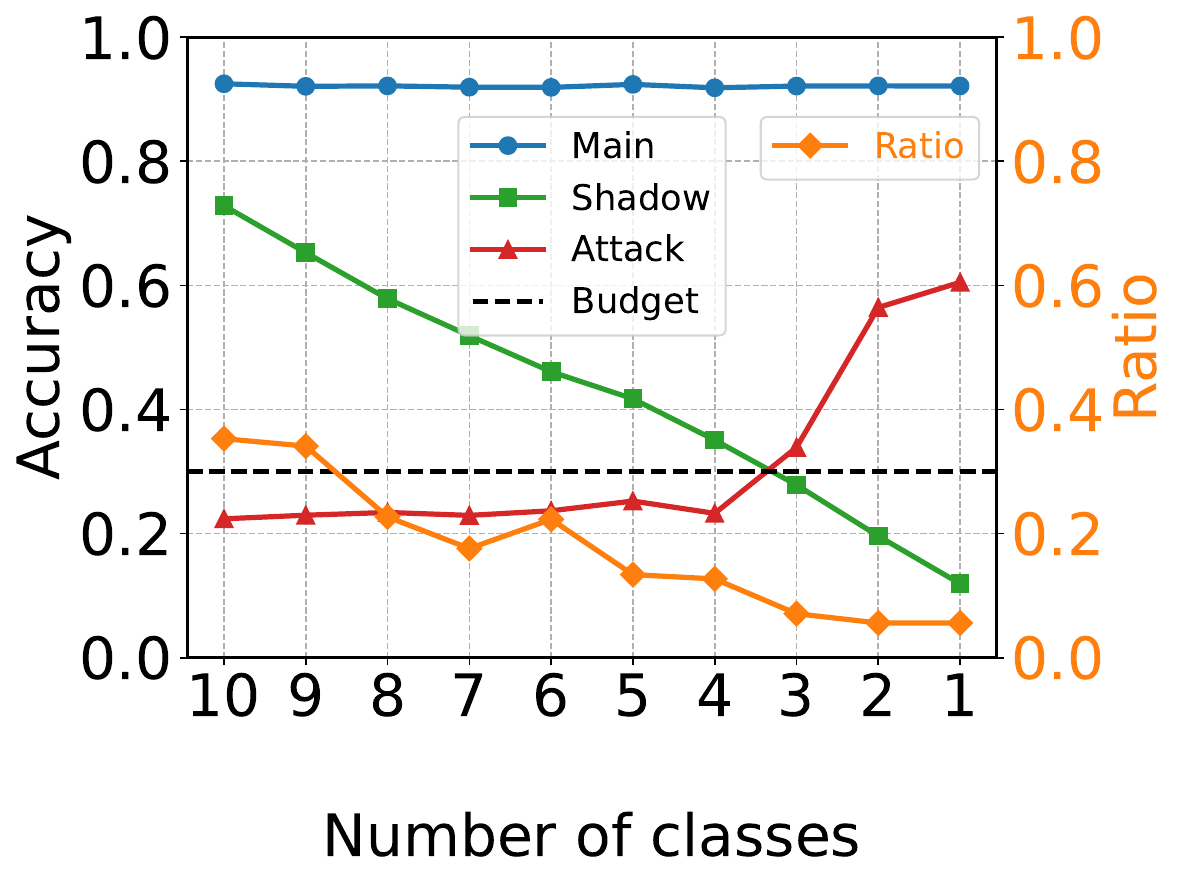}
		\caption{Label non-IID}
		\label{ablation_label_noniid_resnet}
	\end{subfigure}\hfill
	\begin{subfigure}[b]{0.49\linewidth}  
		\includegraphics[width=\linewidth]{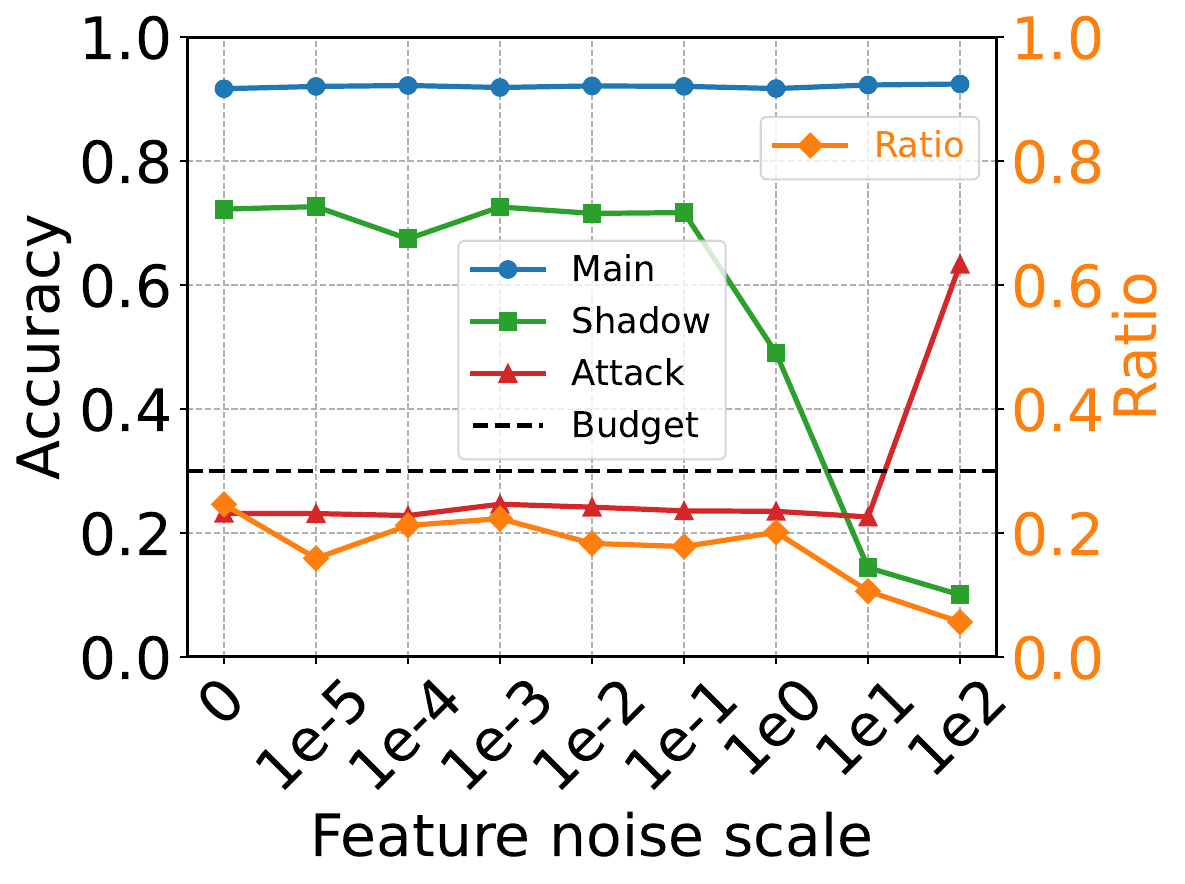}
		\caption{Feature non-IID}
		\label{ablation_feature_noniid_resnet}
	\end{subfigure}\hfill
	\caption{Impact of non-IID auxiliary dataset under the (ResNet18, CF10) configuration.}
	\label{fig:ablation_aux_dataset_distribution}
	\vspace{-6pt}
\end{figure}

\underline{\emph{In-distribution but non-IID.}} 
We assess two non-IID scenarios: label non-IID, where the auxiliary dataset contains only a subset of classes from the original dataset,
and feature non-IID, which involves adding varying levels of Gaussian noise to the auxiliary dataset \cite{zhu2021federated, li2022federated}. 
Experiments are conducted under (ResNet18, CF10) configuration, with the privacy budget fixed at 0.3. 
As observed in Figure \ref{fig:ablation_aux_dataset_distribution}, increasing the non-IID extent of auxiliary dataset (by decreasing the number of classes or increasing the feature noise scale) leads to a gradual decrease in shadow model accuracy, which results in reduced estimated attack accuracy and a corresponding decline in mask ratio.
Notably, even when the auxiliary dataset contains only 4 classes or has high feature noise (scale of $1e1$), resulting in shadow model accuracies of 35.08\% and 14.39\%, respectively, VMask still effectively limits attack accuracy to 26.74\% and 22.5\% (both within the privacy budget), while maintaining main task accuracy comparable to vanilla VFL. 
Similar results were observed under (VGG13, CF10), as detailed in Appendix ~\ref{appendix:ablation}. 

\underline{\emph{Out-of-distribution.}} 
VMask remains effective against MC attack even when the auxiliary dataset originates from a completely different task.
In Table \ref{tab:ablation_different_task}, for example, under the VGG13 configuration, using CF10 as the auxiliary dataset for the CN10 main task yields a shadow model accuracy of 54.53\%, significantly lower than the vanilla VFL model's 84.62\%.
Despite this, VMask successfully identifies critical layers for masking, achieving an attack accuracy of 16.28\%, which is even lower than the 17.79\% attack accuracy from ``Scratch''.
These findings demonstrate that VMask is robust to distribution shifts in the auxiliary dataset.

\noindent \textbf{Layer Selection Strategy.}
We compare the layer selection strategy adopted by VMask with those of its three variants: VMask-AS, VMask-RS, and VMask-ALLS.
As shown in Table \ref{tab:effectiveness_best_model} and Figure \ref{fig:every_epoch_attack}, both VMask and VMask-AS reduce the attack accuracy to the ``Scratch'' level while maintaining high main accuracy. 
In contrast, VMask-RS exhibits significantly higher attack accuracy, indicating its ineffectiveness in protecting label privacy.
Additionally, Table \ref{tab:runtime} demonstrates that VMask requires less running time than both VMask-AS and VMask-ALLS.
These results confirm that the layer selection strategy adopted by VMask is more effective and efficient compared to its variants.

\begin{table}[t] 
	\footnotesize
	\caption{
		\add{
			Impact of out-of-distribution auxiliary dataset originating from a completely different task.
		}
    }
	\label{tab:ablation_different_task}
	\centering
	\resizebox{\linewidth}{!}{%
	\begin{tabular}{c|cc|ccc|ccc}
		\toprule
		\multirow{2}{*}{\add{Model}} & \multicolumn{2}{c|}{\add{Dataset}} & \multicolumn{3}{c|}{\add{Main Accuracy(\%)}} & \multicolumn{3}{c}{\add{Attack Accuracy(\%)}} \\
		& \add{Main} & \add{Auxiliary} & \add{Vanilla} & \add{Shadow} & \add{VMask} & \add{Vanilla} & \add{Scratch} & \add{VMask} \\
		\midrule
		\multirow{2}{*}{\add{LeNet5}}   & \add{M}    & \add{FM}    & \add{98.72} & \add{19.90} & \add{98.72} & \add{80.72} & \add{37.67} & \add{43.37} \\
		                          & \add{FM}   & \add{M}     & \add{90.99} & \add{9.07}  & \add{91.02} & \add{67.34} & \add{45.37} & \add{43.57} \\
		\cmidrule(lr){1-9}
		\multirow{2}{*}{\add{VGG13}}    & \add{CF10} & \add{CN10}  & \add{90.74} & \add{70.79} & \add{90.64} & \add{73.62} & \add{14.46} & \add{15.55} \\
		                          & \add{CN10} & \add{CF10}  & \add{84.62} & \add{54.53} & \add{84.59} & \add{60.92} & \add{17.79} & \add{16.28} \\
		\cmidrule(lr){1-9}
		\multirow{2}{*}{\add{Resnet18}} & \add{CF10} &\add{ CN10}  & \add{92.11} & \add{69.71} & \add{92.05} & \add{72.87} & \add{16.87} & \add{16.22} \\
		                          & \add{CN10} & \add{CF10}  & \add{86.85} & \add{56.40} & \add{86.87} & \add{59.68} & \add{14.67} & \add{16.72} \\
		\bottomrule
	\end{tabular}
	}	
	\vspace{-4pt}
\end{table}

\begin{table}[t] 
	\footnotesize
	\caption{
 Impact of number of participants with $1$ attacker.
    }
	\centering
	\begin{tabular}{cccccc}
		\toprule
		\multirow{2}{*}{\makecell{No.\\Participants}} & \multicolumn{2}{c}{Main Accuracy(\%)} & \multicolumn{3}{c}{Attack Accuracy(\%)} \\
		& Vanilla & VMask & Vanilla & Scratch & VMask \\
		\midrule
		2  & 99.02 & 99.00 & 80.72 & 41.35 & 44.83\\
		4  & 98.97 & 98.92 & 36.71 & 31.14 & 31.79\\
		6  & 98.96 & 98.89 & 32.59 & 28.25 & 28.30\\
		8  & 98.77 & 98.70 & 21.02 & 21.63 & 20.32\\
		10 & 98.74 & 98.66 & 19.82 & 20.52 & 20.53\\
		\bottomrule
	\end{tabular}
	\label{tab:ablation_one_attacker}
	\vspace{-4pt}
\end{table}

\begin{table}[t] 
	\footnotesize
	\caption{
		Impact of number of participants with $K-1$ attacker(s). 
    }
	\centering
	\resizebox{\linewidth}{!}{%
	\begin{tabular}{ccccccc}
		\toprule
		\multirow{2}{*}{\makecell{No.\\Participants}} & \multicolumn{2}{c}{Main Accuracy(\%)} & \multirow{2}{*}{Attacker(s)} & \multicolumn{3}{c}{Attack Accuracy(\%)} \\
		& Vanilla & VMask & & Vanilla & Scratch & VMask \\
		\midrule
		2 & 92.11 & 92.01 & $P_1$ & 72.87 & 16.87 & 17.52 \\
		\cmidrule(lr){1-7}
		\multirow{2}{*}{3} & \multirow{2}{*}{91.39} & \multirow{2}{*}{91.18} & $P_1$ & 52.39 & 15.29 & 16.15 \\
		  & & & $P_2$ & 53.09 & 15.41 & 16.01 \\
		\cmidrule(lr){1-7}
		  \multirow{3}{*}{4} & \multirow{3}{*}{89.61} & \multirow{3}{*}{89.22} & $P_1$ & 52.52 & 17.37 & 17.38 \\
		 & & & $P_2$ & 57.28 & 15.77 & 15.76 \\
		 & & & $P_3$ & 52.64 & 14.70 & 15.12 \\
		\bottomrule
	\end{tabular}
	}
	\label{tab:ablation_multiple_attakers}
	\vspace{-4pt}
\end{table}


\noindent \textbf{Number of Participants.}
The impact of the number of participants on VMask's main task accuracy and attack accuracy is evaluated under (MLP, TM) and (ResNet18, CF10) configurations. 
In the former case, only one passive party acts as the attacker, whereas in the latter case, all passive parties are assumed to be attackers. 
In both scenarios, the privacy budget is fixed at 0.25.

\underline{\emph{Scenario with a Single Attacker.}}
Experiments are conducted with 2, 4, 6, 8, and 10 participants.
As summarized in Table \ref{tab:ablation_one_attacker},  increasing the number of participants slightly reduces the main task accuracy in both vanilla VFL and VMask. 
However, the gap in the main task accuracy between VMask and vanilla VFL remains less than 0.1\% across all cases.
In most cases, the attack accuracy under VMask is lower than that under vanilla VFL and is close to the attack accuracy from ``Scratch''. 
For example, with 6 participants, the attack accuracy from VMask is only 28.30\%, which is lower than the 32.59\% from vanilla VFL and nearly identical to the 28.25\% from ``Scratch''.

\underline{\emph{Scenario with $K-1$ Attacker(s).}}
We also explore a scenario where all passive parties act as attackers. 
As shown in Table \ref{tab:ablation_multiple_attakers}, the difference in main task accuracy between VMask and vanilla VFL is less than 0.5\% across all cases. 
Furthermore, the attack accuracy from VMask is substantially lower than that from vanilla VFL in every case.
For example, in a VFL system with three participants, where $P_1$ and $P_2$ are the attackers and $P_3$ is the active party, the attack accuracies of $P_1$ and $P_2$ from vanilla VFL are 52.39\% and 53.09\%, respectively. 
In contrast, the attack accuracies from VMask are only 16.1\% and 16.01\%, respectively, representing a significant reduction compared to vanilla VFL, and are close to the 15.29\% and 15.41\% accuracies from ``Scratch'' baseline.

\section{Related Work} \label{sec:related_work}
In literature, various methods have been proposed to defend against label inference attacks.
These methods can be categorized into four types: 
(1) Perturbation-based ~\cite{fu2022label, gu2023fedpass, li2022label}
(2) Confusion-based ~\cite{zou2022label, wan2023pslf, ghazi2021labeldp,arazzi2024kdk,he2024labobf}
(3) Regularization-based ~\cite{sun2022label, duan2023privascissors, zou2023mutual, zheng2022making}
(4) Cryptography-based ~\cite{zhou2022toward, zhang2020additively, wang2023beyond, qiu2023vfedsec, fu2022blindfl, cai2022secure}.

\textbf{Perturbation-based.}
Defense methods in this category introduce perturbations to the intermediate information during training. 
Fu et al. \cite{fu2022label} propose methods such as Noisy Gradient (NG), which applies differential privacy principles \cite{dwork2006differential} by adding Gaussian or Laplacian noise; Compressed Gradient (CG), which zeroes out smaller gradient elements; and Discrete Gradient (DG), which discretizes gradients into intervals. 
FedPass \cite{gu2023fedpass} utilizes the private passport technique \cite{fan2021passport} to  dynamically apply perturbations based on the model's training state. 
Marvell \cite{li2022label}  adds noise to gradients to maintain identical distributions for different sample classes.

\textbf{Confusion-based.}
These methods aim to protect private labels by directly making it more confused.
Zou et al. \cite{zou2022label} develop a confusion autoencoder that utilizes entropy regularization to obscure labels while preserving task accuracy. 
LabelDP \cite{ghazi2021labeldp} and \cite{wan2023pslf} uses a randomized response technique to flip labels randomly for training. 
KD$k$ \cite{arazzi2024kdk} combines knowledge distillation and $k$-anonymity to increase the uncertainty into the raw labels.


\textbf{Regularization-based.}
These defense methods integrate an additional loss term into the optimization objective function to reduce correlations between intermediate outputs and true labels. 
Zou et al. \cite{zou2023mutual} introduced the Mutual Information Regularization Defense (MID), which limits information about private labels in exchanged outputs.
Sun et al. \cite{sun2022label} propose to decrease label inference accuracy by minimizing the distance correlation (dCor) \cite{szekely2007dcor} between feature embeddings and private labels. 

\textbf{Cryptography-based.}
This category employs cryptographic techniques like HE and MPC to design secure training protocols. 
Zhang et al. \cite{zhang2020additively} present a purely HE-based Asymmetrically Collaborative Machine Learning (ACML) framework, splitting the model into an unencrypted feature extractor and a partially encrypted classifier, ensuring secure information exchange. 
Wang et al. \cite{wang2023beyond} introduce a purely SS-based Dispersed Training (DT) framework, where model parameters are secret shared, effectively preventing inference attacks. 
Hybrid approaches combining SS and HE are explored by \cite{zhou2022toward}, \cite{cai2022secure}, and \cite{fu2022blindfl}. 
The SPNN \cite{zhou2022toward} framework partitions computation graph to protect label-related layers using SS and HE, ensuring privacy and scalability. 
Similarly, BlindFL \cite{fu2022blindfl} also divides the model and apply  HE and SS on the input layer, while training other layers in plaintext to bypass heavy cryptographic computations.
The SFA \cite{cai2022secure} framework designs a secure forward aggregation protocol with removable masks to protect private data.

While numerous approaches have been proposed in the literature for label protection, they often struggle to achieve a favorable privacy-utility trade-off or become impractical due to high computational and communication costs. 
In contrast, our work introduces a novel approach that effectively balances label privacy and model utility while maintaining efficiency.

\section{Discussion and Future Work} \label{sec:discussion}

\textbf{VMask Design Alternatives.}  
\ul{\emph{(1) Integrating more efficient MPC engine.}}  
The VMask framework currently uses the classic additive SS scheme for training masked layers due to its simplicity and ease of implementation. 
However, VMask is agnostic to the specific MPC engine employed. 
In future work, we plan to incorporate more efficient MPC frameworks, such as ABY3 \cite{mohassel18aby3} and Falcon \cite{wagh2021falcon}, to further enhance VMask's efficiency.  
\ul{\emph{(2) Masking non-linear layers.}}  
Currently, VMask focuses on masking linear layers (e.g., FC, Conv) because SS ensures the integrity of linear computations, thereby preserving main task accuracy. 
However, these layers often contain many parameters, resulting in high computation and communication overhead. 
Looking forward, we aim to explore masking non-linear layers (e.g., BatchNorm), which generally contain fewer parameters. 
The forward and backward propagation of non-linear layers using SS, however, requires techniques like linear approximation, potentially affecting model performance. Balancing model utility and training efficiency here is a promising future direction.

\textbf{Security Against Malicious Attackers.}
The training protocol in VMask framework is secure against semi-honest attackers.
However, when faced with a malicious attacker who  deliberately deviates from the protocol (e.g., bypassing the layer masking module), the critical layers could be exposed, comprising VMask's privacy guarantees.
To mitigate such risks, from a practical engineering perspective, the defender could compile the VMask protocol into a binary executable, making the framework a black box to the attacker and hiding its internal state.
This ensures that the critical layers remain masked.
Although the attacker could still modify the inputs (features) and outputs (embeddings) of the executable, they would lack necessary optimization objectives to effectively guide these modifications for successful label inference. 
Furthermore, indiscriminately altering outputs could even degrade main task accuracy, which would be counterproductive for the attacker.
\discuss{
    In future work, zero-knowledge proof techniques \cite{zkp} could be applied to the layer masking module.
    This would ensure that the attacker's behavior can be verified, guaranteeing that the critical layers remain securely masked.
}


\textbf{Combination with Existing Defenses.}  
VMask is designed to mitigate the most powerful MC attack, addressing the urgent need for practical defenses against this threat.  
Since VMask does not interfere with feature embeddings or their gradients, it is orthogonal to existing defenses that target other attack surfaces, such as Marvell \cite{li2022label} and dCor \cite{kang2022framework, sun2022label}.  
Therefore, in practice, VFL systems should adopt a defense-in-depth strategy by combining VMask with existing defenses—leveraging specialized techniques to counter specific vulnerabilities rather than pursuing an impractical ``one-size-fits-all'' solution, which may lead to unnecessary complexity or over-engineering.

\section{Conclusion} \label{sec:conclusion}
In this paper, we propose VMask, a label privacy protection framework for VFL designed to defend against the model completion attack from the perspective of layer masking.
The key insight of our approach is to decrease the correlation between input data and intermediate output of the attacker's model by applying secret sharing to mask layer parameters.
We develop a critical layer selection strategy for masking to reduce the computation and communication overhead. 
Additionally, we introduce a tunable privacy budget to enable flexible control over the levels of label privacy.
Comprehensive experiments verify that VMask effectively protects label privacy while remaining efficient.



\bibliographystyle{IEEEtran}
\bibliography{reference}

\appendix
\subsection{Algorithm Details}

\subsubsection{Model Completion Attack}  \label{appendix:mc_attack}
As shown in Algorithm ~\ref{algo:mc_attack}, the attacker first randomly initializes an inference head model $f(\theta_H)$ (Line 2) and then constructs an attack model by concatenating the inference head model with the well-trained bottom model $f(\theta)$ (Line 3). 
The attacker then fine-tunes the attack model with the attack dataset $D_A$ (Lines 5-10), and finally outputs the best attack accuracy (i.e., label inference accuracy) $\epsilon'$ (Line 11).

\begin{algorithm}[h!]
    \small
    \caption{Model Completion Attack}
    \label{algo:mc_attack}
    \SetKwInOut{Input}{Input}
    \SetKwInOut{Output}{Output}
    \Input{Attack dataset $D_A$, bottom model parameters $\theta$,  fine-tuning epochs $T'$, learning rate $\eta$.}
    \Output{Attack accuracy $\epsilon'$.}

    Split $D_A$ into $D_{A}^{train}$ and $D_A^{test}$ \;
    Randomly initialize inference head model $f(\theta_H)$ \;
    Construct attack model $f_A(\theta_A) = f(\theta) \circ f(\theta_H)$ \;
    Set best attack accuracy $\epsilon' \gets 0$ \;
    \For{each epoch $t\in[1, T']$}{
        $X, Y \gets D_{A}^{train}$ \;
        $loss = \mathcal{L}\left(f_A(X; \theta_A), Y\right)$ \;
        $\theta_A \gets \theta_A - \eta \frac{\partial loss}{\partial \theta_{A}}$ \;
        $\epsilon \gets \texttt{Eval}(D_{A}^{test}; \theta_A)$ \;  \tcp{\texttt{Eval}() calculates classification accuracy}
        $\epsilon' \gets \text{max}(\epsilon', \epsilon)$ \;
    }
   \Return{$\epsilon'$} 
\end{algorithm}

\subsubsection{VMask Framework} \label{appendix:vmask_framework}
The relationship among the modules and procedures in VMask framework is detailed in Algorithm \ref{algo:framework}, where the \textbf{LayerMasking} (Line 5 and Line 13) function corresponds to Algorithm \ref{algo:layer_masking} in Section \ref{sec:vmask_layer_masking}, the \textbf{SecureModelUpdate} (Line 9) and \textbf{ShadowModelUpdate} (Line 10) functions correspond to Algorithm \ref{algo:secure_model_update} and Algorithm \ref{algo:shadow_model_update} in Appendix \ref{appendix:secure_model_update} and \ref{appendix:shadow_model_update}, and the \textbf{LayerSelection} (Line 12) function corresponds to Algorithm \ref{algo:layer_selection} in Section \ref{sec:layer_selection}.
Before the main VFL training loop, we initialize the masked layer indices $U$, plaintext layer indices $V$, and the accumulated gradient norm $G$ for each passive party $P_k$, where \(k \in [1, K-1]\) (Lines 2-6).

\begin{algorithm}[t]
    \small
    \caption{The VMask Framework}
    \label{algo:framework}
    \SetKwInOut{Input}{Input}
    \SetKwInOut{Output}{Output}
    \Input{Datasets $D_1, \cdots, D_K$, auxiliary datasets $D_1^{aux}, \cdots, D_{K-1}^{aux}$, privacy budget $\epsilon$, VFL training epochs $T$, MC attack training epochs $T'$, learning rate $\eta$, number of bottom model layers $L$, number of labeled samples $M$, noise standard variance $\sigma$.}
    \Output{Well trained model parameters $\theta_1, \theta_2, \cdots, \theta_K, \theta_T$.}
    
    \tcc{Initialization}
    Randomly initialize bottom model parameters $\theta_1,\cdots,\theta_K$, shadow model parameters $\theta_{s,1},\cdots,\theta_{s,K-1}$ and top model parameters $\theta_T$ \;
    \For{each $k\in[1,K-1]$ in parallel}{
        $U_{k}^{0} \gets \emptyset, U_{k}^{1} \gets \{1\}$ \tcp{masked layer indices}
        $V_{k}^{0} \gets [L], V_{k}^{1} \gets [L] - \{1\}$ \tcp{plaintext layer indices}
        $\theta_k \gets$ \textbf{LayerMasking}($U_{k}^{1}, U_{k}^{0}, V_{k}^{1}, V_{k}^{0}, \theta_k, k, \sigma$) \;
        Set historical accumulated gradient norm $G_k = 0$ \;
    }
    \tcc{Training loop}
    $X_K, Y \gets D_K$ \;
    \For{each epoch $t \in [1, T]$}{
        $\theta_1\cdots \theta_K, \theta_T \gets$ \textbf{SecureModelUpdate}($D_1\cdots D_K, \theta_1\cdots \theta_K, \theta_T, U_1\cdots U_{K-1}, L, \eta$) \;
        $\theta_{s,1}\cdots\theta_{s,K-1}, \nabla\theta_{s,1}\cdots \nabla\theta_{s,K-1} \gets$ \textbf{ShadowModelUpdate}($D_{1}^{aux}\cdots D_{K-1}^{aux}, \theta_{s_1}\cdots\theta_{s,K-1}, \theta_T, Y, \eta$) \;
        \For{each $k\in[1,K-1]$ in parallel}{
            $U_{k}^{t+1}, V_{k}^{t+1}, G_k \gets$ \textbf{LayerSelection}($D_{k}^{aux}, \theta_{s,k},\nabla\theta_{s,k}, U_{k}^{t}, \epsilon, G_k, L, M, Y, T',\eta$) \;
            $\theta_k \gets$ \textbf{LayerMasking}($U_{k}^{t+1}, U_{k}^{t}, V_{k}^{t+1}, V_{k}^{t}, \theta_k, k, \sigma$) 
        }
    }
    \Return{$\theta_1,\cdots,\theta_K,\theta_T$}
\end{algorithm}

\subsubsection{Secure Model Update Procedure} \label{appendix:secure_model_update}
As detailed in Algorithm \ref{algo:secure_model_update}, the secure model update procedure includes secure forward and secure backward sub-procedures.
During the secure forward phase (Lines 1-13), for each passive party \(P_k\), where \(k \in [1, K-1]\), and for each layer parameter \(\theta_k^j\), where \(j \in [1, L]\) of \(P_k\)'s bottom model \(\theta_k\), we check if $\theta_k^j$ has been masked.
If masked (Lines 4-8), the layer input \(z_k\) is split into two shares, \(\llbracket z_k \rrbracket_k\) and \(\llbracket z_k \rrbracket_K\), and distributed to \(P_k\) and the active party \(P_K\), respectively (Line 5). 
Each input share is then independently fed to the shared layer parameters to obtain the corresponding output share (Lines 6-7), which is subsequently reconstructed to serve as the input for the next layer (Line 8). 
If \(\theta_k^j\) is not masked, the layer input \(z_k\) is directly fed into it, yielding the plaintext output for the subsequent layer (Lines 9-10). 
After all passive parties have produced their bottom model embeddings, the active party aggregates all these embeddings to get $Z$ (Line 12) and then feeds $Z$ into the top model to compute the loss (Line 13).

In the secure backward phase, the gradient of the active party's bottom model parameter \(\theta_K\) and the top model parameter \(\theta_T\) are both calculated in plaintext. 
These parameters are updated using SGD (Line 14). 
Subsequently, for each passive party \(P_k\), we first compute the gradient of each bottom model's embedding \(\nabla z_k\) (Line 16).
For each layer parameter \(\theta_k^j\) of \(P_k\)'s bottom model, we also determine whether this layer has been masked. 
If \(\theta_k^j\) is masked, it is updated in a secret-shared manner.
We first share the input gradient (Line 19), then  backpropogate the gradient in parallel (Lines 20-23), and finally reconstruct the gradient before the next layer (Line 24).
If \(\theta_k^j\) is not masked, we directly backpropagate the gradient to the previous layer in plaintext (Lines 25-27).
After the secure model update procedure, all VFL models, including \(\theta_1, \ldots, \theta_K, \theta_T\), are updated and returned (Line 28).

\subsubsection{Shadow Model Update Procedure} \label{appendix:shadow_model_update}
The shadow model update procedure involves training shadow models \(\theta_{s,1}, \ldots, \theta_{s,K-1}\) with auxiliary datasets in plaintext by the active party alone. 
As depicted in Algorithm \ref{algo:shadow_model_update}, the top model parameters \(\theta_T\) are fixed initially (Line 1). 
During the forward pass, each shadow model \(\theta_{s,k}\), where \(k \in [1,K-1]\), processes its corresponding auxiliary dataset \(D_k^{aux}\) to generate the embedding \(z_{s,k}\) (Lines 2-3). 
These embeddings are then combined and fed into the frozen top model \(f(\theta_T)\) to calculate the loss (Lines 4-5). 
In the backward pass, the gradient \(\nabla \theta_{s,k}\) for each shadow model is computed using the chain rule (Line 7). 
Subsequently, the parameters of the shadow models are updated (Line 8). 
The procedure returns the updated shadow models and their respective gradients, which the layer selection module then evaluates to determine which layers should be selected for masking (Line 9).


\begin{table*}[t!]
    \caption{Model architectures, dataset statistics, VFL training parameters, MC attack settings, and configurations of VMask.}
	\centering
    \resizebox{\linewidth}{!}{%
	\begin{tabular}{ccc||cccc||ccc||cc||cc}
	  \toprule
       \makecell{Bottom\\Model} & \makecell{Top\\Model} & \makecell{Inference\\Head} & \makecell{Dataset\\(abbreviated name)} & \makecell{No.\\Training\\Samples} &\makecell{No.\\Test\\Samples} & \makecell{No.\\Classes} & \makecell{Training\\Epochs} & \makecell{Batch\\Size} & \makecell{Learning\\Rate} & \makecell{No.\\Known \\Labels \\Per\\ Class}  & \makecell{MC\\Attack\\Training\\Epochs} & \makecell{Privacy\\Budget} & \makecell{Auxiliary\\Dataset\\Size\\(ratio)} \\
	  \midrule
	  \multirow{3}{*}{MLP3} & \multirow{3}{*}{MLP2} & \multirow{3}{*}{MLP2}   & TabMNIST (TM) & 60,000 & 10,000 & 10 & 50 &128 &0.1 & 4 & 50&  0.25& 640 (1.1\%)\\
      & & & TabFMNIST (TFM) & 60,000 & 10,000 & 10 &50 &128 & 0.1& 4  & 50&0.25& 640 (1.1\%)\\
      & & & CRITEO (CR)   & 480,000 & 120,000 & 2  &50 &256 & 0.1& 50 & 50&0.25& 8,192 (1.7\%)\\
    \cmidrule(lr){1-14}

	  \multirow{3}{*}{LeNet5} & \multirow{3}{*}{MLP2} & \multirow{3}{*}{MLP2}  & MNIST (M) & 60,000 & 10,000 & 10 & 50 &128&0.1& 4  & 50&0.25& 640 (1.1\%)\\
      & & & FMNIST (FM) & 60,000 & 10,000 & 10 &50 &128&0.1& 4  & 50&0.25& 640 (1.1\%)\\
      & & & SVHN   &73,257 & 26,032 & 10 &50 & 128&0.1&4  & 50&0.25& 640 (0.9\%)\\
      \cmidrule(lr){1-14}

	  \multirow{4}{*}{VGG13} & \multirow{4}{*}{MLP2} & \multirow{4}{*}{MLP2} & CIFAR10 (CF10) & 50,000 & 10,000 & 10 & 50 & 128&0.1&4  & 50 &0.25& 2,560 (5.1\%)\\
       & & & CIFAR100 (CF100) & 50,000 & 10,000 & 100 & 50&128&0.1& 4  & 50&0.25& 2,600 (5.2\%)\\
       & & & CINIC10 (CN10)  & 180,000 & 90,000 & 10  &50 &256&0.1& 4  & 50&0.25& 9,000 (5.0\%)\\
       & & & TinyImageNet (TI) & 100,000 & 10,000 & 200 & 50 &128&0.1& 4   &50&0.25& 5,000 (5.0\%)\\
       \cmidrule(lr){1-14}

      \multirow{4}{*}{ResNet18} & \multirow{4}{*}{MLP2} & \multirow{4}{*}{MLP2}  & CIFAR10 (CF10) & 50,000 & 10,000 & 10 & 50 &128&0.1& 4   & 50 &0.25& 2,560 (5.1\%)\\
       & & & CIFAR100 (CF100) & 50,000 & 10,000 & 100 &50 &128&0.1& 4   & 50&0.25& 2,600 (5.2\%)\\
       & & & CINIC10 (CN10)  & 180,000 & 90,000 & 10  &50 &256&0.1& 4    & 50&0.25& 9,000 (5.0\%)\\
       & & & TinyImageNet (TI) & 100,000 & 10,000 & 200 &50 &128&0.1& 4   & 50&0.25& 5,000 (5.0\%)\\
       \cmidrule(lr){1-14}

       \multirow{3}{*}{DistilBERT} & \multirow{3}{*}{MLP2} & \multirow{3}{*}{MLP2}  & TREC & 5,452 & 500 & 6 & 20 &6&5e-5 & 10   & 20 &0.25& 270 (5.0\%)\\
        & & & AG's NEWS (NEWS) & 30,000 & 1,900 & 4 &20 &6&5e-5& 10   & 20&0.25& 1,500 (5.0\%)\\
        & & & IMDB   & 25,000 & 25,000 & 2 &20 &6&5e-5& 10   & 20&0.25& 1,250 (5.0\%)\\
	  \bottomrule
	\end{tabular}
    }
	\label{tab:models_and_datasets}
\end{table*}

\begin{algorithm}[t!]
    \small
    \caption{Secure Model Update}
    \label{algo:secure_model_update}
    \SetKwInOut{Input}{Input}
    \SetKwInOut{Output}{Output}
    \Input{Datasets $D_1,\cdots,D_K$, model parameters $\theta_1,\cdots,\theta_K, \theta_T$, masked layer index set $U_1,\cdots,U_{K-1}$, number of bottom model layers $L$, learning rate $\eta$.}
    \Output{Updated model parameters $\theta_1,\cdots \theta_K, \theta_T$.}
    \tcc{Secure forward}
    \For{each $k\in[1,K-1]$ in parallel}{
        $X_k \gets D_k$, $z_k \gets X_k$ \;
        \For{each layer $j\in[1,L]$}{
            \eIf{$j\in U_k$}{
                $\llbracket z_k \rrbracket_k, \llbracket z_k \rrbracket_K \gets \mathtt{Share}(z_k)$ \;
                \For{each $i$ from  $(k, K)$ in parallel}{
                    $\llbracket z_k \rrbracket_i \gets f(\llbracket z_k \rrbracket_i ; \llbracket \theta_k^j \rrbracket_i)$ \;
                }
                $z_k \gets \mathtt{Reconstruct}(\llbracket z_k \rrbracket_k, \llbracket z_k \rrbracket_K)$ \;
            }{
                $z_k \gets f(z_k ; \theta_k^j)$ \;
            }
        }
    }
    $X_K, Y \gets D_K$ \;
    $Z \gets z_1 \oplus \cdots \oplus z_{K-1} \oplus f(X_K ; \theta_K)$ \;
    $loss \gets \mathcal{L}(f(Z ; \theta_T), Y)$ \;
    
    \tcc{Secure backward}
    $\theta_T \gets \theta_T - \eta \frac{\partial loss}{\partial \theta_T}, \theta_K \gets \theta_K - \eta \frac{\partial loss}{\partial \theta_K}$ \;
    \For{each $k\in[1,K-1]$ in parallel}{
        $\nabla z_k \gets \frac{\partial loss}{\partial Z}$ \;
        \For{each layer $j\in[1,L]$}{
            \eIf{$j\in U_k$}{
                $\llbracket \nabla z_k \rrbracket_k, \llbracket \nabla z_k \rrbracket_K \gets \mathtt{Share}(\nabla z_k)$ \;
                \For{each $i$ from  $(k, K)$ in parallel}{
                    $\llbracket \nabla \theta_k^j \rrbracket_i \gets \frac{\partial f(\llbracket z_k \rrbracket_i; \llbracket \theta_k^j \rrbracket_i)}{\partial \llbracket \theta_k^j \rrbracket_i} \llbracket \nabla z_k \rrbracket_i$ \;
                    $\llbracket \nabla z_k \rrbracket_i \gets \llbracket \nabla z_k \rrbracket_i \frac{\partial f(\llbracket z_k \rrbracket_i; \llbracket \theta_k^j \rrbracket_i)}{\partial \llbracket z_k \rrbracket_i}$ \;
                    $\llbracket \theta_k^j \rrbracket_i \gets \llbracket \theta_k^j \rrbracket_i - \eta  \llbracket \nabla\theta_k^j \rrbracket_i$
                }
                $\nabla z_k \gets \mathtt{Reconstruct}(\llbracket \nabla z_k \rrbracket_k, \llbracket \nabla z_k \rrbracket_K)$ \;
            }{
                $\nabla \theta_k^j \gets \frac{\partial f(z_k;\theta_k^j)}{\partial \theta_k^j} \nabla z_k, \nabla z_k \gets \nabla z_k \frac{\partial f(z_k;\theta_k^j)}{\partial z_k}$ \;
                $\theta_k^j \gets \theta_k^j - \eta \nabla \theta_k^j$ \;
            }
        }
    }
    \Return{$\theta_1, \cdots, \theta_K, \theta_T$}
\end{algorithm}

\begin{algorithm}
    \small
    \caption{Shadow Model Update}
    \label{algo:shadow_model_update}
    \SetKwInOut{Input}{Input}
    \SetKwInOut{Output}{Output}
    \Input{Auxiliary datasets $D_1^{aux},\cdots,D_{K-1}^{aux}$,  shadow model parameters $\theta_{s,1},\cdots,\theta_{s,K-1}$, top model parameters $\theta_T$, ground truth label $Y$,  learning rate $\eta$.}
    \Output{Updated shadow model parameters $\theta_{s,1},\cdots,\theta_{s,K-1}$, gradients of shadow models $\nabla\theta_{s,1},\cdots,\nabla\theta_{s,K-1}$, }

    Frozen top model parameters $\theta_T$ \;
    \tcc{Forward}
    \For{each $k \in [1, K-1]$ in parallel}{
        $z_{s,k} = f(D_k^{aux} ; \theta_{s,k})$ \;
    }
    $Z_s = z_{s,1} \oplus z_{s,2} \oplus \cdots \oplus z_{s,K-1}$ \;
    $loss = \mathcal{L}\left(f(Z_s ; \theta_T), Y\right)$ \;
    \tcc{Backward}
    \For{each $k \in [1, K-1]$ in parallel}{
        $\nabla \theta_{s,k} \gets \frac{\partial loss}{\partial \theta_{s,k}} = \frac{\partial loss}{\partial Z_s} \frac{\partial Z_s}{\partial \theta_{s,k}}$ \;
        $\theta_{s,k} \gets \theta_{s,k} - \eta \nabla \theta_{s,k}$ \;
    }
    \Return{$\theta_{s,1},\cdots,\theta_{s,K-1}$ and $\nabla\theta_{s,1},\cdots,\nabla\theta_{s,K-1}$}
\end{algorithm}

\subsection{Experimental Setup Details} \label{appendix:configs}

\textbf{Details about Models and Datasets.}  
For both the active and passive parties' bottom models, we utilize five different model architectures: multi-layer perceptron (MLP3), LeNet5 \cite{lecun1998lenet}, VGG13 \cite{simonyan2014vgg}, ResNet18 \cite{he2016resnet}, and the Transformer-based DistilBERT \cite{sanh2019distilbert}. 
For the top model and the attacker's inference head model, we employ MLP2.
Each type of bottom model architecture is evaluated on multiple datasets, as illustrated in Table \ref{tab:models_and_datasets}:
\begin{itemize}
    \item MLP3: We utilize the TabMNIST (abbreviated as TM), TabFMNIST (TFM), and CRITEO (CR) \cite{criteo_dataset} datasets. TabMNIST and TabFMNIST are constructed by flattening the original \(28 \times 28\) grayscale images from the MNIST \cite{mnist_dataset} and FMNIST \cite{fmnist_dataset} datasets into 784-dimensional pixel vectors. The CRITEO dataset consists of 13 numerical features and 26 categorical features, commonly used for binary click-through rate prediction. For our experiments, we randomly select 600,000 samples from the original dataset.
    
    \item LeNet5: We utilize the MNIST (M), FMNIST (FM), and SVHN \cite{svhn_dataset} datasets, all designed for 10-class classification tasks. MNIST and FMNIST consist of grayscale images, while SVHN comprises colored images.
    
    \item VGG13 and ResNet18: We employ the CIFAR10 (CF10), CIFAR100 (CF100) \cite{cifar10_dataset}, CINIC10 (CN10) \cite{cinic10_dataset}, and TinyImageNet (TI) \cite{tiny_imagenet_dataset} datasets. CIFAR10 and CIFAR100 each contain 60,000 \(32 \times 32\) color images. CIFAR10 has 10 classes, while CIFAR100 includes 100 classes. CINIC10 serves as a substitute for CIFAR10, constructed by extending CIFAR10 with additional down-sampled ImageNet images. TinyImageNet contains 110,000 images across 200 classes (550 images per class), each downsized to \(64 \times 64\) color images, with 500 training images and 50 test images per class.
    
    \item DistilBERT: We use datasets for common text classification tasks, including TREC \cite{trec_dataset} for question classification, AG's News (NEWS) \cite{ag_news_dataset} for topic classification, and IMDB \cite{imdb_dataset} for sentiment analysis. For the AG's News dataset, we randomly select 30,000 training samples and 1,900 test samples from the original dataset for simplification.
\end{itemize}

\textbf{Details about Hyperparameters.}  
\ul{\emph{(1) VFL Training.}}
For MLP3, LeNet5, VGG13, and ResNet18, both the VFL training epochs and the MC attack fine-tuning epochs are set to 50.
For DistilBERT, these are set to 20.
Additionally, the VFL training batch size and learning rate are provided in Table \ref{tab:models_and_datasets}.

\ul{\emph{(2) MC Attack.}}
In the MC attack experiments, the number of known labels per class assigned to the attacker follows the specifications in Table \ref{tab:models_and_datasets}, except for the experiments detailed in Section \ref{sec:tunable_privacy_protection} and the ablation studies on \emph{Auxiliary Dataset Size} and \emph{Auxiliary Dataset Distribution}, where 50 labeled samples per class are assigned to the attacker for both the ResNet18 and VGG models.

\ul{\emph{(3) VMask.}}
For VMask, the noise standard variance \(\sigma\) (as introduced in Algorithm \ref{algo:layer_masking}) is set to 0.01, and the locally simulated MC attack training epochs are fixed at 20 across all experiments. 
Details about the privacy budget, auxiliary dataset size, and the ratio of the auxiliary dataset size to the passive party's original dataset size are provided in Table \ref{tab:models_and_datasets}.
We consider all linear layers in MLP3, LeNet5, VGG13, and ResNet18 for selection, whereas in DistilBERT (which contains six attention blocks in total), we restrict VMask to selecting only the linear layers in the first two attention blocks.

\ul{\emph{(4) Other Defense Methods.}}
For other defense methods, the hyperparameters are configured as follows.
For NG, CG, and DG, we set noise scale to \(1 \times 10^{-3}\), preserved ratio to 0.2, and the number of discrete bins to 60. 
For FedPass, the mean and variance of the Gaussian distribution used for key generation are set to -1.0 and 0.5, respectively. 
In LabelDP, the privacy budget is set to 3. 
For KD\(k\), the smoothing parameter \(\epsilon\) is fixed at 0.45, with \(k\) set to 3 for \(k\)-anonymity, in accordance with the original paper. 
Lastly, for dCor and MID, the distance correlation weight and the mutual information weight are set to 0.15 and \(5 \times 10^{-5}\), respectively.

\subsection{Additional Experimental Results}
\subsubsection{VMask's Effectiveness on CINIC10 Dataset} \label{appendix:effectiveness_cinic10}
The evaluation results are presented in Table \ref{tab:appendix_effectiveness_cinic10}.
VMask continues to achieve the best privacy-utility trade-off compared to other defense methods.
For the VGG13 and ResNet18 models, the averaged VFL model accuracy drop is only 0.04\% and 0.07\%, respectively, which is significantly lower than other methods.
For instance, the averaged main task accuracy drop from DG is much higher at 15.35\%.
Moreover, the averaged attack accuracy from VMask is only 2.96\% higher than the ``Scratch'' baseline in ResNet18 model.
Notably, in VGG13 model, the attack accuracy is even 0.32\% lower than the ``Scratch'' baseline.
This suggests that, for the attacker, the bottom model trained with VMask is even harder to fine-tune than a completely randomized bottom model.

\begin{table}
	\caption{
		Comparison of the effectiveness of defense methods in defending against MC attack while maintaining main task accuracy, evaluated under the (VGG13, CN10) and (ResNet18, CN10) configurations.
	}
	\label{tab:appendix_effectiveness_cinic10}
	\centering
	\resizebox{\linewidth}{!}{%
	\begin{threeparttable}[t]
		\footnotesize
		\begin{tabular}{c||c||c|c||c|c}
			\toprule
			\multirow{2}{*}{Metric} & \multirow{2}{*}{Method}  &  \multicolumn{1}{c|}{VGG13} & \multirow{2}{*}{\makecell{Avg.\\ Diff.}} & \multicolumn{1}{c|}{ResNet18} & \multirow{2}{*}{\makecell{Avg.\\ Diff.}} \\
		  	& & CN10 & & CN10 &  \\
			\midrule
			\multirow{14}{*}{\makecell{Main\\Accuracy(\%)}} & Alone   & 77.50 & -7.12 & 80.83 & -6.02 \\
			& Vanilla  & 84.62 &0 & 86.85 & 0\\
			
			\cmidrule(r){2-6}
			& NG \cite{fu2022label}       & 84.57 & -0.12 & 86.78 & -0.07\\
			& CG \cite{fu2022label}        & 78.28 & -6.34  &80.84 & -6.01\\
			& DG \cite{fu2022label}        &69.27  & -15.35 &80.88 & -5.97\\
			& FedPass \cite{gu2023fedpass}  &84.47 & -0.15 &86.66 & -0.19 \\
			& LabelDP \cite{ghazi2021labeldp}   &82.12 &-2.50 &84.18 &-2.67\\
			& \add{KD$k$} \cite{arazzi2024kdk}     & \add{79.21} & -5.41&\add{82.61} &-4.24 \\
			& dCor \cite{sun2022label}     & 78.27 &  -6.35& 81.14 & -5.71\\
			& MID \cite{zou2023mutual}     &79.83 & -4.79&81.85 & -5.00 \\

			\cmidrule(r){2-6}
            & VMask-RS  & 83.86 &-0.76 & 85.98 & -0.87\\
            & \textbf{VMask-AS}  & 84.67 & \textbf{+0.05} & 86.74 &  \textbf{-0.11}\\
            & \textbf{VMask-ALLS} &  84.60 &  \textbf{-0.02} & 86.77 & \textbf{-0.08} \\
			& \textbf{VMask} & 84.58 & \textbf{-0.04} & 86.78 & \textbf{-0.07} \\

			\hhline{======}

			\multirow{14}{*}{\makecell{Attack \\Accuracy(\%)}} & Scratch &17.79 & 0& 14.67&0\\  
			& Vanilla  & 60.92 & +43.13 & 59.68&+45.01\\  
			
			\cmidrule(r){2-6}
			& NG \cite{fu2022label}     &58.67 & +40.88 &57.23 & +42.56\\
			& CG \cite{fu2022label}      &20.51 &+2.72 &23.32 &+8.65\\
			& DG \cite{fu2022label}      &17.37 & -0.42&17.35 & +2.68 \\
			& FedPass \cite{gu2023fedpass}  &59.86 & +42.07&57.61 &+42.94\\
			& LabelDP \cite{ghazi2021labeldp}  &+36.51& 54.30& 52.75&+38.08\\
			& \add{KD$k$} \cite{arazzi2024kdk}      &\add{48.39} & +30.60&\add{49.76} &+35.09 \\
			& dCor \cite{sun2022label}    & 18.84 &+1.05 &20.95 & +6.28\\
			& MID \cite{zou2023mutual}     &27.08 & +9.29&24.33 & +9.66\\

			\cmidrule(r){2-6}
            & VMask-RS    & 20.06 & +2.27 & 26.80 & +12.13\\
            & \textbf{VMask-AS}    & 16.50 &  \textbf{-1.29} & 15.68 & \textbf{+1.01} \\
            & \textbf{VMask-ALLS}  & 11.70 & \textbf{-6.09} & 11.65 & \textbf{-3.02}\\
			& \textbf{VMask}       & 17.47 & \textbf{-0.32} & 17.63 & \textbf{+2.96}  \\

			\bottomrule
		\end{tabular}
		
	\end{threeparttable}
	}

\end{table}

\subsubsection{Visualization of Feature Embedding  Under Different Privacy Budget} \label{appendix:vis_embed}
\begin{figure}[t]
	\centering
	\begin{subfigure}[b]{0.15\textwidth} 
		\includegraphics[width=\textwidth]{figures/embed_cifar10_resnet18.pdf}
		\caption{Vanilla}
		\label{embedding_vmask_vanilla}
	\end{subfigure}
	\hfill
	\begin{subfigure}[b]{0.15\textwidth} 
		\includegraphics[width=\textwidth]{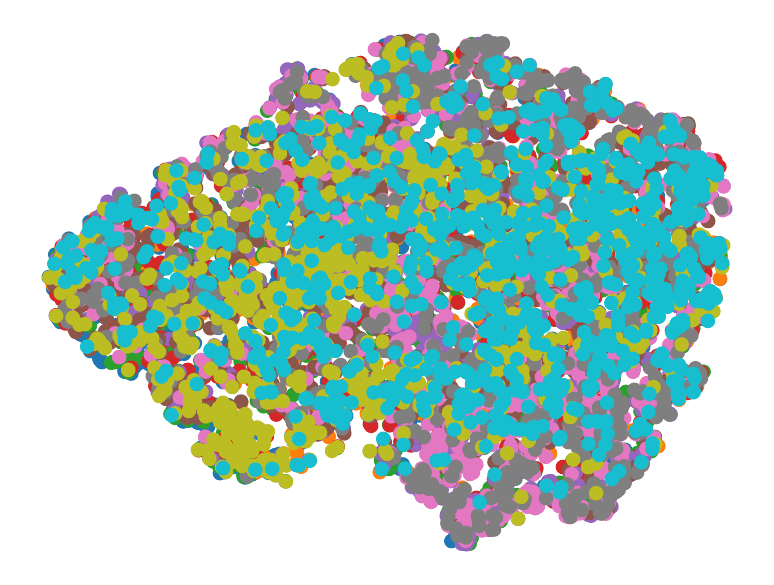}
		\caption{$\epsilon=0.6$}
		\label{embedding_vmask_epsilon_1}
	\end{subfigure}
	\hfill
	\begin{subfigure}[b]{0.15\textwidth}
		\includegraphics[width=\textwidth]{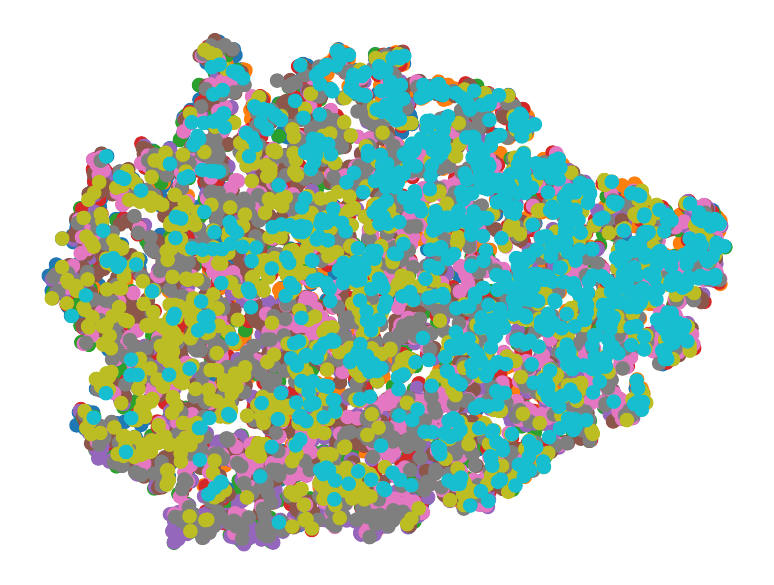}
		\caption{$\epsilon=0.5$}
		\label{embedding_vmask_epsilon_2}
	\end{subfigure}
	\hfill
	\begin{subfigure}[b]{0.15\textwidth}
		\includegraphics[width=\textwidth]{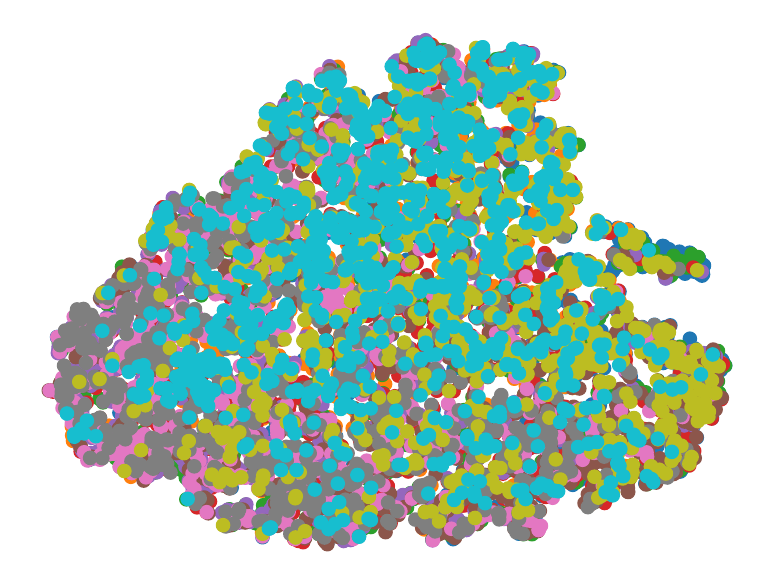}
		\caption{$\epsilon=0.4$}
		\label{embedding_vmask_epsilon_3}
	\end{subfigure}
	\hfill
	\begin{subfigure}[b]{0.15\textwidth}
		\includegraphics[width=\textwidth]{figures/embed_cifar10_resnet18_vmask_0.3_4000_layer1,3,6.pdf}
		\caption{$\epsilon=0.3$}
		\label{embedding_vmask_epsilon_4}
	\end{subfigure}
	\hfill
	\begin{subfigure}[b]{0.15\textwidth}
		\includegraphics[width=\textwidth]{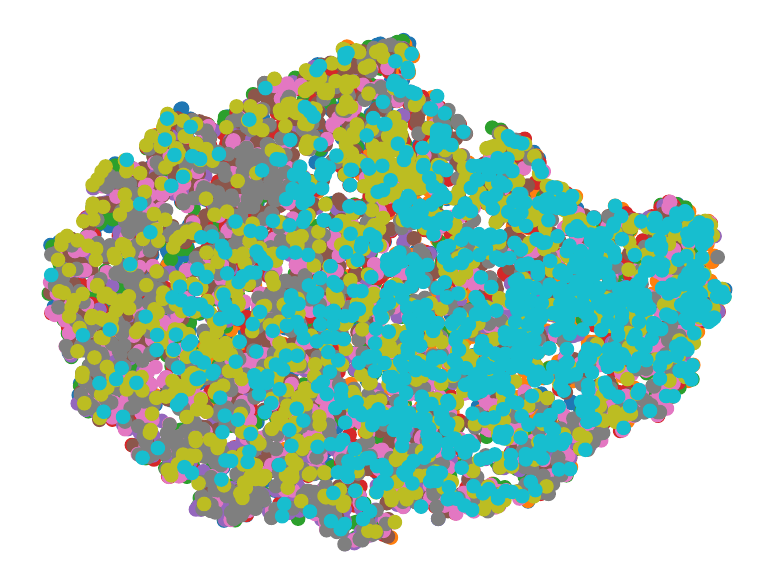}
		\caption{$\epsilon=0.2$}
		\label{embedding_vmask_epsilon_5}
	\end{subfigure}

	\caption{Feature embedding distribution of VMask with respect to different privacy budgets under the (ResNet18, CF10) configuration.}
	\label{fig:embedding_vmask}
\end{figure}

We conduct experiments in the case of (ResNet18, CIFAR10). We visualize the embedding distribution with respect to different privacy budgets of VMask, as shown in Figure \ref{fig:embedding_vmask}. 
As a comparison, we also visualize the embedding distribution under vanilla VFL, shown in Figure \ref{embedding_vmask_vanilla}. 
We observe that the bottom model under vanilla VFL learns better in separating samples from each class (each class cluster is well-separated). 
However, under VMask, even with a relatively high privacy budget (e.g., \( \epsilon = 0.6 \)), the bottom mo
del fails to learn a good feature representation, and the clusters of each class overlap with each other. 
Such a masked bottom model is difficult for the attacker to fine-tune with limited labeled samples.

\subsubsection{Additional Ablation Study Results}
\label{appendix:ablation}



\begin{figure}
    \captionsetup[subfigure]{aboveskip=-0.5pt,belowskip=-0.5pt}
	\centering
	\begin{minipage}{1.0\linewidth}
		\centering
		\includegraphics[width=0.9\linewidth, trim=50 500 50 0, clip]{figures/ablation_aux_samples_legend.pdf}
	\end{minipage}

	\vspace{-2pt}

    \begin{minipage}{1.0\linewidth}
        \centering
        \includegraphics[width=0.7\linewidth]{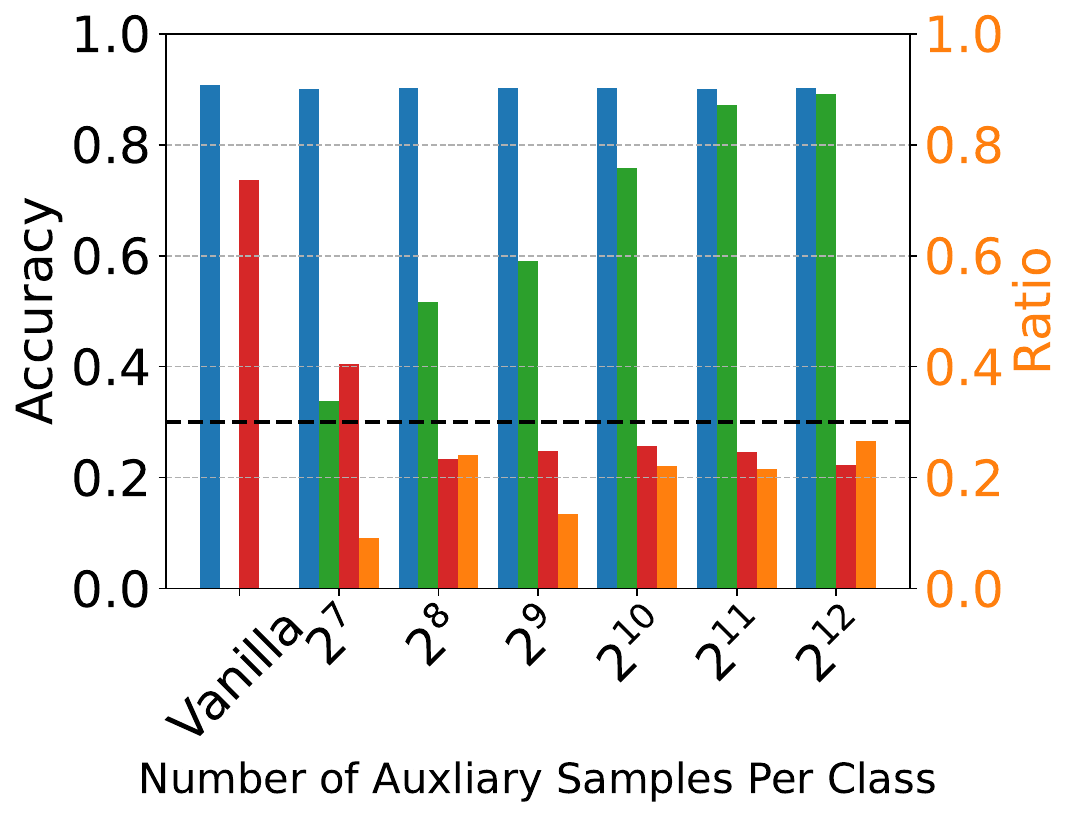}
    \end{minipage}
    
    \caption{Impact of auxiliary dataset size under the (VGG13, CF10) configuration.}
    \label{fig:ablation_appendix_number_aux_samples}
\end{figure}


\begin{figure}[t]
	\centering
	\begin{subfigure}[b]{0.235\textwidth} 
		\includegraphics[width=\textwidth]{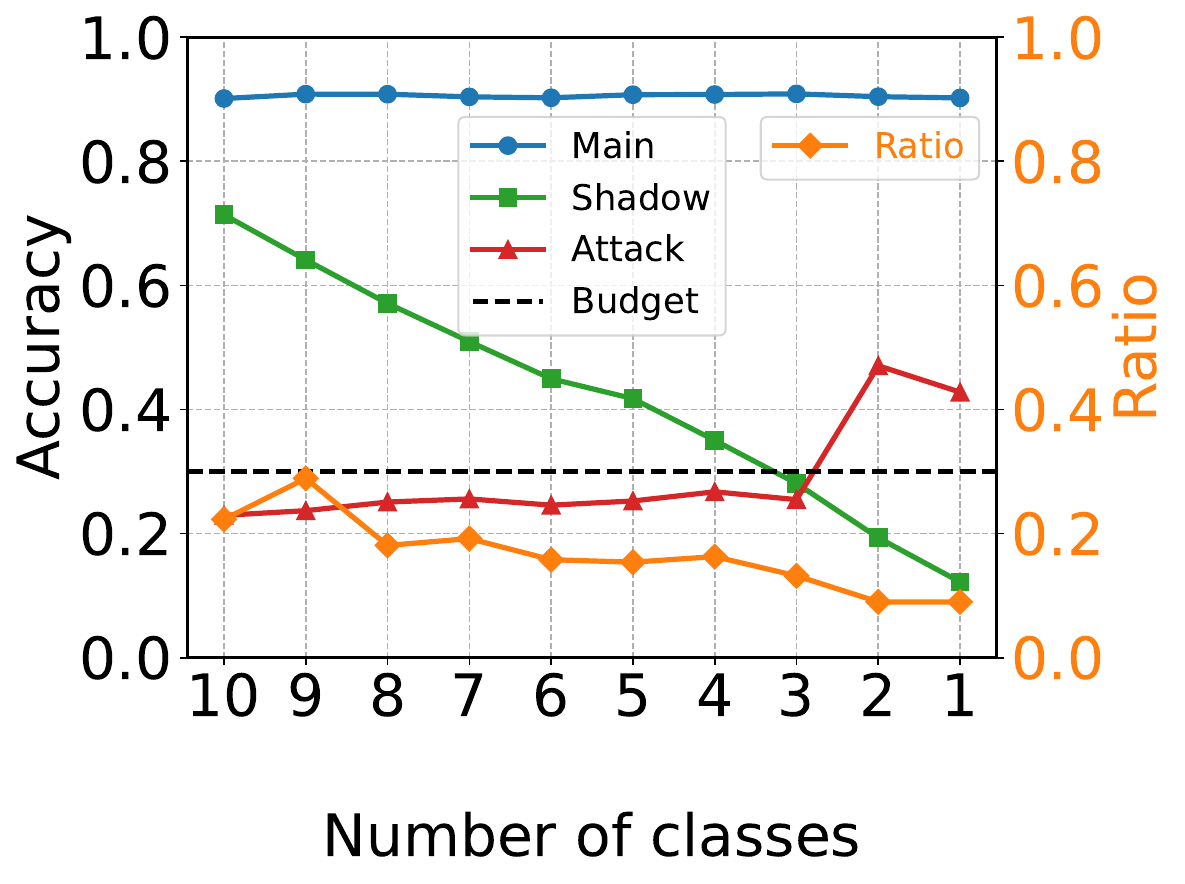}
		\caption{Label non-IID}
		\label{ablation_label_noniid_vgg}
	\end{subfigure}\hfill
	\begin{subfigure}[b]{0.23\textwidth}  
		\includegraphics[width=\textwidth]{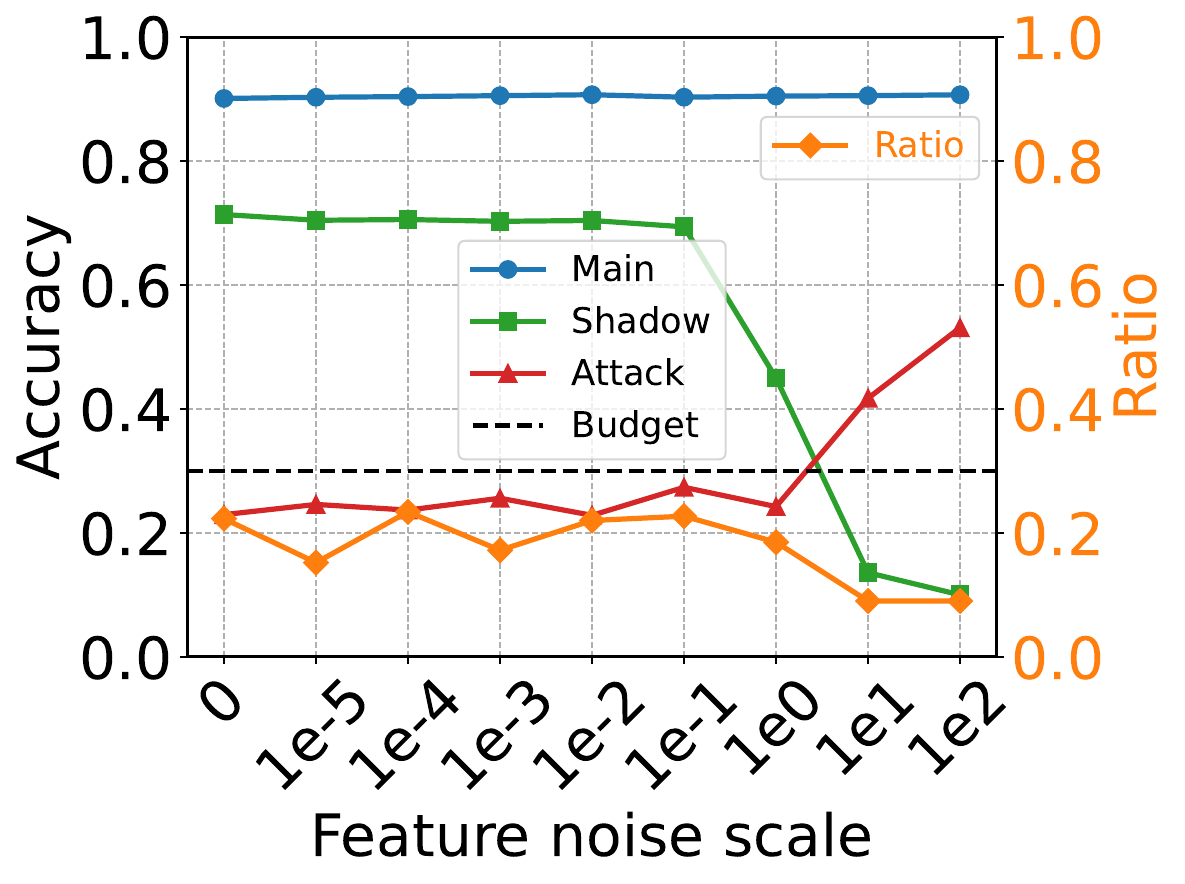}
		\caption{Feature non-IID}
		\label{ablation_feature_noniid_vgg}
	\end{subfigure}\hfill
    \caption{Impact of non-IID auxiliary dataset  under the (VGG13, CF10) configuration.}
	\label{ablation_appendix_aux_dataset_distribution}
\end{figure}

We provide additional ablation study results conducted under other configurations in this section.


\textbf{Impact of Auxiliary Dataset Size.}
Experiments are conducted under the (VGG13, CF10) configuration with the privacy budget fixed at 0.3.  
As illustrated in Figure ~\ref{fig:ablation_appendix_number_aux_samples}, the trends in main task accuracy, attack accuracy, shadow model accuracy, and mask ratio align with the conclusions drawn in Section \ref{sec:ablation}.
Specifically, when the number of auxiliary samples per class reaches $2^8$, the auxiliary dataset constitutes only 5\% of the original passive party's dataset (50,000 samples), and the attack accuracy is reduced to 0.23, which is lower than the 0.3 privacy budget.
These results further confirm that the MC attack accuracy can be effectively controlled within the privacy budget, even with a small auxiliary dataset.

\textbf{Impact of Auxiliary Dataset Distribution.}
Experiments are conducted under the (VGG13, CF10) configuration with the privacy budget fixed at 0.3. 
As shown in Figure \ref{ablation_appendix_aux_dataset_distribution}, the results are consistent with those from the (ResNet18, CF10) configuration. 
As the degree of non-IID increases, the shadow model accuracy gradually decreases in both the label non-IID and feature non-IID scenarios.
Despite this, VMask remains effective in identifying critical layers for masking, successfully controlling the attack accuracy within the defined privacy budget. 
For instance, in the label non-IID scenario, even with only 3 classes in the auxiliary dataset, the attack accuracy is 25.47\%, which is below the privacy budget of 0.3.

\subsection{Rationale of Layer Selection Criterion}
\label{appendix:vis_accumu_grad_norm}
We utilize the accumulative gradient norm as a criterion for selecting critical layers. 
Below, we provide both a theoretical proof and empirical experimental results to demonstrate the Rationale of this criterion.

\subsubsection{Proof}
For a passive party $P_k, k\in[1, K-1]$, we define the accumulative gradient norm of the $j$-th layer $\theta_k^j$ of its bottom model $\theta_k$ as 
\begin{equation}
    G_k^j = \sum_{t=1}^{T} |\nabla \theta_{k,t}^{j}|,
\end{equation}
where $T$ is the total number of training epochs.
For simplicity, we denote $g_{k,t}^j = |\nabla \theta_{k,t}^{j}|$.
At epoch $t$, the change in the layer parameters $\theta_{k}^{j}$ is $\Delta \theta_{k}^{j}$, and the impact on the loss function is 
\begin{equation}
    \Delta L_t \approx g_{k,t}^j \Delta \theta_{k}^{j}.
\end{equation}
Then the accumulative change in the loss function due to layer parameters $\theta_{k}^{j}$ over the training history is

\begin{equation}
    \label{eq:proof}
    \begin{aligned}
        \Delta L_{\text{total}} = \sum_{t=1}^{T} \Delta L_t &\approx \sum_{t=1}^{T} g_{k,t}^j \Delta \theta_{k}^j \\
        &= T \Delta \theta_{k}^j \sum_{t=1}^{T} g_{k,t}^j \\
        &= T \Delta \theta_{k}^j G_k^j.
    \end{aligned}
\end{equation}

From Equation \ref{eq:proof}, it is evident that a layer parameter $\theta_k^j$ with a larger accumulative gradient norm $G_k^j$ has a greater impact on the changes in the loss function over the entire training process.
Therefore, the larger the accumulative gradient norm, the more critical the layer parameters.
When such a layer is masked, it significantly impacts the reduction of the MC attack accuracy.
 
\begin{figure}[t]
	\centering
	\begin{subfigure}[b]{0.23\textwidth} 
		\includegraphics[width=\textwidth]{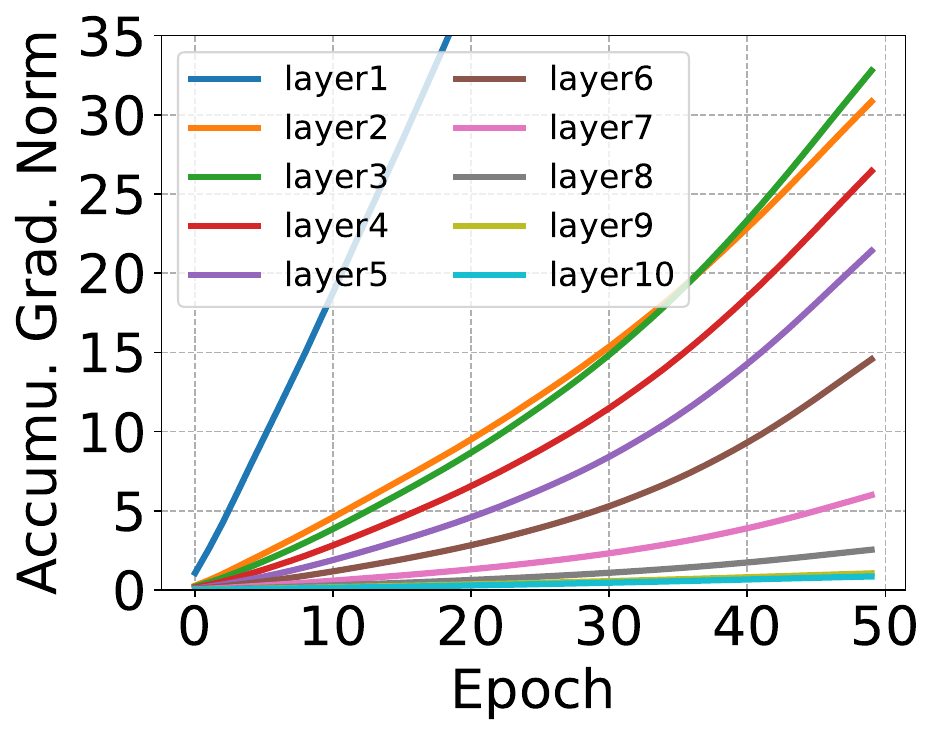}
		\caption{VGG13, CF10}
		\label{grad_norm_vgg13}
	\end{subfigure}\hfill
	\begin{subfigure}[b]{0.23\textwidth}  
		\includegraphics[width=\textwidth]{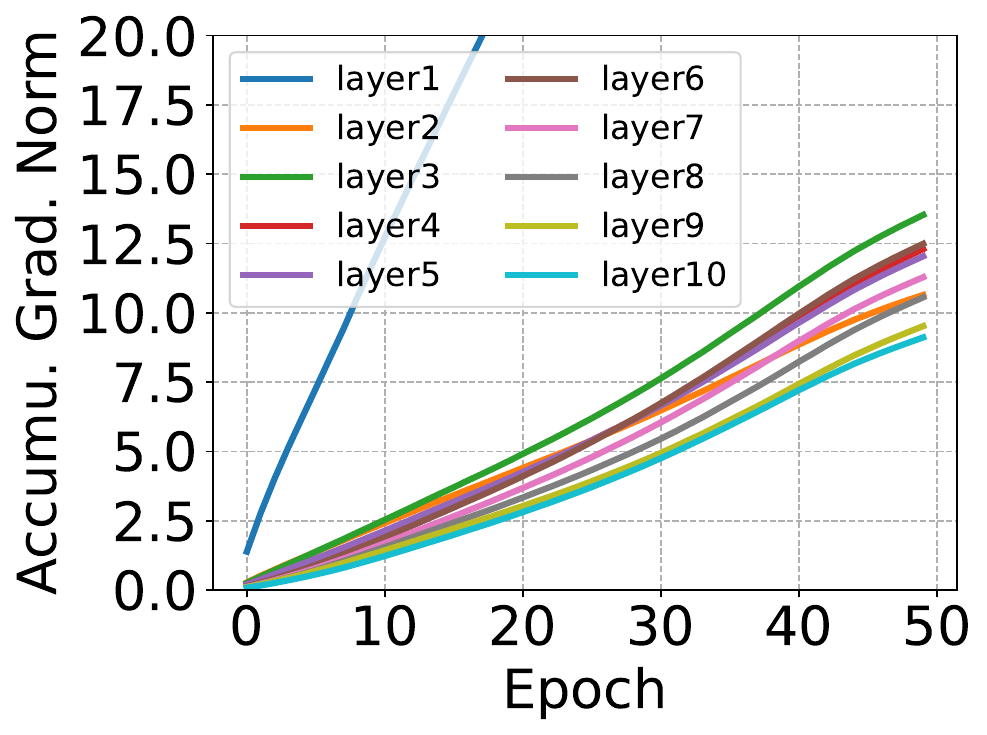}
		\caption{ResNet18, CF10}
		\label{grad_norm_resnet18}
	\end{subfigure}\hfill
    \caption{Visualization of the accumulated gradient norm for each layer across every training epoch.}
	\label{fig:accumu_grad_norm}
\end{figure}

\subsubsection{Empirical Experimental Results}
We conduct experiments in a two-party VFL setting using two configurations: (VGG13, CIFAR10) and (ResNet18, CIFAR10). 
We plot the accumulated gradient norm of the first ten layers of the passive party's bottom model, as shown in Figure \ref{fig:accumu_grad_norm}. 
We observe that the top two layers with the largest accumulated gradient norms are non-adjacent. 
For both VGG13 and ResNet18, the top two layers are Layer 1 and Layer 3. 
This finding demonstrates that using the accumulated gradient norm as the layer selection criterion allows us to effectively select non-adjacent layers for masking, thereby reducing attack accuracy more effectively, as discussed in Observation 3 in Section \ref{sec:layer_masking}. 
Additionally, we note that the accumulated gradient norm of the first layer is significantly larger than that of all other layers. 
This observation justifies our choice to mask the first layer during the initial epoch at the initialization stage.

\end{document}